\documentclass{aa}
\usepackage{CJK}
\usepackage{xcolor}
\usepackage{enumitem}
\usepackage{verbatim}
\usepackage{amsmath}
\usepackage[percent]{overpic}
\usepackage[]{geometry}
\usepackage[verbose]{placeins}

\newcommand{\XMM}{ XMM-{\em Newton}\xspace}
\newcommand{\Chandra}{{\em Chandra}\xspace}
\newcommand{\eROSITA}{{eROSITA}\xspace}
\newcommand{\Gaia}{{\em Gaia}\xspace}
\newcommand{\SRG}{{\em SRG}\xspace}
\newcommand{\egs}{erg~cm$^{-2}$~s$^{-1}$}
\newcommand{\NH}{$N_\textrm{H}$\xspace}
\newcommand{\arcdeg}{\mbox{$^\circ$}\xspace}

\renewcommand*\arcmin{\ensuremath{^\prime}\xspace}
\renewcommand*\arcsec{\ensuremath{^{\prime\prime}}\xspace}

\newcommand{\A}[1]{\AA{}}
\newcommand{\TL}[1]{#1}
\newcommand{\Edit}[1]{#1}
\newcommand{\Final}[1]{#1}

\begin{document}
\begin{CJK*}{UTF8}{gkai}

\title{The \eROSITA extragalactic CalPV serendipitous catalog\thanks{The catalog is available at the CDS via anonymous ftp to cdsarc.u-strasbg.fr (130.79.128.5) or via http://cdsarc.u-strasbg.fr/viz-bin/qcat?J/A+A/}}
\author{Teng~Liu (刘腾)\inst{\ref{in:mpe}}
Andrea~Merloni\inst{\ref{in:mpe}}\and
Julien~Wolf\inst{\ref{in:mpe}}\and
Mara~Salvato\inst{\ref{in:mpe}}\and
Thomas~H.~Reiprich\inst{\ref{in:bonn}}\and
Johan~Comparat\inst{\ref{in:mpe}}\and
Riccardo~Arcodia\inst{\ref{in:mpe}}\and
Georg~Lamer\inst{\ref{in:aip}}\and
Antonis~Georgakakis\inst{\ref{in:athens}}\and
Tom~Dwelly\inst{\ref{in:mpe}}\and
Jeremy~Sanders\inst{\ref{in:mpe}}\and
Johannes~Buchner\inst{\ref{in:mpe}}\and
Frank~Haberl\inst{\ref{in:mpe}}\and
Miriam~E.~Ramos-Ceja\inst{\ref{in:mpe}}\and
J\"orn~Wilms\inst{\ref{in:remeis}}\and
Kirpal~Nandra\inst{\ref{in:mpe}}\and
Hermann~Brunner\inst{\ref{in:mpe}}\and
Marcella~Brusa\inst{\ref{in:bologna1},\ref{in:bologna2}}\and
Axel~Schwope\inst{\ref{in:aip}}\and
Jan~Robrade\inst{\ref{in:hamburg}}\and
Michael~Freyberg\inst{\ref{in:mpe}}\and
Thomas~Boller\inst{\ref{in:mpe}}\and
Chandreyee~Maitra\inst{\ref{in:mpe}}\and
Angie~Veronica\inst{\ref{in:bonn}}\and
Adam~Malyali\inst{\ref{in:mpe}}
}
\institute{Max-Planck-Institut f\"ur extraterrestrische Physik, Giessenbachstra{\ss}e 1, D-85748 Garching bei M\"unchen, Germany\\ \email{Teng Liu<liu@mpe.mpg.de>}\label{in:mpe}
\and Argelander-Institut f\"ur Astronomie, Rheinische Friedrich-Wilhelms-Universit\"at Bonn, Auf dem H\"ugel 71, 53121 Bonn, Germany\label{in:bonn}
\and Leibniz-Institut f\"ur Astrophysik, An der Sternwarte 16, 14482 Potsdam, Germany\label{in:aip}
\and Institute for Astronomy and Astrophysics, National Observatory of Athens, V. Paulou and I. Metaxa 11532, Greece\label{in:athens}
\and Dr.~Karl Remeis-Sternwarte \& Erlangen Centre for Astroparticle Physics, Sternwartstr.~7, 96049 Bamberg, Germany \label{in:remeis}
\and Dipartimento di Fisica e Astronomia "Augusto Righi", Universit\`a di Bologna,  via Gobetti 93/2,  40129 Bologna, Italy\label{in:bologna1}
\and INAF - Osservatorio di Astrofisica e Scienza dello Spazio di Bologna, via Gobetti 93/3,  40129 Bologna, Italy\label{in:bologna2}
\and Hamburger Sternwarte, University of Hamburg, Gojenbergsweg 112, 21029 Hamburg, Germany\label{in:hamburg}
}

  \date{} 
  \abstract
  {The \eROSITA X-ray telescope on board the \Final{Spectrum-Roentgen-Gamma (\SRG)} observatory performed \Final{calibration and performance verification} (CalPV) observations between September 2019 and December 2019, ahead of the planned four-year all-sky surveys. Most of them were deep, pointing-mode observations.}
   {We present here the X-ray catalog detected from the set of extra-galactic CalPV observations released to the public by the German eROSITA consortium, and the multiband counterparts of these X-ray sources.}
   {We \Final{developed} a source detection method optimized for point-like X-ray sources by including extended X-ray emission in the background measurement. The multiband counterparts \Final{were} identified using a Bayesian method from the CatWISE catalog.}
   {Combining 11 CalPV fields, we present a catalog containing $9515$ X-ray sources, whose X-ray fluxes \Final{were} measured through spectral fitting. CatWISE counterparts are presented for 77\% of the sources. Significant variabilities are found in $99$ of the sources, which are also presented with this paper. Most of these fields show similar number counts of point sources as typical extragalactic fields, and a few harbor particular stellar populations.
     }
   {}

\titlerunning{Extragalactic CalPV Catalog}
\authorrunning{Liu et al.}
   \maketitle
\end{CJK*}

\section{Introduction}
The Russian-German Spectrum-Roentgen-Gamma (\SRG) observatory \citep{Sunyaev2021} was successfully launched on 13 July 2019, and subsequently brought into a large halo orbit around the Lagrangian point L2 of the Sun-Earth system.
\SRG's largest instrument is the \eROSITA X-ray telescope \citep{Predehl2021}, the main mission of which, \Final{that is to say}, a deep all-sky survey, started in December 2019 and will take four years to scan the full sky eight times.
Ahead of this main phase of the mission, the commissioning, calibration, and performance verification (CalPV) programs were executed.
A variety of targets were observed during the CalPV phase, including galaxy clusters, star clusters, active galactic nuclei (AGN), X-ray binaries, supernova remnants (SNR), \Final{among others}.
Thanks to the large field of view (FOV; diameter $\sim 1\arcdeg$) of \eROSITA, the CalPV observations result in a rich archive of X-ray data.

For two contiguous-field surveys in the performance verification (PV) phase, \Final{that is}, the \eROSITA Final Equatorial Depth Survey (eFEDS; $\sim 20\arcdeg \times 7\arcdeg$) and the Eta Chamaeleontis (star cluster) survey ($\sim 5\arcdeg\times 5\arcdeg$), the X-ray catalogs are presented in \citet{Brunner2021} and \citet{Robrade2021}, respectively.
In this work, we present the serendipitous X-ray survey based on the extragalactic (Galactic latitude $>$17$\arcdeg$) CalPV observations whose data rights are attributed to the \eROSITA German consortium, \Final{that is they are} either located in the German hemisphere (Galactic longitude $>$180$\arcdeg$) or \Final{correspond} to a project led by the \eROSITA German consortium.
The X-ray catalog detected from the galactic (Galactic latitude $<$17$\arcdeg$) CalPV observations \Final{will be} presented in Lamer et al. (in prep.).

\TL{This paper is organized as follows. Sect.~\ref{sec:data} introduces the data and the astrometric correction to the data. Sect.~\ref{sec:method} describes our source detection method that is optimized for the detection of point sources. Sect.~\ref{sec:results} presents the catalog and its properties, including source fluxes, variabilities, counterpart identification, and point-source number counts in particularly selected subsurvey regions. Sect.~\ref{sec:discussion} discusses the AGN \Final{and} star populations in all the fields through multiband counterparts of the sources and the source detection efficiency through a 100ks simulation of a pointing-mode \eROSITA observation. At last, a summary is given in Sect.~\ref{sec:conclusion}.}

\section{Data}
\label{sec:data}

\begin{figure*}[t]
\begin{center}
  \begin{minipage}{\textwidth}
  \centering
  \begin{overpic}[width=0.285\textwidth]{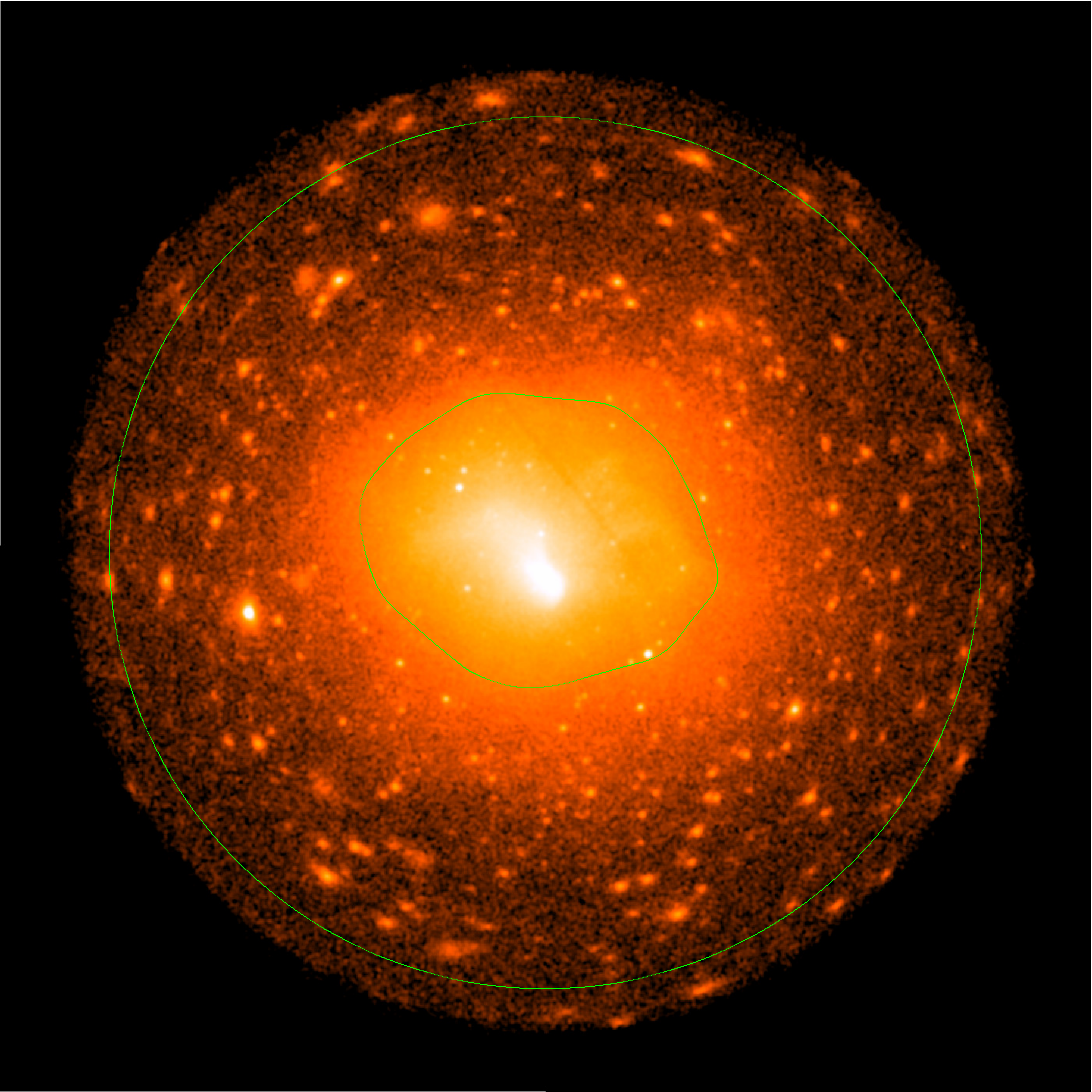}
    \put (2,2) {\large\color{white} A3266}
  \end{overpic}
  \hspace*{-0.07in}
  \begin{overpic}[width=0.285\textwidth]{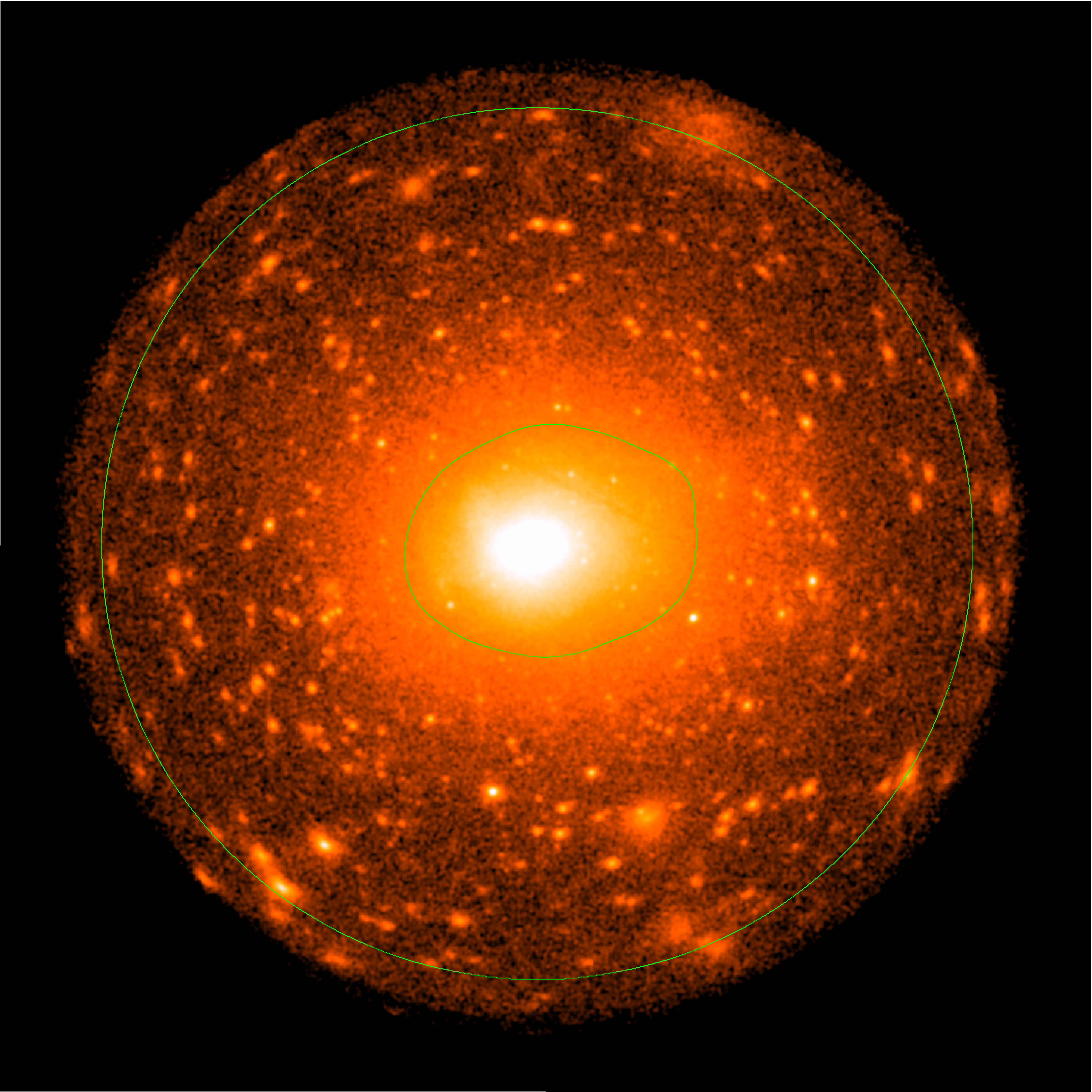}
    \put (2,2) {\large\color{white} A3158}
  \end{overpic}
  \hspace*{-0.07in}
  \begin{overpic}[width=0.38\textwidth]{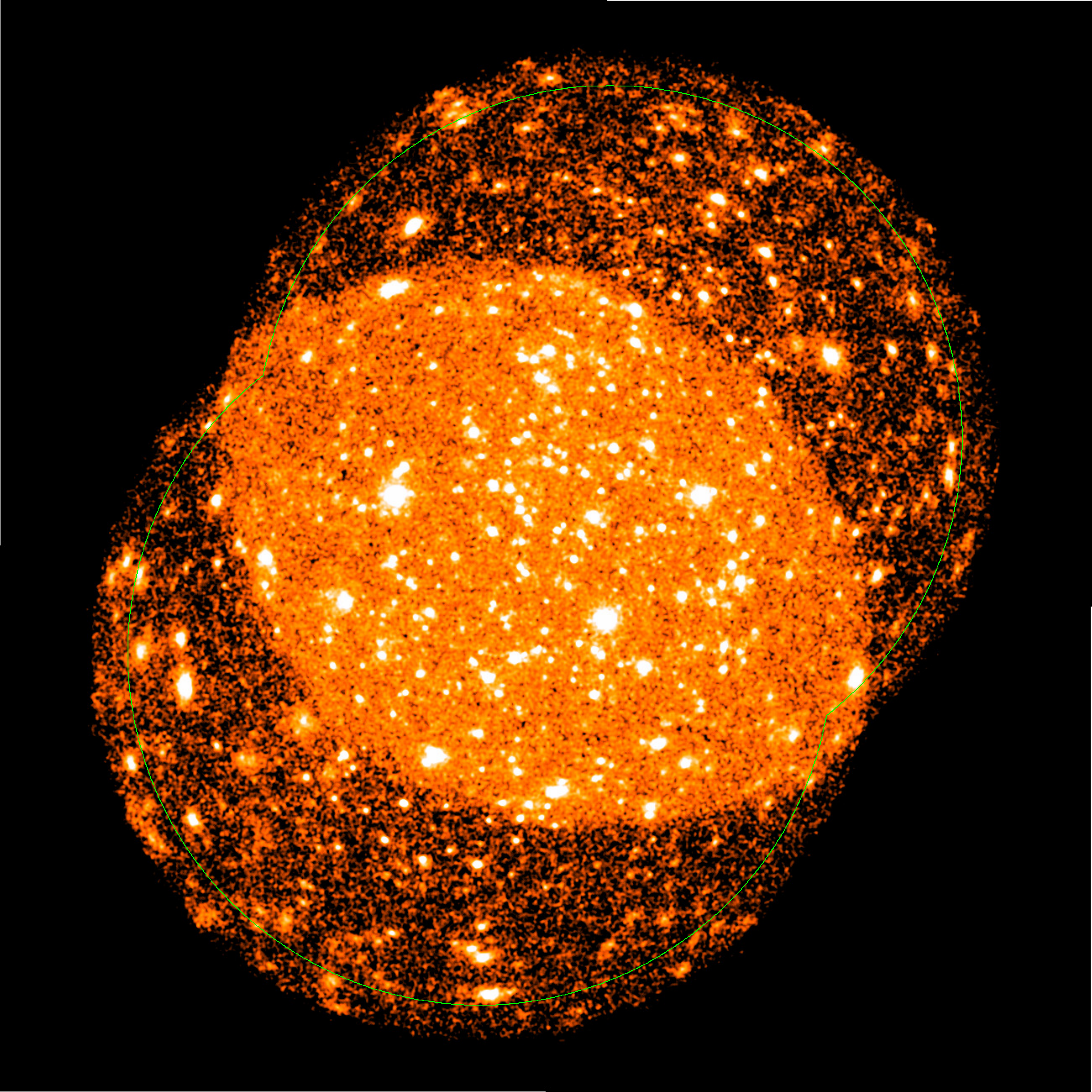}
    \put (2,2) {\large\color{white} J2334}
  \end{overpic}
  \end{minipage}
  \hspace{0.5in}

  \begin{minipage}{\textwidth}
  \centering
  \begin{overpic}[width=0.285\textwidth]{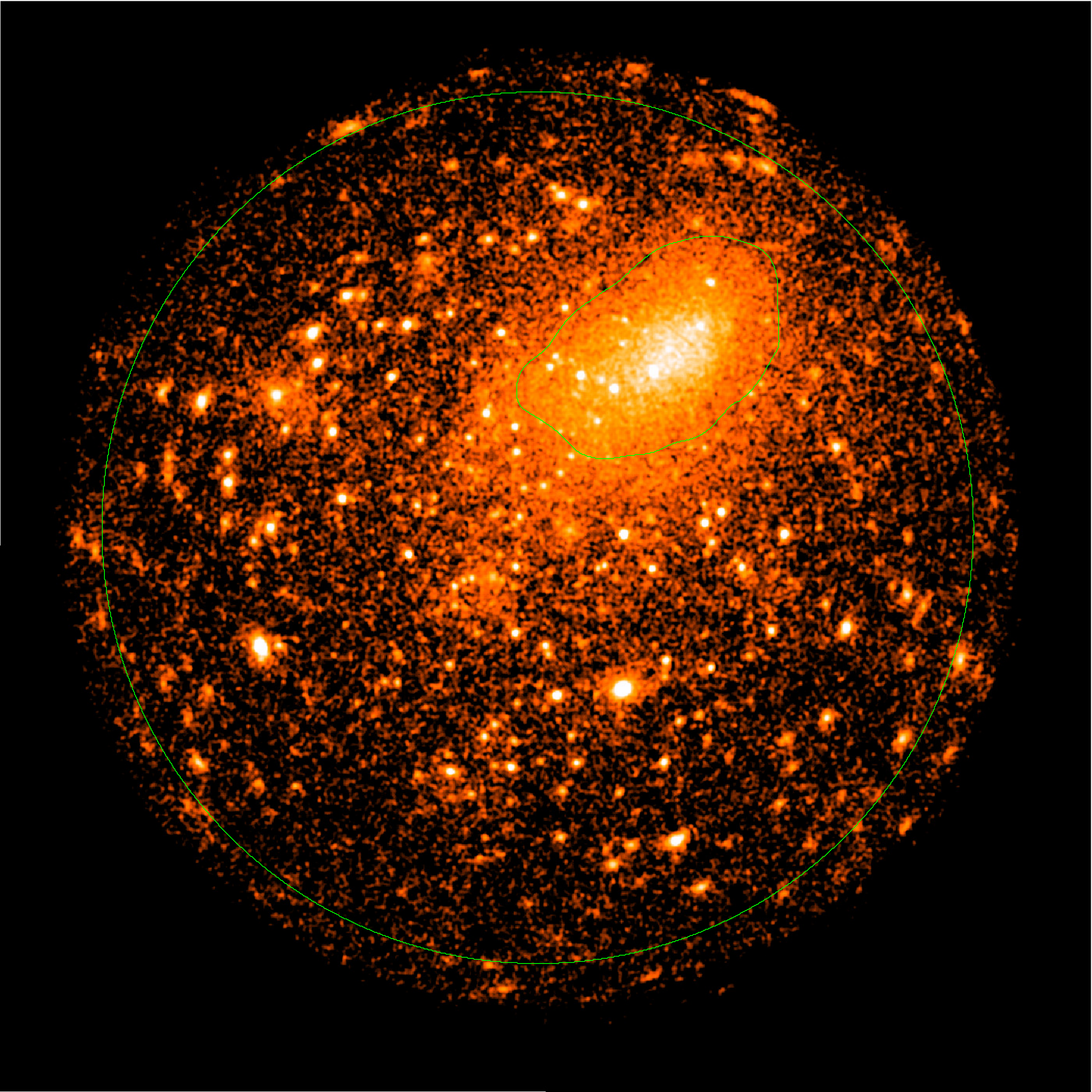}
    \put (2,2) {\large\color{white} 1H0707}
  \end{overpic}
  \hspace*{-0.07in}
  \begin{overpic}[width=0.285\textwidth]{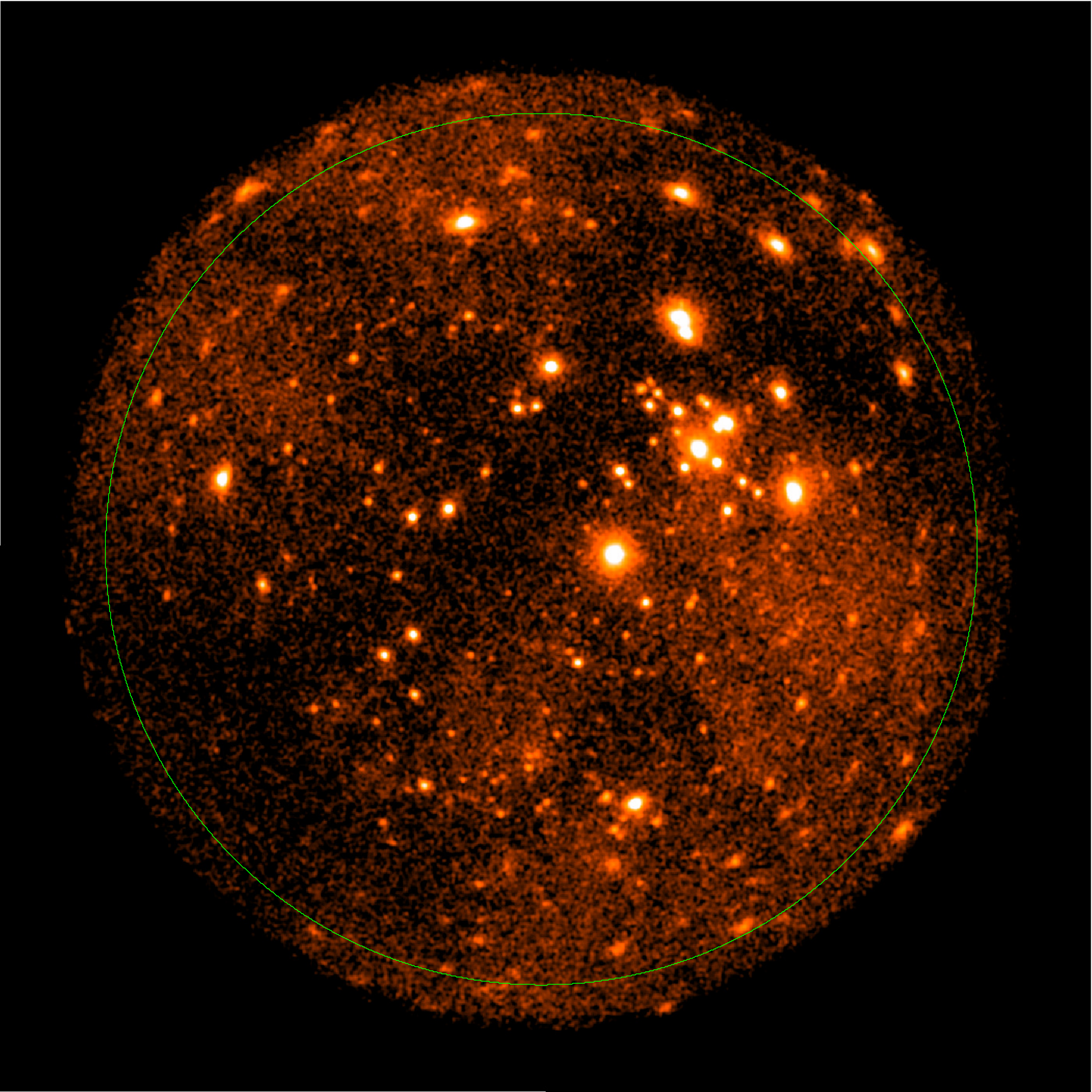}
    \put (2,2) {\large\color{white} H2213}
  \end{overpic}
  \hspace*{-0.07in}
  \begin{overpic}[width=0.285\textwidth]{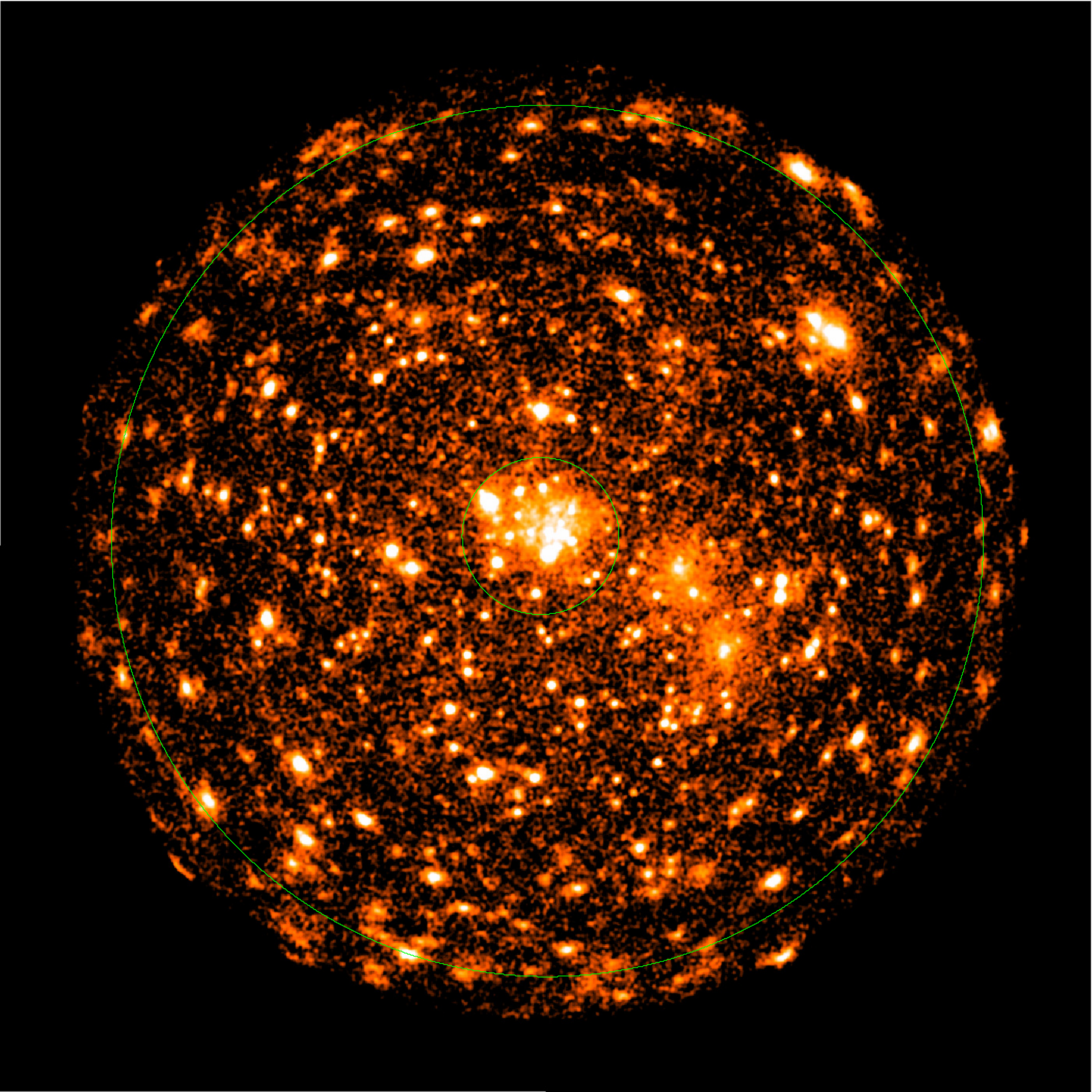}
    \put (2,2) {\large\color{white} N7793}
  \end{overpic}
  \end{minipage}
  \hspace{0.5in}

  \begin{minipage}{\textwidth}
  \centering
  \raisebox{0.0\textwidth}{
  \begin{overpic}[width=0.38\textwidth]{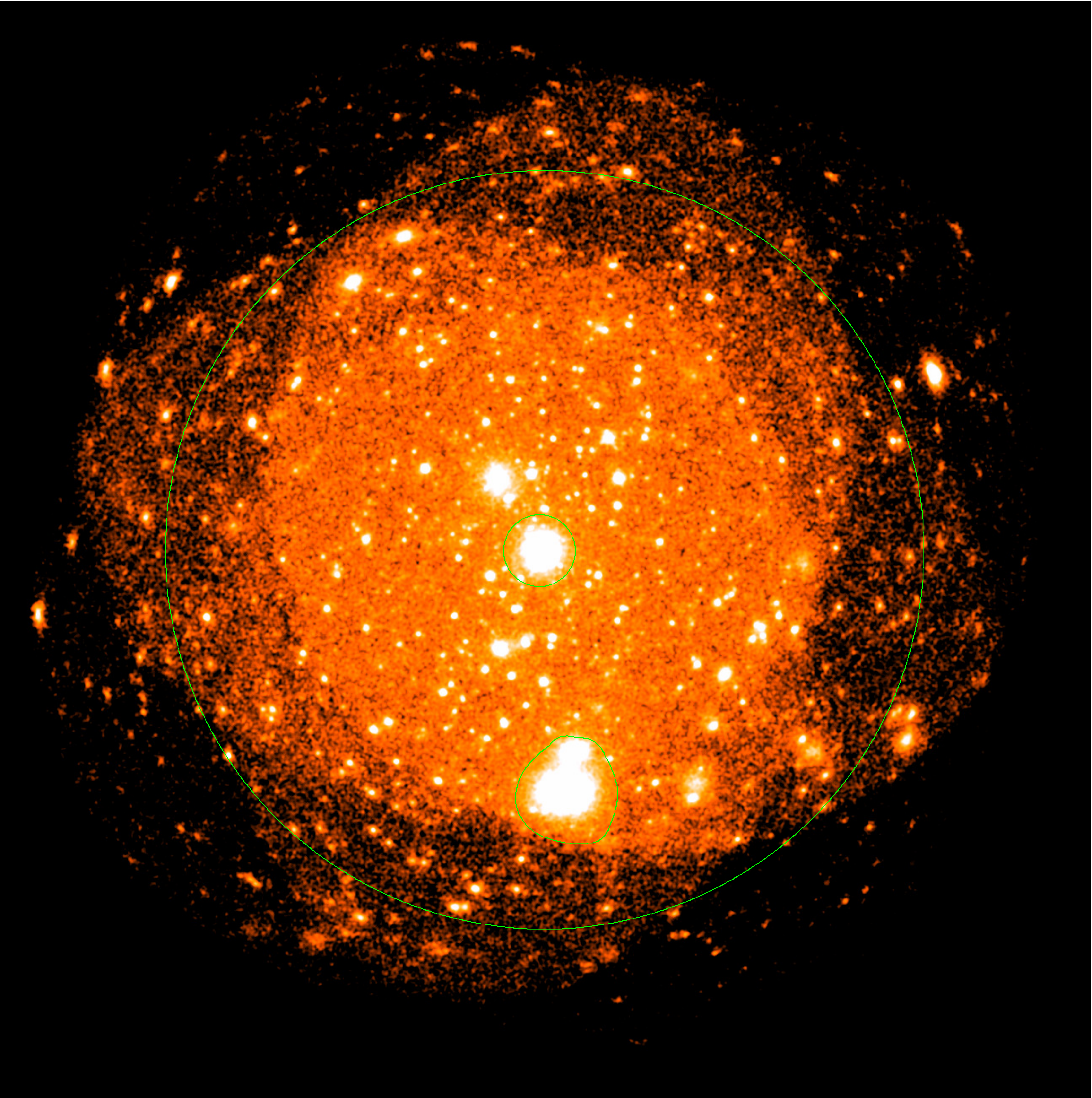}
    \put (2,2) {\large\color{white} 47Tuc}
  \end{overpic}}
  \hspace*{-0.12in}
  \raisebox{0.095\textwidth}{
  \begin{overpic}[width=0.285\textwidth]{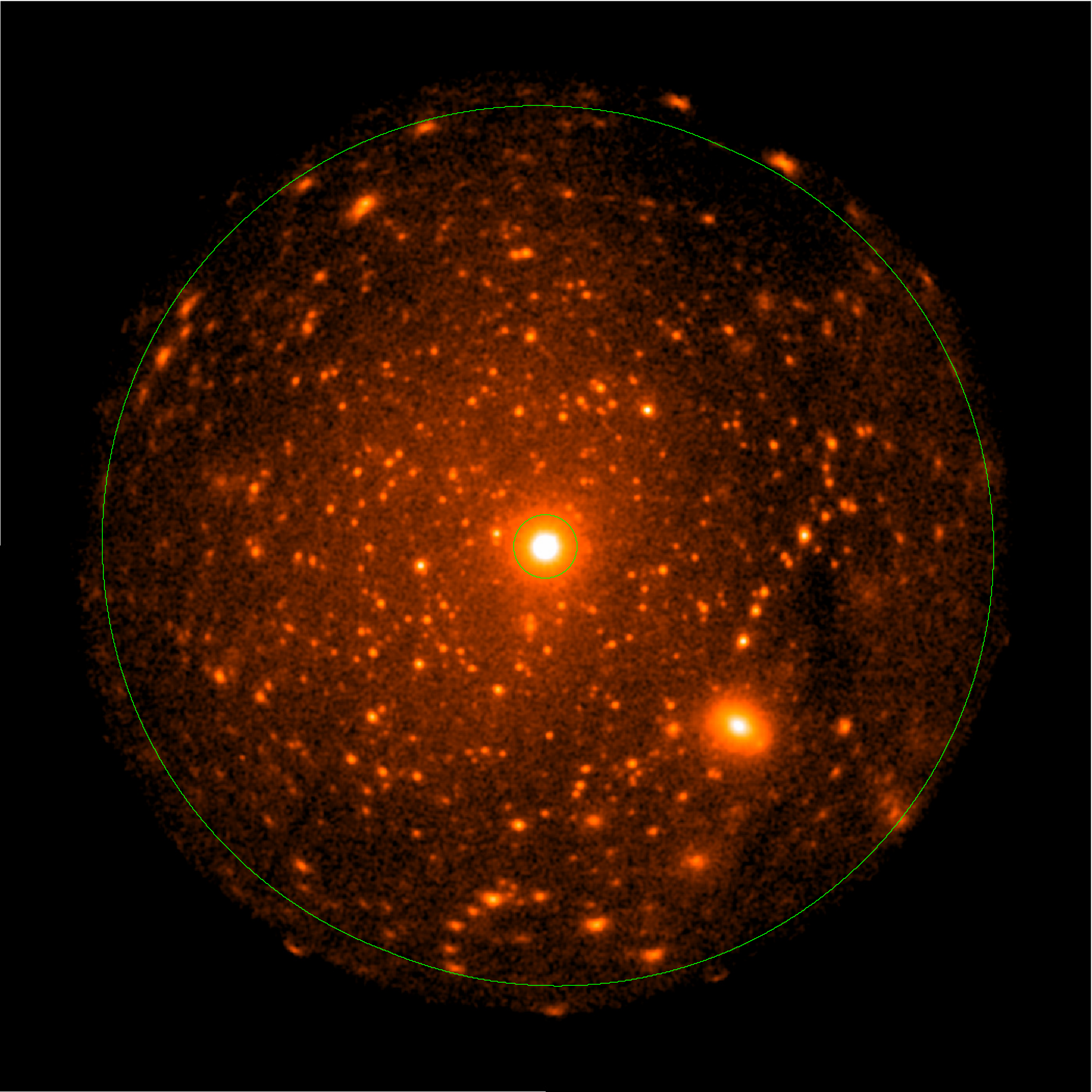}
    \put (2,2) {\large\color{white} J1856}
  \end{overpic}}
  \hspace*{-0.12in}
  \raisebox{0.095\textwidth}{
  \begin{overpic}[width=0.285\textwidth]{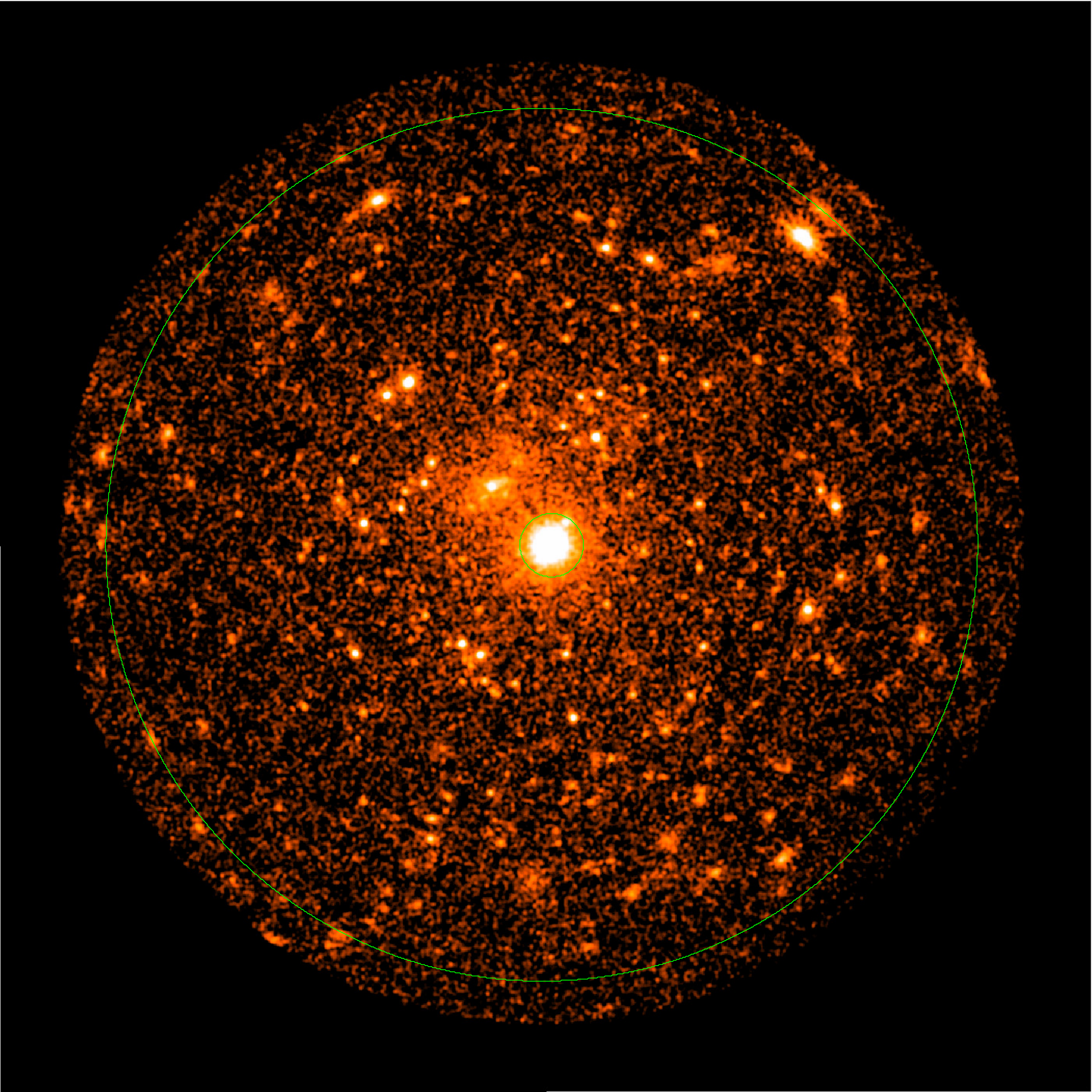}
    \put (2,2) {\large\color{white} 3C390}
  \end{overpic}}
  \end{minipage}
  \caption{0.2-2.3 keV photon images of nine of the 11 fields with the same scale.
    The green lines show the selected regions for \Edit{a subsurvey} that is relatively flat and stochastic (Sect.~\ref{sec:pntsurvey})\Edit{, using the outer boundaries to exclude shallow field borders and using the inner, small regions to exclude bright sources.}
  }
\label{fig:images1}
\end{center}
\end{figure*}

\begin{figure*}[htbp]
  \centering
  \begin{center}
    \begin{overpic}[width=0.49\textwidth]{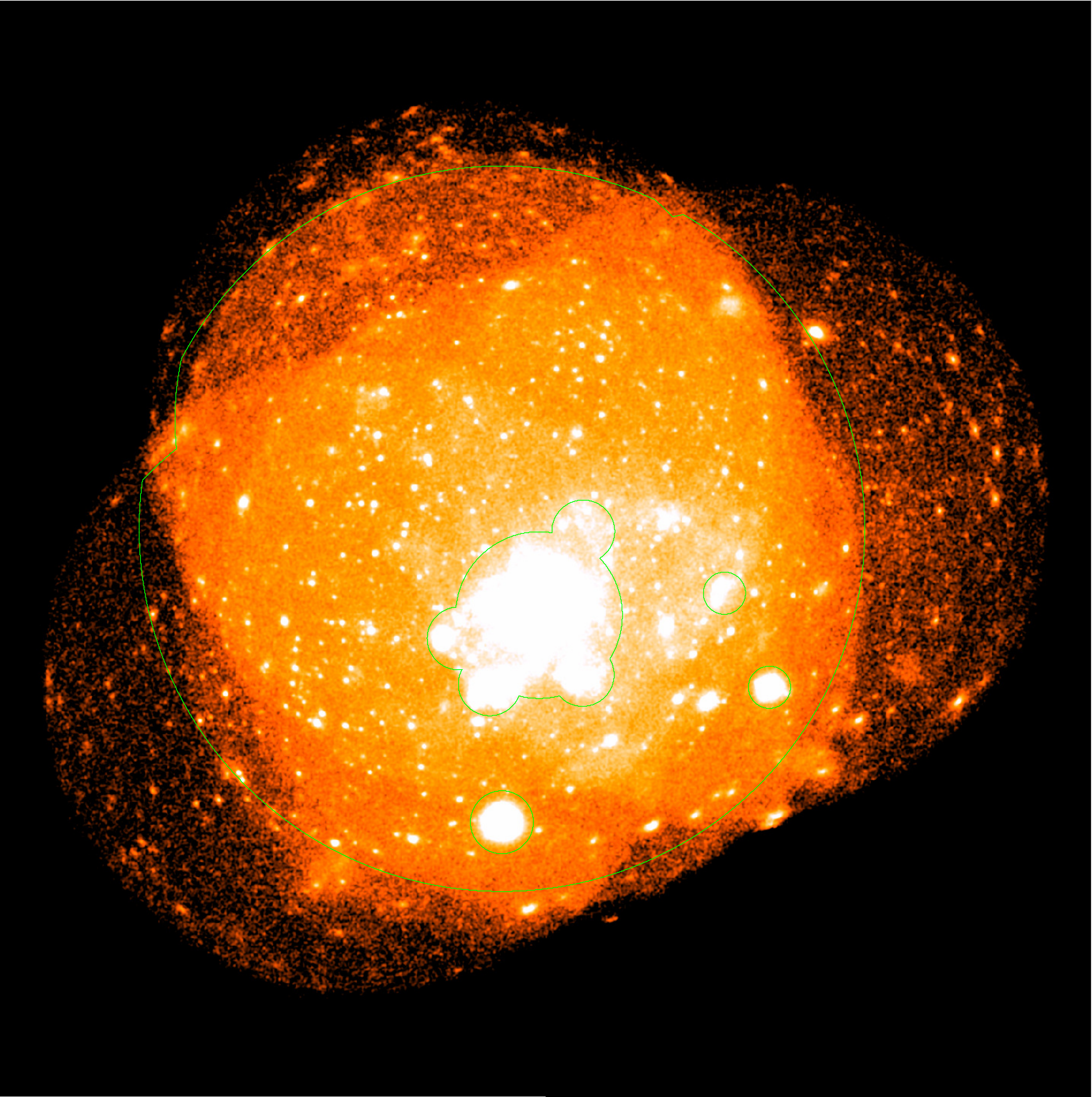}
      \put (2,2) {\large\color{white} ES0102}
    \end{overpic}
    \begin{overpic}[width=0.49\textwidth]{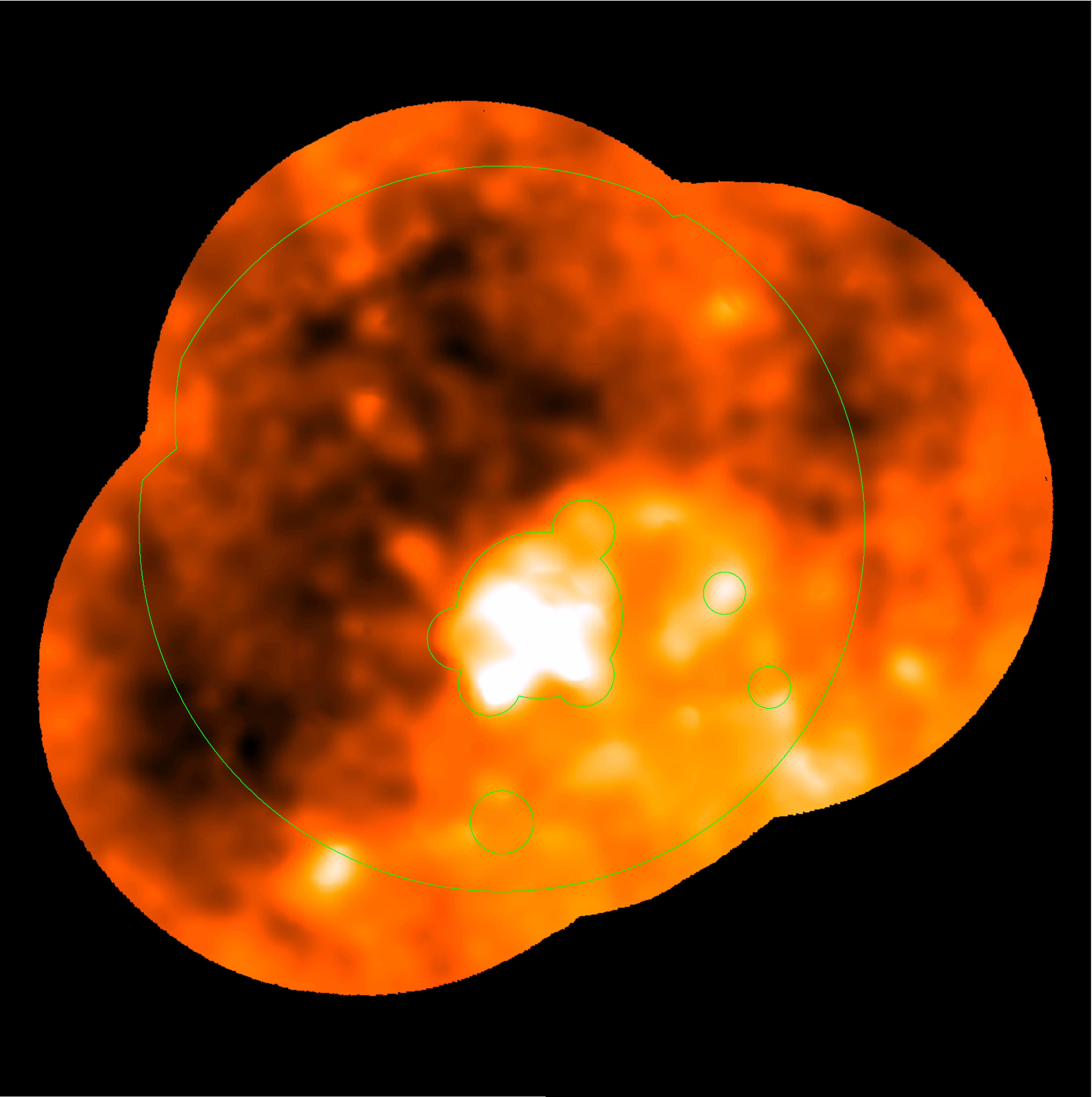}
      \put (2,2) {\large\color{white} ES0102}
    \end{overpic}
  \begin{overpic}[width=0.49\textwidth]{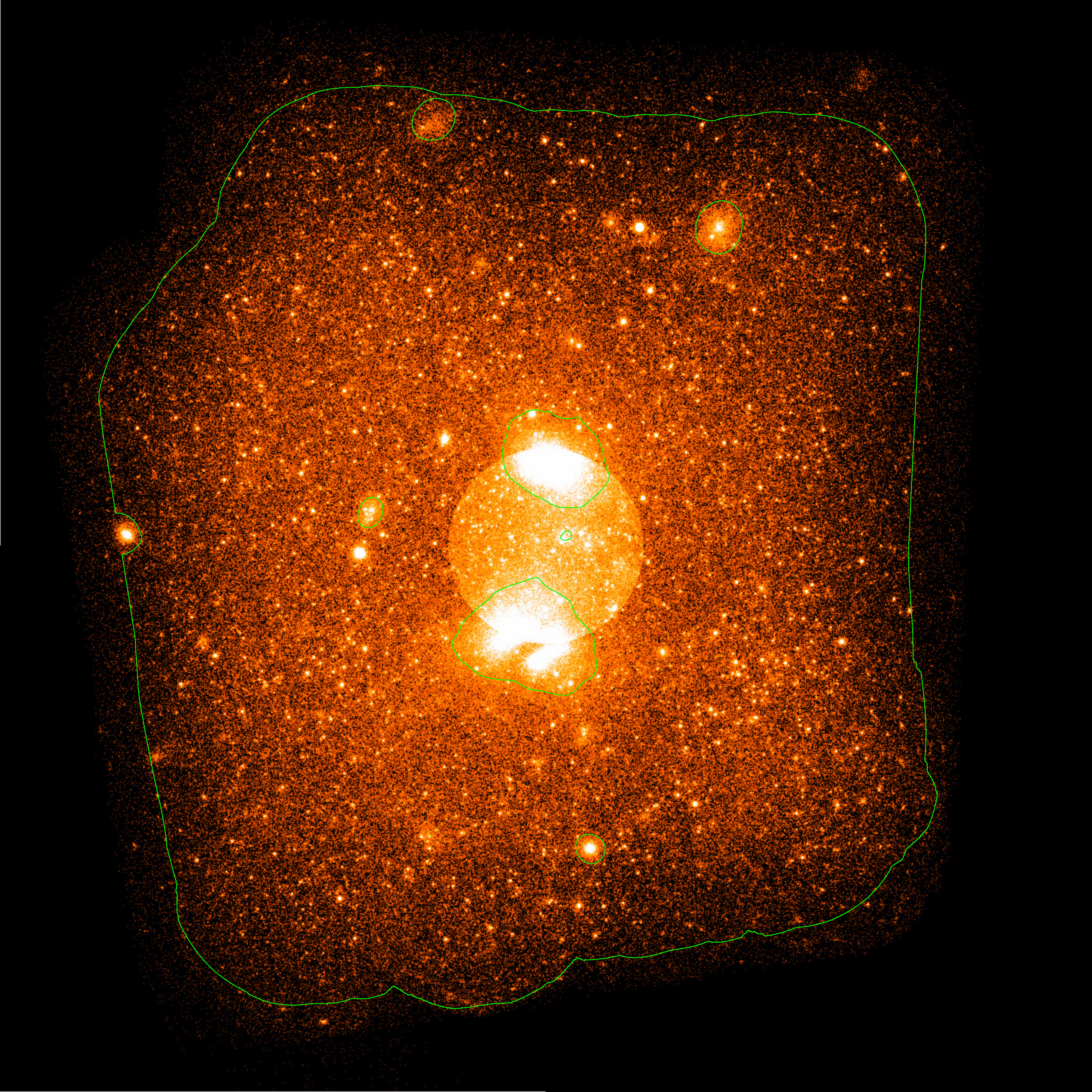}
    \put (2,2) {\large\color{white} A3391}
  \end{overpic}
  \begin{overpic}[width=0.49\textwidth]{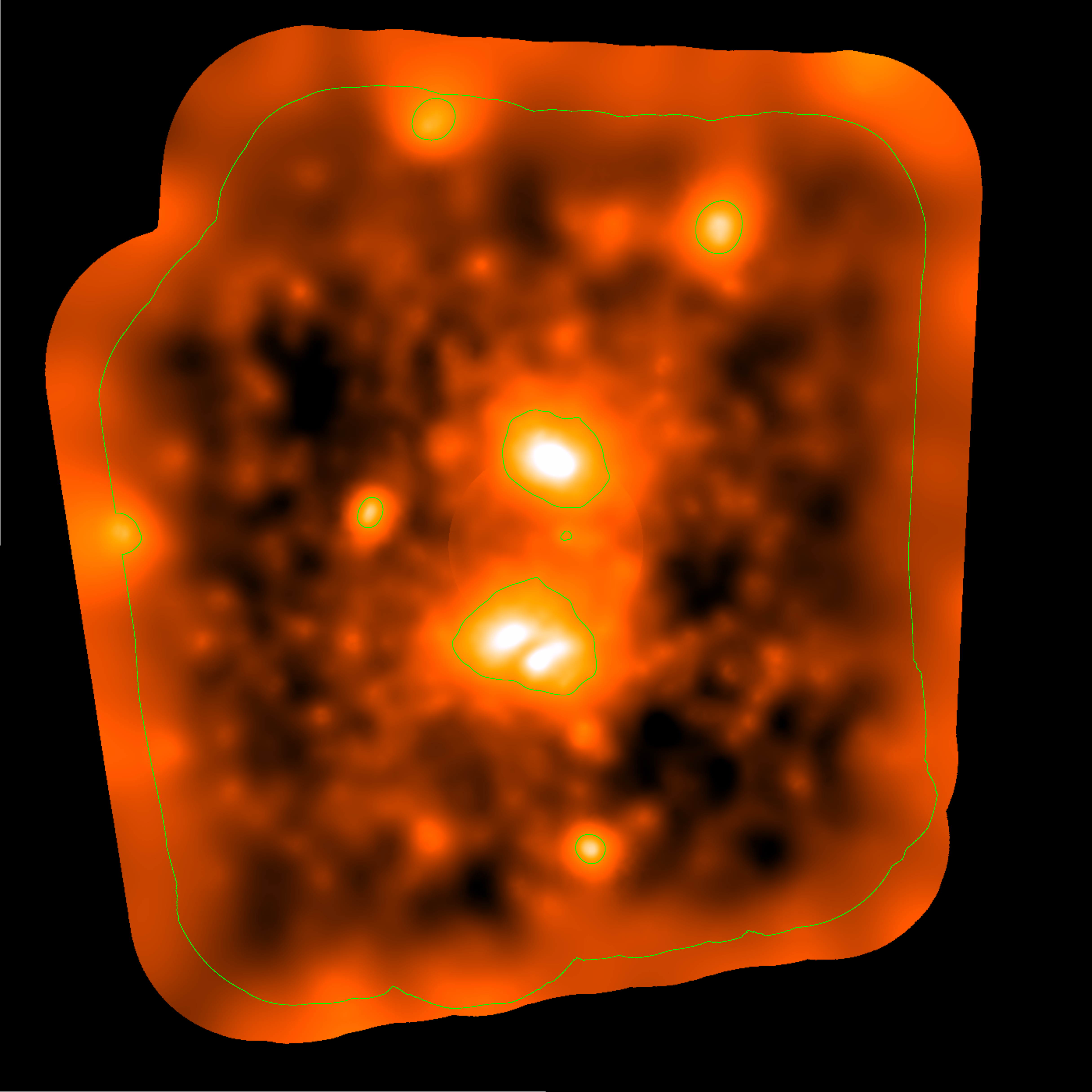}
    \put (2,2) {\large\color{white} A3391}
  \end{overpic}
  \caption{0.2--2.3~keV images (left) and 0.6--2.3~keV background \Final{maps} divided by exposure \Final{maps} (right) for the ES0102 (upper) and A3391 (lower) fields.
    \Edit{The same as in Fig.~\ref{fig:images1}, the green lines display the subsurvey regions (Sect.\ref{sec:pntsurvey}). The A3391 photon image shows a bright circle at the center, which corresponds to a long-exposure pointing-mode observation (300014; see Table.~\ref{tab:obs}).}
    \label{fig:images2}}
  \end{center}
\end{figure*}

\begin{table*}[htbp]
\centering
\caption{11 \eROSITA CalPV fields \Final{analyzed} in this work}
\begin{tabular}{p{4cm}rrrrrrrrrr}
\hline\hline
Field name&Time&Area&RA& DEC& Lon& Lat & Type & \texttt{SYS\_ERR} &Area$_s$ & $\log$\NH \\
\small{and target description} &&&&&&&&&&\\
   &ks &deg$^2$& deg & deg & deg & deg&  & arcsec & deg$^2$ &cm$^{-2}$\\
\hline
  A3391   &28.7	&14.9	&  96.65 & -54.06 & 262.78 & -25.19 &  PV &  0  & 11.5 & 20.84\\
\small{galaxy clusters \object{Abell\ 3391}, \object{Abell\ 3395}} &&&&&&&&&&\\
\cline{1-1}
  1H0707  &21.3	&0.9	& 107.31 & -49.39 & 260.04 & -17.53 &  PV &  1.28 & 0.6 & 20.77\\
\small{AGN \object{1H\ 0707-495}, galaxy cluster \object{Abell\ 3408}} &&&&&&&&&&\\
\cline{1-1}
  A3266   &61.9	&0.9	&  67.82 & -61.42 & 272.10 & -40.16 &  CAL & 1.40 & 0.6 & 20.33\\
\small{galaxy cluster \object{Abell\ 3266}} &&&&&&&&&&\\
\cline{1-1}
  A3158   &67.2	&0.9	&  55.70 & -53.63 & 265.05 & -48.95 &  CAL & 1.64 & 0.6 & 20.09\\
\small{galaxy cluster \object{Abell\ 3158}} &&&&&&&&&&\\
\cline{1-1}
  N7793   &51.2	&0.9	& 359.46 & -32.62 &   4.39 & -77.16 &  PV & 2.07 & 0.7 & 20.04\\
\small{ULX \object{NGC\ 7793\ P13}} &&&&&&&&&&\\
\cline{1-1}
  J1856   &166.2 	&0.9	& 284.15 & -37.91 & 358.60 & -17.22 &  CAL & 1.93 & 0.7 & 21.02\\
\small{neutron star \object{1RXS J185635.1-375433}} &&&&&&&&&&\\
\cline{1-1}
  3C390   &12.1	&0.8	& 280.57 & 79.77  & 111.44 &  27.07 & COM & 1.70 & 0.7 & 20.67\\
\small{AGN \object{3C390.3}} &&&&&&&&&&\\
\cline{1-1}
  H2213   &44.0	&0.9	& 285.61 & -37.12 & 359.81 & -18.01 &  PV & 2.37 & 0.7 & 21.26\\
\small{dark cloud \object{TGU\ H2213\ P1}} &&&&&&&&&&\\
\cline{1-1}
  J2334   &71.5	&1.2	& 353.64 & -47.37 & 334.59 & -64.83 &  CAL & 2.17 & 1.0 & 20.06\\
\small{white dwarf \object{RE\ J2334-471}} &&&&&&&&&&\\
\cline{1-1}
  47Tuc   &111.3 	&1.4	&   6.00 & -72.08 & 305.90 & -44.89 &  CAL & 1.91 & 0.8 & 20.78\\
\small{globular cluster \object{47\ Tuc}} &&&&&&&&&&\\
\cline{1-1}
  ES0102  &311.4 	&1.7	& 16.03  & -71.95 & 301.54 & -45.14 & CAL & 2.44 & 1.0 & 21.72\\
\small{SNR \object{1ES 0102-72.2} (in SMC)} &&&&&&&&&&\\
\hline
\end{tabular}
  \tablefoot{The columns are: 1) field name and brief description of the target; 2) effective unvignetted exposure time averaged among the 7 TMs, i.e., a sum of the exposure values in all the active cameras (not necessarily 7) divided by 7; 3) field area in degree$^2$; 4-5) field position (J2000); 6-7) field position in galactic coordinates; 8) type of observation, including COM (commissioning), CAL (calibration), and PV (performance verification); 9) additional systematic positional uncertainties (Sect.\ref{sec:ctp}); 10) area of selected region for the subsurvey (Sect.\ref{sec:pntsurvey}); 11) median value of the total Galactic absorption column density (Figure~\ref{fig:Exp_NH}).

  \label{tab:fields}
}
\end{table*}

\begin{table*}[htbp]
\centering
\caption{32 observations \Final{analyzed} in this work}
\begin{tabular}{llrrrrlrrr}
\hline\hline
Field&Obs ID &RA  & DEC& Time&Date& TM & $\Delta$RA & $\Delta$DEC & $\Delta\Theta$ \\
 name&       &deg & deg&  ks &    &    & arcsec     &  arcsec     & arcsec \\
\hline
A3391&	300014    &	96.6568&	-54.0524&	8.3&	2019-10-07&	5,6,7&	$0.18_{-0.35}^{+0.31}$&	$-0.94_{-0.27}^{+0.25}$&	$-0.01_{-0.02}^{+0.01}$\\
A3391&	300005$^s$&	96.6989&	-53.8210&	1.0&	2019-10-08&	5,6,7&	$0.59_{-0.48}^{+0.50}$&	$-1.74_{-0.39}^{+0.39}$&	$-0.01_{-0.02}^{+0.02}$\\
A3391&	300006$^s$&	96.7149&	-53.8080&	1.0&	2019-10-09&	5,6,7&  $0.38_{-0.39}^{+0.48}$&	$-2.70_{-0.64}^{+0.45}$&	$-0.01_{-0.02}^{+0.02}$\\
A3391&	300016$^s$&	97.1387&	-54.3858&	2.2&	2019-10-17&	1-7&	$\mathbf{-2.76_{-0.04}^{+0.04}|}$ & $\mathbf{0.78_{-0.04}^{+0.04}|}$        & \bf{reference}\\
\hline
1H0707&	300003&	107.3127&	-49.3800&	11.9&	2019-10-11&	5,6,7  &	$\mathbf{-1.25_{-0.10}^{+0.10}|}$       & $\mathbf{-0.46_{-0.10}^{+0.10}|}$     & \\
\hline
A3266&	700154&	67.8202&	-61.4269&	31.6&	2019-11-11&	1-7	  &	  $\mathbf{1.78_{-0.10}^{+0.09}|}$        & $\mathbf{6.01_{-0.08}^{+0.09}|}$      & \\
\hline
A3158&	700177&	55.7146&	-53.6270&	34.8&	2019-11-21&	1-7	  &	  $\mathbf{3.32_{-0.09}^{+0.09}|}$        & $\mathbf{2.26_{-0.08}^{+0.09}|}$      & \\
\hline
N7793&	300011&	359.4624&	-32.6090&	27.4&	2019-11-18&	1-6	  &	  $\mathbf{3.91_{-0.12}^{+0.09}|}$        & $\mathbf{-0.52_{-0.10}^{+0.09}|}$     & \\
\hline
J1856&	700008&	284.1596&	-37.9058&	38.6&	2019-10-24&	1-7&	$0.78_{-0.12}^{+0.12}$&	$-2.02_{-0.12}^{+0.13}$&	$-0.01_{-0.01}^{+0.00}$\\
J1856&	710001&	284.1332&	-37.9139&	16.2&	2020-04-01&	1-7&	$-3.60_{-0.14}^{+0.15}$&	$-1.29_{-0.16}^{+0.14}$&	$0.01_{-0.01}^{+0.01}$\\
J1856&	720002&	284.1600&	-37.9047&	30.1&	2020-10-07&	1-7&	$\mathbf{-0.13_{-0.10}^{+0.09}|}$ & $\mathbf{1.67_{-0.12}^{+0.11}|}$        & \bf{reference}\\
\hline
3C390&	900060&	280.5922&	79.7712&	1.6&	2019-09-24&	6&	$-0.47_{-0.45}^{+0.47}$&	$0.31_{-0.25}^{+0.23}$&	$0.18_{-0.18}^{+0.38}$\\
3C390&	900068&	280.5920&	79.7719&	1.5&	2019-09-28&	6&	$0.56_{-0.35}^{+0.34}$&	$-1.80_{-0.29}^{+0.31}$&	$0.07_{-0.12}^{+0.19}$\\
3C390&	900069&	280.5920&	79.7721&	1.7&	2019-09-29&	6&	$0.70_{-0.25}^{+0.26}$&	$-0.24_{-0.33}^{+0.30}$&	$0.05_{-0.19}^{+0.23}$\\
3C390&	900070&	280.5912&	79.7724&	1.7&	2019-09-30&	6&	$\mathbf{0.79_{-0.12}^{+0.13}|}$  & $\mathbf{-2.85_{-0.16}^{+0.14}|}$       & \bf{reference}\\
\hline
H2213&	300017&	285.6054&	-37.1228&	25.5&	2019-10-26&	1-7	  &	 $\mathbf{0.33_{-0.17}^{+0.17}|}$        & $\mathbf{0.74_{-0.14}^{+0.16}|}$      & \\
\hline
J2334&	700180&	353.5144&	-47.2260&	18.9&	2019-11-19&	1-7&	$-0.02_{-0.24}^{+0.27}$&	$0.65_{-0.17}^{+0.20}$&	$-0.00_{-0.01}^{+0.00}$\\
J2334&	700181&	353.7685&	-47.5110&	18.5&	2019-11-20&	1-7&	$\mathbf{2.87_{-0.09}^{+0.08}|}$        & $\mathbf{-0.62_{-0.10}^{+0.11}|}$     & \bf{reference}\\
\hline
47Tuc&	700012&	6.5318&	-72.1758&	1.5&	2019-09-28&	6      &	$0.27_{-0.31}^{+0.30}$&	$0.79_{-0.30}^{+0.34}$&	$0.04_{-0.06}^{+0.09}$\\
47Tuc&	700163&	6.5264&	-72.1604&	11.1&	2019-11-01&	1,2,4-7&	$\mathbf{3.61_{-0.10}^{+0.09}|}$        & $\mathbf{0.65_{-0.10}^{+0.10}|}$      & \bf{reference}\\
47Tuc&	700011&	5.5181&	-71.9681&	11.3&	2019-11-01&	1,2,4-7&	$-1.45_{-0.25}^{+0.35}$&	$2.26_{-0.37}^{+0.32}$&	$0.03_{-0.09}^{+0.13}$\\
47Tuc&	700013&	6.3215&	-71.9086&	10.9&	2019-11-02&	1,2,4-7&	$-0.08_{-0.29}^{+0.30}$&	$3.37_{-0.20}^{+0.19}$&	$0.01_{-0.03}^{+0.04}$\\
47Tuc&	700014&	5.7088&	-72.2212&	11.9&	2019-11-02&	1-7    &	$-1.39_{-0.22}^{+0.19}$&	$0.19_{-0.40}^{+0.35}$&	$-0.02_{-0.04}^{+0.02}$\\
47Tuc&	700173&	5.5198&	-71.9692&	3.0&	2019-11-19&	1-7    &	$-0.37_{-0.44}^{+0.45}$&	$0.80_{-0.54}^{+0.46}$&	$0.02_{-0.05}^{+0.07}$\\
47Tuc&	700174&	6.5349&	-72.1609&	3.8&	2019-11-19&	1-7    &	$-0.77_{-0.24}^{+0.25}$&	$-0.21_{-0.22}^{+0.20}$&	$-0.03_{-0.05}^{+0.03}$\\
47Tuc&	700175&	6.3350&	-71.9095&	3.8&	2019-11-19&	1-7    &	$0.91_{-0.36}^{+0.35}$&	$-0.06_{-0.24}^{+0.21}$&	$0.00_{-0.03}^{+0.04}$\\
\hline
ES0102&	700001&	16.0279&	-72.0127&	27.6&	2019-11-07&	1-7    &	$\mathbf{1.74_{-0.09}^{+0.10}|}$        & $\mathbf{3.66_{-0.10}^{+0.11}|}$      & \bf{reference}\\
ES0102&	700002&	15.0568&	-71.8692&	29.6&	2019-11-08&	1-7    &	$0.17_{-0.23}^{+0.19}$&	$1.11_{-0.20}^{+0.20}$&	$0.01_{-0.02}^{+0.03}$\\
ES0102&	700003&	17.0144&	-72.1514&	30.2&	2019-11-09&	1-7    &	$0.46_{-0.24}^{+0.22}$&	$-1.52_{-0.29}^{+0.21}$&	$0.04_{-0.04}^{+0.04}$\\
ES0102&	700004&	16.3022&	-71.8317&	27.6&	2019-11-09&	1-7    &	$0.73_{-0.18}^{+0.19}$&	$0.95_{-0.21}^{+0.19}$&	$0.02_{-0.09}^{+0.09}$\\
ES0102&	700005&	16.4384&	-71.7413&	28.0&	2019-11-10&	1-7    &	$2.41_{-0.20}^{+0.22}$&	$0.50_{-0.27}^{+0.27}$&	$-0.00_{-0.08}^{+0.05}$\\
ES0102&	710000&	16.2691&	-72.0570&	16.0&	2020-06-18&	1-4,6,7&	$-5.11_{-0.27}^{+0.27}$&	$-1.47_{-0.22}^{+0.22}$&	$0.16_{-0.10}^{+0.11}$\\
\hline
\end{tabular}
\tablefoot{The columns are: 1) field name; 2) observation ID; 3-4) pointing coordinates (J2000); 5) 0.2--2.3~keV vignetted exposure time, averaged in the field and among the 7 TMs; 6) date of the observation starting time; 7) active telescope modules; 8-10) relative astrometric correction on the pointing RA,DEC and the roll angle $\Theta$ with 1-$\sigma$ uncertainties.\\
  The mark $^s$ on the observation ID indicates scanning mode.\\
  The $\Delta$RA is already multiplied by $\cos(\mathrm{DEC})$, where DEC is the Declination of the field center (Table~\ref{tab:fields}). The $\Delta$RA, $\Delta$DEC \Edit{in bold font and} with a suffix $|$ mark are absolute corrections of each field with respect to WISE counterparts (Sect.\ref{sec:ctp}); and the others are relative corrections of each observation (Sect.\ref{sec:astrometric}) with respect to the reference observation (flagged as ``reference'' in the $\Delta\Theta$ column).
  \label{tab:obs}
  }
\end{table*}

This serendipitous survey combines 11 extragalactic fields, which target various types of astronomical objects for various purposes.
They are listed in Tables~\ref{tab:fields} and \ref{tab:obs}.
An image of each field is displayed in Figures~\ref{fig:images1} and \ref{fig:images2}.
We give each field a name by truncating the name of its target object (Table~\ref{tab:fields}).
The bright AGN 3C390.3 was observed for commissioning (COM) and calibration of the ART-XC telescope; six fields were observed for calibration (CAL) \Final{purposes} targeting well-known, bright X-ray sources; and the other four fields were performance verification (PV) observations designed with the scientific goals of investigating galaxy cluster (Abell 3391/3395, PI Thomas Reiprich), AGN (1H~0707-495, PI Thomas Boller), ULX (NGC~7793~P13, PI J\"orn Wilms), and Galactic dark cloud (TGU~H2213~P1, PI Michael Freyberg).
Table~\ref{tab:obs} lists the 32 observations of these 11 fields.
All the observations were in pointing mode except \Final{for} three observations of A3391.
Thanks to these scanning-mode observations, the A3391 field has an area larger than the sum of all the other fields.
Seven fields have a single, almost-fixed pointing coordinate, thus their footprints correspond just to the \eROSITA FOV size.
Three fields (J2334, 47Tuc, ES0102) have multiple overlapping pointings, which result in larger area coverage.
\eROSITA is composed of seven almost-identical telescope modules (TMs), namely, TM1--7. Not all the TMs were active during the CalPV observations, as also reported in Table~\ref{tab:obs}.
\TL{
Detailed studies of some interesting astronomical objects in these fields have been published in several papers, including 
Abell~3391/3395 \citep{Reiprich2021,Veronica2021},
Abell~3266 \citep{Sanders2021},
Abell~3158 \citep{Whelan2021},
Abell~3408 \citep{Iljenkarevic2021},
47~Tuc \citep{Saeedi2021},
1H~0707-495 \citep{Boller2021},
high-mass X-ray binaries (HMXBs) in the Small Magellanic Cloud (SMC) \citep{Haberl2021}.}

The data were processed with the \eROSITA Science Analysis Software System \citep[eSASS;][]{Brunner2021}.
We use the tasks in the \eROSITA early data release (EDR) version of eSASS (eSASSusers\_201009), except for the source detection task \texttt{ermldet}, for which an updated version in eSASSusers\_211214 is used.
The telemetry data are first processed to create calibrated event files.
Then using the eSASS task \texttt{evtool}, we extract the events in specific energy ranges and create images with a resolution of $4\arcsec$ per pixel, which is smaller than the physical pixel size of \eROSITA ($9\farcs6$). Both vignetted exposure maps in each band and an unvignetted exposure map are created using the task \texttt{expmap}. A source detection mask is created with the task \texttt{ermask} based on the 0.6--2.3\,keV band vignetted exposure map applying a minimum cut at 1\% of the maximum value. Adopting such a low threshold, only the outermost border of each field is excluded (comprising $1\%\sim5\%$ of the area).

\subsection{Astrometric correction}
\label{sec:astrometric}
Astrometric correction can be either applied to the X-ray events before source detection, or to the catalog source after the detection process.
In this work, we perform post-hoc astrometric corrections to the X-ray catalog, thus a prior astrometric correction on the X-ray events is not needed when a field has a single pointing-mode observation.
For fields with multiple pointing-mode observations, we perform relative astrometric corrections on the events before merging the data.
For each observation, we perform source detection in a single 0.2--2.3~keV band, which is the most sensitive band of \eROSITA \citep{Liu2021_sim}, using the method as described below in Sect.\ref{sec:bkg3}.
Such a single-band source detection is only used for astrometric corrections. The formal source detection (Sect.\ref{sec:method}) is not based on this single band.
\TL{From the multiple observations of one field, we choose one observation as a reference, and calculate the required corrections for all other observations with respect to the catalog of this reference observation using the method described below.}

We exclude sources at off-axis angles $>$25\arcmin because, with the butterfly-shaped PSF at the border of the FOV \citep{Dennerl2020}, the measured source position has a much larger uncertainty than at the center of the FOV.
We consider a frame transformation with both a slight shift ($\Delta$RA,$\Delta$DEC) and a slight rotation ($\Delta\Theta$), and measure these three parameters simultaneously by minimizing the following cost function
\begin{equation}
  \sum_i{\frac{\delta_i^2}{\sigma_i^2(1+A_i/A_2)^2}},
  \label{equ1}
\end{equation}
where, for each source $i$, $\delta_i$ is the separation between the X-ray measured source position and the reference position, $\sigma_i$ is the X-ray measured positional uncertainty, $A_i$ is the off-axis angle in units of arcmin, and $A_2=20\arcmin$.
We also measure confidence intervals for $\Delta$RA,$\Delta$DEC, and $\Delta\Theta$ through 500-folds bootstrapping.
The results are listed in Table~\ref{tab:obs}. 
We find that, in most cases, the relative correction is very small and negligible in comparison to \eROSITA's physical pixel size $9\farcs6$, confirming that \eROSITA has an excellent pointing accuracy.
Although a slight offset of a few arcseconds is needed only in a few cases, we apply the $\Delta$RA, $\Delta$DEC, and $\Delta\Theta$ corrections to all the events before merging the multiple observations of one field.
The source detection below is performed on the merged event file where applicable.
\TL{This method is also used later in Sect.~\ref{sec:ctp} for absolute astrometric corrections to the catalogs, where the WISE counterparts are used as a reference.}

\section{Source detection and photometry}
\label{sec:method}
\subsection{Source detection procedure}
\begin{figure*}[hptb]
\centering
\includegraphics[width=\textwidth]{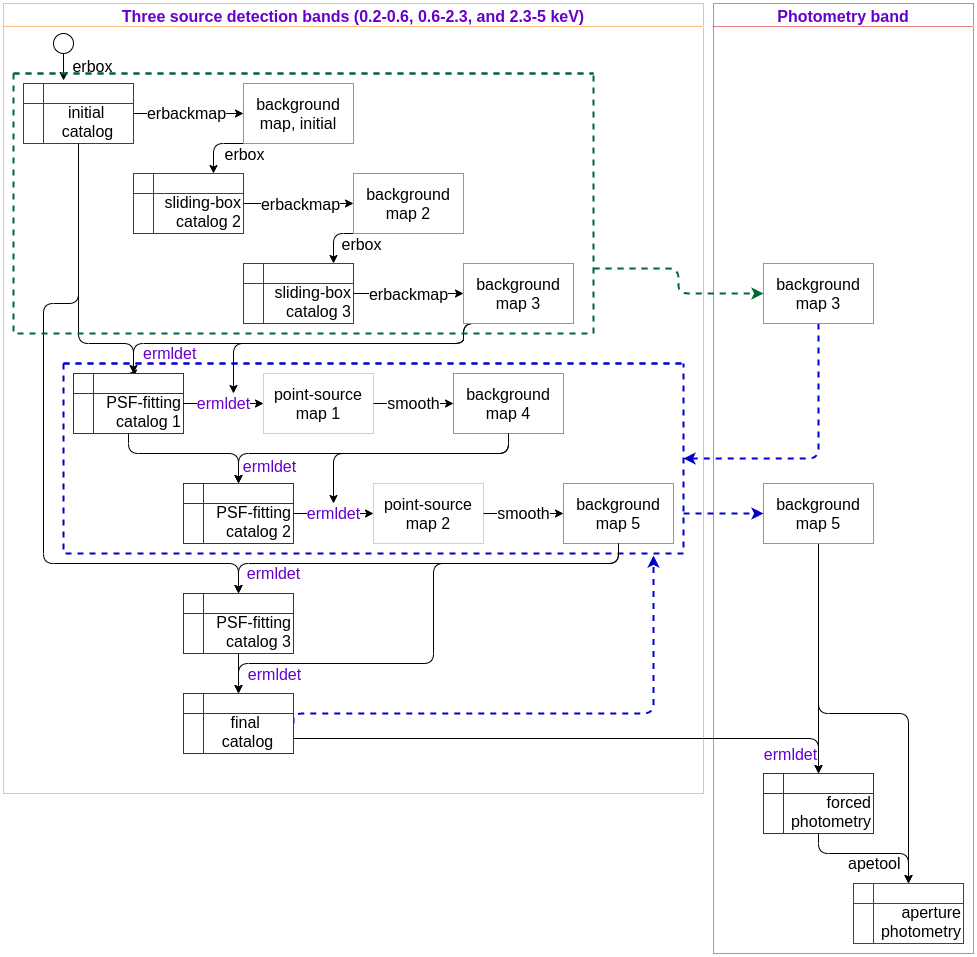}
\caption{Flowchart of source detection and photometry procedures.
\label{fig:flowchart}}
\end{figure*}

Our core algorithm for source detection is fitting the image of a source with the point spread function (PSF) model using the eSASS task \texttt{ermldet}.
We perform simultaneous PSF-fitting in three energy bands 0.2--0.6, 0.6--2.3, and 2.3--5~keV.
The PSF-fitting requires a seed catalog, which contains all the potential source candidates, and a background map for each band.
The overall source detection procedure is designed around this PSF-fitting, as illustrated by the orange box in Fig.~\ref{fig:flowchart}. This procedure is composed of three steps, as described below.

\subsubsection{A standard procedure}
\label{sec:bkg3}
As a first step, we create a background map following the method used in the eFEDS survey \citep{Brunner2021}, which is illustrated in the green dashed box in Fig.~\ref{fig:flowchart}.
First, we run the eSASS sliding-box detection task \texttt{erbox}\footnote{We do not repeat the sliding-box search with rebinned images with \texttt{erbox}.} in local mode (without background map) adopting a detection likelihood threshold of 6 and a box size of 7 pixels (task parameter \texttt{boxsize}$=$3) to create an initial catalog.
This initial catalog provides a list of seed sources, including a large number of spurious sources.
With such a temporary catalog, we can mask out the sources and adaptively smooth the image to create a background map using the eSASS task \texttt{erbackmap}, adopting a source selection likelihood threshold of 6, a source mask-out radius at a brightness of 0.001 counts per pixel, and requiring a final signal-to-noise ratio of 40.
\Final{We note} that such background maps based on a significantly overpopulated catalog are thus biased low, since overdensities due to background fluctuation are also masked out as sources.
With these background maps prepared, we can then run the sliding-box detection in ``map'' mode, adopting a detection likelihood \Edit{threshold} of 4 and a box size of 7 pixels, to create an updated catalog and use it to update the background map (``background map 2'').
We then repeat this step of updating the sliding-box catalog and then updating the background map,\Edit{ adopting the same setting as used above, to create a ``background map 3''.
Now, this background map is} determined by the settings adopted in the background map creation and the sliding-box detection.
A further iteration is not needed since its \Edit{effect} on the result is negligible.
When creating background maps, the vignetted exposure map is used for the two soft bands (0.2--0.6 and 0.6--2.3~keV), while an un-vignetted exposure map is used for the hard band (2.3--5~keV), since the hard-band background is dominated by the un-vignetted particle background component.

We can now input the initial catalog and the background map into \texttt{ermldet} for PSF-fitting.
We run photon-mode PSF-fitting adopting an extraction radius (\texttt{cutrad}) of 15 pixels (1\arcmin), a multiple sources search radius (\texttt{multrad}) of 30 pixels (2\arcmin), a detection likelihood threshold (\texttt{likemin}) of 5, an extent likelihood threshold (\texttt{extlikemin}) of 6, an extent range between 3 and 30 pixels, a maximum of 5 sources per simultaneous fitting, and allowing a source to be split into a maximum number of 2 sources.
By now, we have already a PSF-fitting selected catalog, which can be formatted using the eSASS task \texttt{catprep}.
We call the source detection procedure above the \textit{``standard procedure''}, which is similar to that used in eFEDS \citep{Brunner2021}.
However, after the standard procedure, we perform several additional steps before creating the final catalog.

\subsubsection{Updating the background map}
\label{sec:bkg5}

The process of source detection is, in essence, that of recovering the dichotomy between source and background signals.
Ideally, source and background are defined consistently and determined simultaneously.
However, in practice, it is usually the case that background is measured first, and only after that sources are found based on the measured background.
In the case of our standard source detection method described above, the background used in PSF-fitting is determined by a sliding-box detected catalog that is completely independent of the output catalog.
This is the first reason why we need to update the background maps -- to recreate the background maps based on the PSF-fitting detected catalog.

Background can be defined on different levels. \Final{To distinguish} the \Final{very large-scale} structures in the Galactic hot gas on the map of the entire sky, \Final{such as} the \eROSITA bubbles \citep{Predehl2020}, the relevant background is the component that is relatively uniform across the sky. For general astronomical \Final{(distance)} X-ray sources, \Final{such as} AGN or galaxy clusters, the Galactic hot gas is just a component of the background. For point sources, underlying diffuse \Final{emissions} from galaxy clusters should also be included as a background component.
This is the second reason why we need to update the background maps -- we define background explicitly as the background of point sources.

Our background-creating method is illustrated in the blue dashed box in Fig.~\ref{fig:flowchart}.
Using the initial catalog and the ``background map 3'' created in the standard source detection procedure, we have created a PSF-fitting catalog (``PSF-fitting catalog 1'').
We feed this catalog back into \texttt{ermldet} for forced PSF-fitting, adopting fixed source positions and taking all the sources as point sources.
Genuine point sources will be fitted well; while for extended sources, only the core will be fitted with the PSF model, and the outer part that makes the source look extended will not be modeled.
Comparing the best-fit model image (source plus background) with the background map, we mask out the regions where the former is 10\% higher than the latter.
Then we smooth the masked image and the exposure map adaptively adopting the same scales to reach \Final{a signal to noise ratio (S/N)} of 20 in the smoothed image using the code of \citet{Sanders2006}.
The smoothing scales are chosen to include at least 400 counts.
We divide the smoothed image by the smoothed exposure map and then multiply it by the original exposure map to create an updated background map (``background map 4'').
Feeding the ``PSF-fitting catalog 1'' and ``background map 4'' to \texttt{ermldet}, we can update the catalog creating the ``PSF-fitting catalog 2''.
Then using ``PSF-fitting catalog 2'', we repeat the step above of forced PSF-fitting, source masking, and image smoothing, to create our final background map (``background map 5'').
This background map is based on a PSF-fitting catalog that is almost identical to the final catalog.
For extended sources, like galaxy clusters, only the emission in the core region and the emission of point sources inside them are excluded from the background map, the residual diffuse emission is considered as a background component.

Taking the fields ES0102 and A3391 as examples, we plot the final background maps in Fig.~\ref{fig:images2}.
Most of the diffuse \Final{emissions} of galaxy clusters in the A3391 field are included in the background.
There are a few spurious point sources detected in some bright clusters. Therefore, a small fraction of cluster emission is excluded before creating the background map by smoothing, and the cluster emission in the background map is incomplete in the sense of flux measurement.
In the case of diffuse supernova remnant emission in the field ES0102, the majority of the diffuse emission can be retained in the background map. However, the current version of \texttt{ermldet} assumes a beta model for extended sources \citep{Brunner2021}, which cannot describe irregular-shaped diffuse gas when it has a sharp edge or appears as a ring. Such diffuse emission results in a clump of point sources, which are excluded from the background map.
Such clumps of spurious point sources will be detected in the final catalog and have to be removed manually.

\subsubsection{Final PSF-fitting}
\label{sec:finalfit}
With the final background map at hand, we feed the initial catalog to \texttt{ermldet} for PSF-fitting, creating the ``PSF-fitting catalog 3''. 
Following the source detection method used in the XMM-RM survey \citep{Liu2020}, rather than taking ``PSF-fitting catalog 3'' as the final one, we feed it back to \texttt{ermldet} for a final PSF-fitting to create the final catalog.
The reason for this \Final{additional} step is that the initial catalog contains many spurious sources near the real ones, which might be not well handled by the multiple-source-fitting algorithm.
In this final PSF-fitting, instead, the input catalog is only slightly larger than the output.

The PSF-fitting measures a few important parameters for each source.
By comparing the best-fit source model with a zero-flux (pure background) model, \texttt{ermldet} calculates a detection likelihood (\texttt{DET\_LIKE}) for each source, defined as $\texttt{DET\_LIKE}=-\ln P$, where P is the probability of the source being caused by a random background fluctuation. By comparing the extent model (beta function) with a $\delta$ function, \texttt{ermldet} also calculates an extent likelihood \texttt{EXT\_LIKE} for each source, which is defined \Final{as} corresponding to the probability of a source being unresolved rather than extended. Sources with extent likelihood above 6 (\texttt{extlikemin}) are fitted with the PSF-convolved beta model and the others are considered as point sources and fitted with the PSF model, setting \texttt{EXT\_LIKE}$=$0. By minimizing the C-statistic, \texttt{ermldet} measures the source position, extent, and count rate, together with the $68\%$ confidence intervals. The total-band count rates \texttt{ML\_RATE\_0}, counts \texttt{ML\_CTS\_0} and fluxes \texttt{ML\_FLUX\_0} have the errors of the single-band rates, counts, and fluxes added in quadrature, treating them as Gaussian errors. The  \texttt{ermldet} algorithm is described in more details in \citet{Brunner2021}.

\subsection{Source photometry}
\label{sec:phot}
\Edit{After the catalog is created using the procedure above, we apply first forced PSF-fitting (position-fixed) photometry and then aperture photometry for the detected sources in the 0.5--2 and 2.3--5~keV bands, as illustrated in the red box in Fig.~\ref{fig:flowchart}.
The forced PSF-fitting provides modeling of overlapping sources, which helps in the second step of aperture photometry with subtracting contamination of nearby sources.}
\Edit{The photometry bands are different from the three source detection bands because the 0.5--2~keV band is more commonly used in previous X-ray surveys using other facilities, \Final{such as} \XMM and \Chandra.}
First, we create a background map for each band using the ``standard'' procedure (green dashed box) as described in Sect.\ref{sec:bkg3}.
Then we input this background map and the final catalog detected in the three bands into the background updating procedure (blue dashed box) as described in Sect.\ref{sec:bkg5}, taking the final catalog as the ``PSF-fitting catalog 1'' in this procedure, to create the final background map.
Eventually, with the final background map, we perform PSF-fitting for the sources in the final catalog, with the source position and extent fixed.
Such \Final{a} forced PSF-fitting only provides a measurement of \Final{the} count rate in a particular band.
Then we input the catalog and source image model from the forced PSF-fitting to the eSASS task \texttt{apetool} to do aperture photometry within a radius corresponding to 60\% EEF.
The \texttt{apetool} measures the source and background counts in the circular aperture and calculates from them a probability (\texttt{APE\_POIS}) of the source being background fluctuation. This probability can be expressed in terms of an aperture-photometry likelihood \texttt{APE\_LIKE}$=-\ln$\texttt{APE\_POIS}.
In addition, \texttt{apetool} also creates a sensitivity map corresponding to a given \texttt{APE\_LIKE} threshold.
We adopt a threshold of \texttt{APE\_LIKE}$>10$, with an exception for the shallower, scanning-mode data of A3391 (increased to 15, see Sect.\ref{sec:popu}).

\begin{figure*}[!hptb]
\centering
\includegraphics[width=0.8\textwidth]{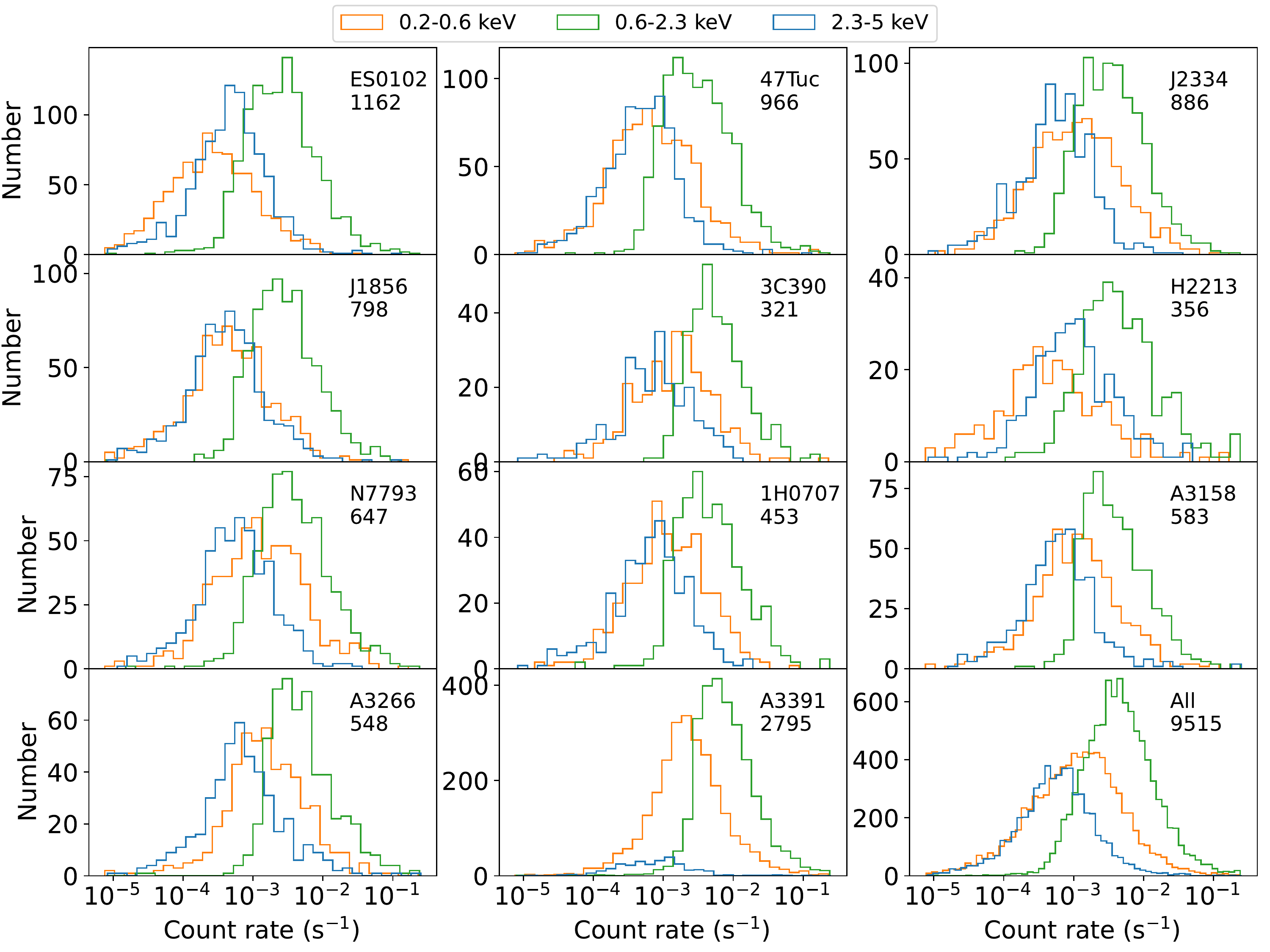}
\caption{0.2--0.6 (orange), 0.6--2.3 (green), and 2.3--5~keV (blue) count rate distributions of the sources for each field and for the whole catalog (lower right). The number of sources are also printed in each panel.
\label{fig:ratehist}}
\end{figure*}

\section{Results}
\label{sec:results}
\subsection{The catalog}

We compile a catalog of $9515$ sources detected in the 11 fields using the method described above in Sect.\ref{sec:method}.
This method detects both point sources and extended sources, but it is only optimized for the detection of point sources. Since we assumed a beta model profile for extended sources, diffuse emission with irregular shape cannot be well fitted and thus results in a crowd of spurious sources. Because of PSF uncertainty and signal fluctuation, ultra-bright point sources also give rise to a crowd of spurious sources in the outer wing.
Such spurious sources are an inherent weakness of the PSF-fitting method in dealing with bright sources, either point or extended. Fortunately, such cases are rare in the X-ray sky. By visual inspection, one can easily identify such spurious sources as a clump, often including both point-like and extended sources, crowding around an ultra-bright source, and thus remove them manually.
In the J1856 field, we removed two clumps of spurious sources at the positions of the two brightest sources in this field, \Final{that is}, the target J185635.1-375433 and a bright extended source RX~J1855.5-3806.
The ES0102 field has a large number of bright SNR and most of them have irregular shapes. Such diffuse emission results in many clumps of spurious sources, we manually remove them from the catalog. \Final{We note} that all these removed sources are located in the mask region (Fig.~\ref{fig:images2}) that is excluded from the point-source subsurvey, and thus have been irrelevant in the number counts analysis anyway.

For each source, count rates are \Edit{both measured during the source detection in the three source-detection bands (0.2--0.6, 0.6--2.3, 2.3--5~keV) and during post-hoc forced photometry in the two forced-photometry bands (0.5--2, 2.3--5~keV).}
The number of sources in each field and the count rate distributions in each field are displayed in Fig.~\ref{fig:ratehist}.

\begin{figure*}[!hptb]
\centering
\includegraphics[width=0.40\textwidth]{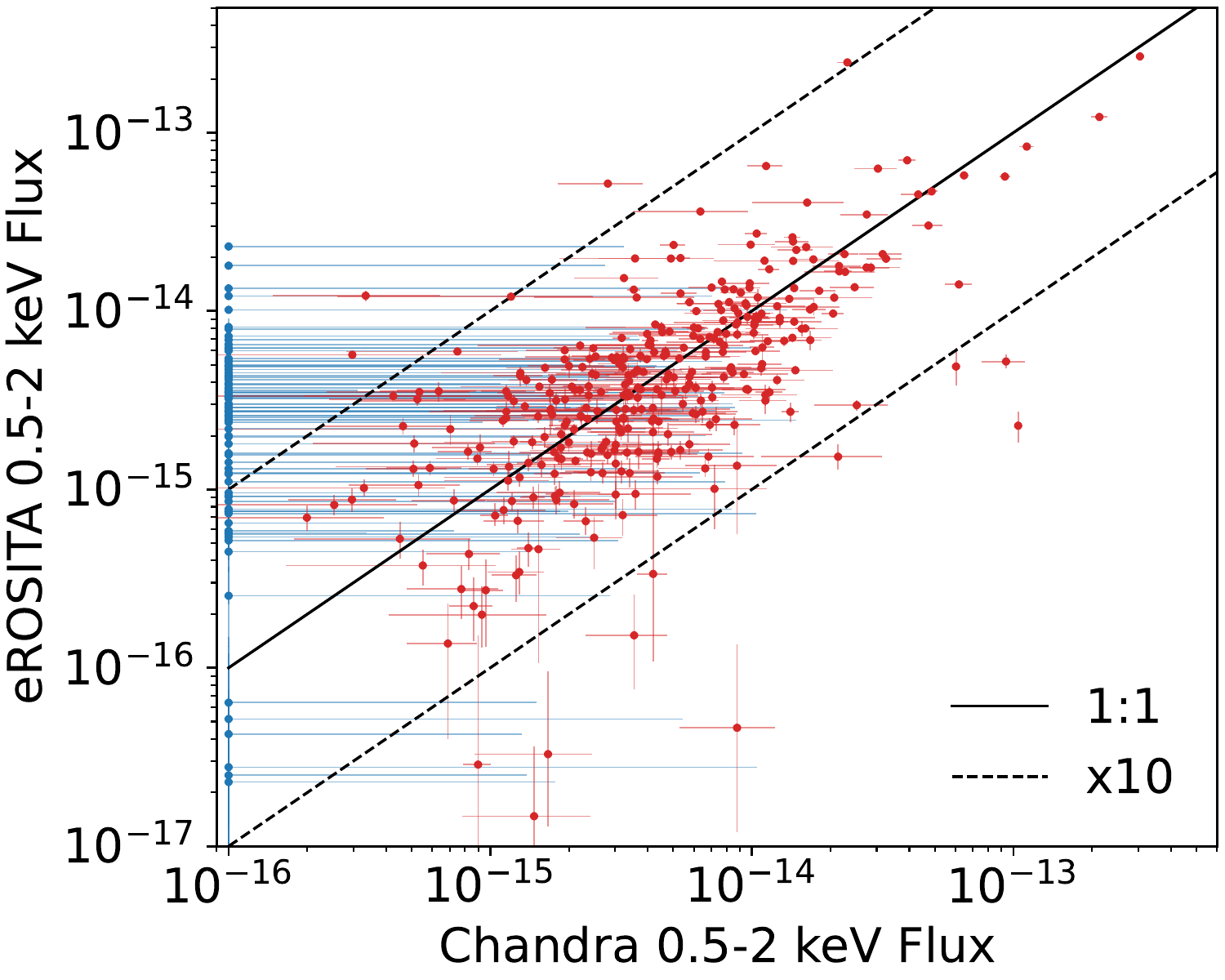}
\includegraphics[width=0.40\textwidth]{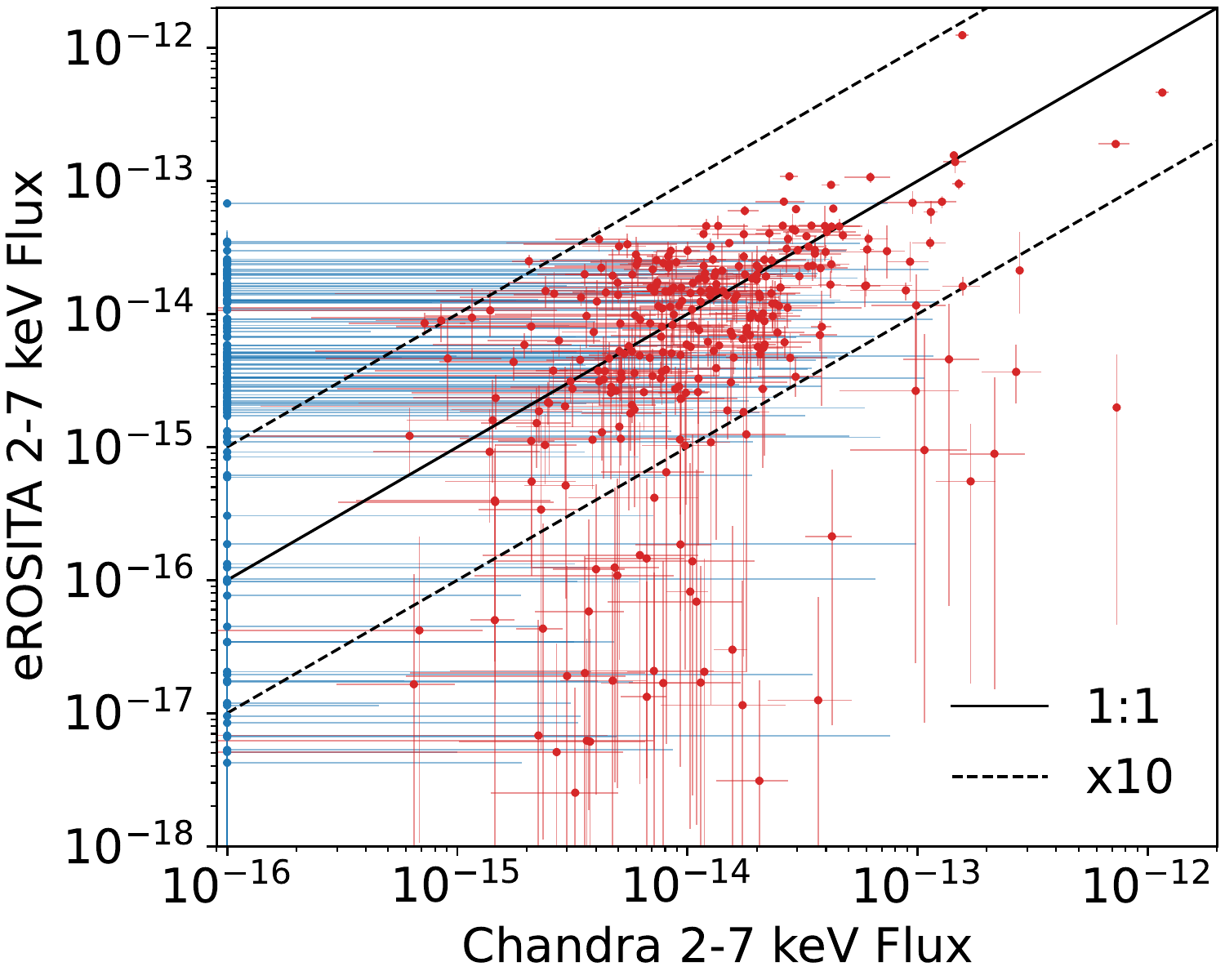}
\includegraphics[width=0.40\textwidth]{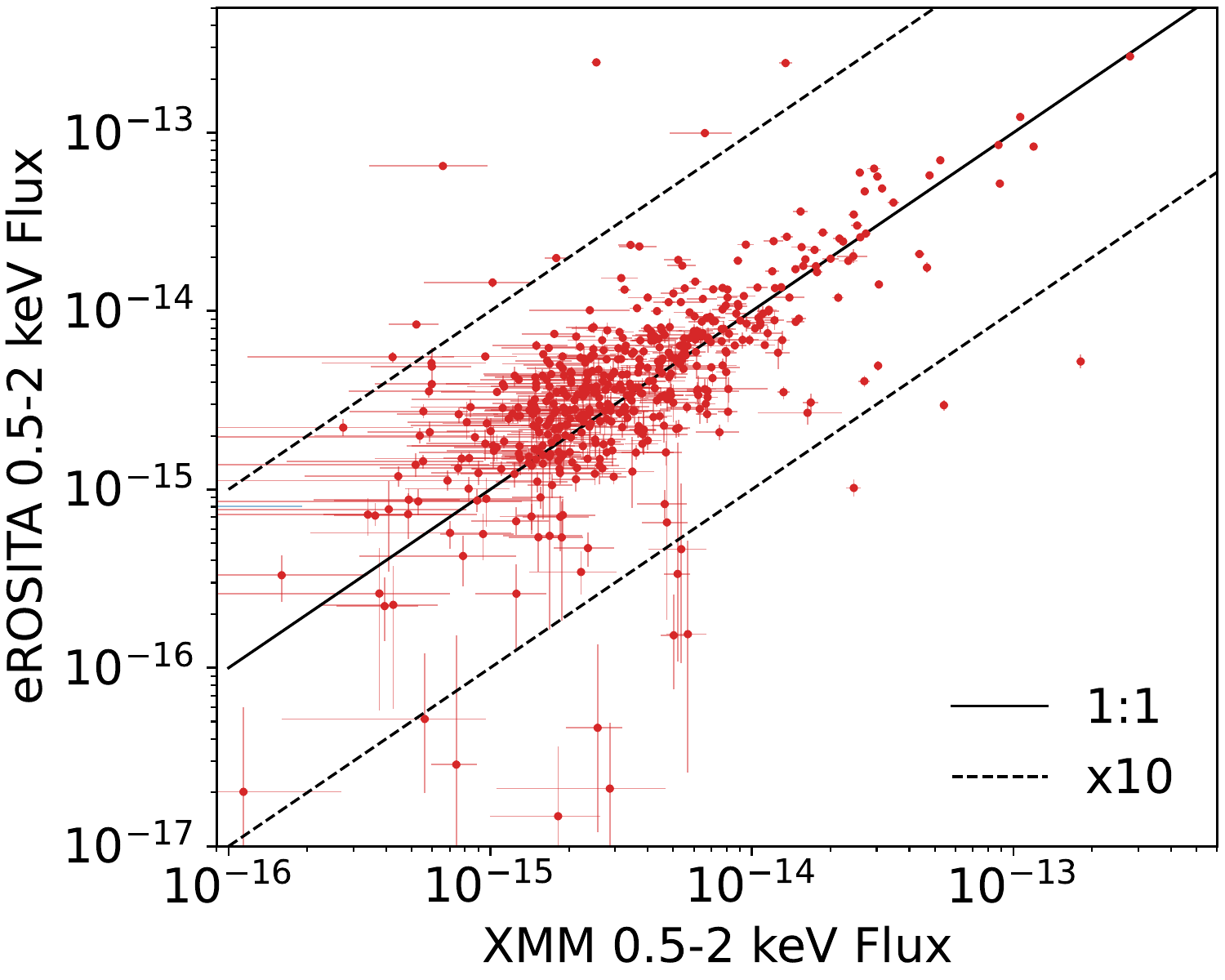}
\includegraphics[width=0.40\textwidth]{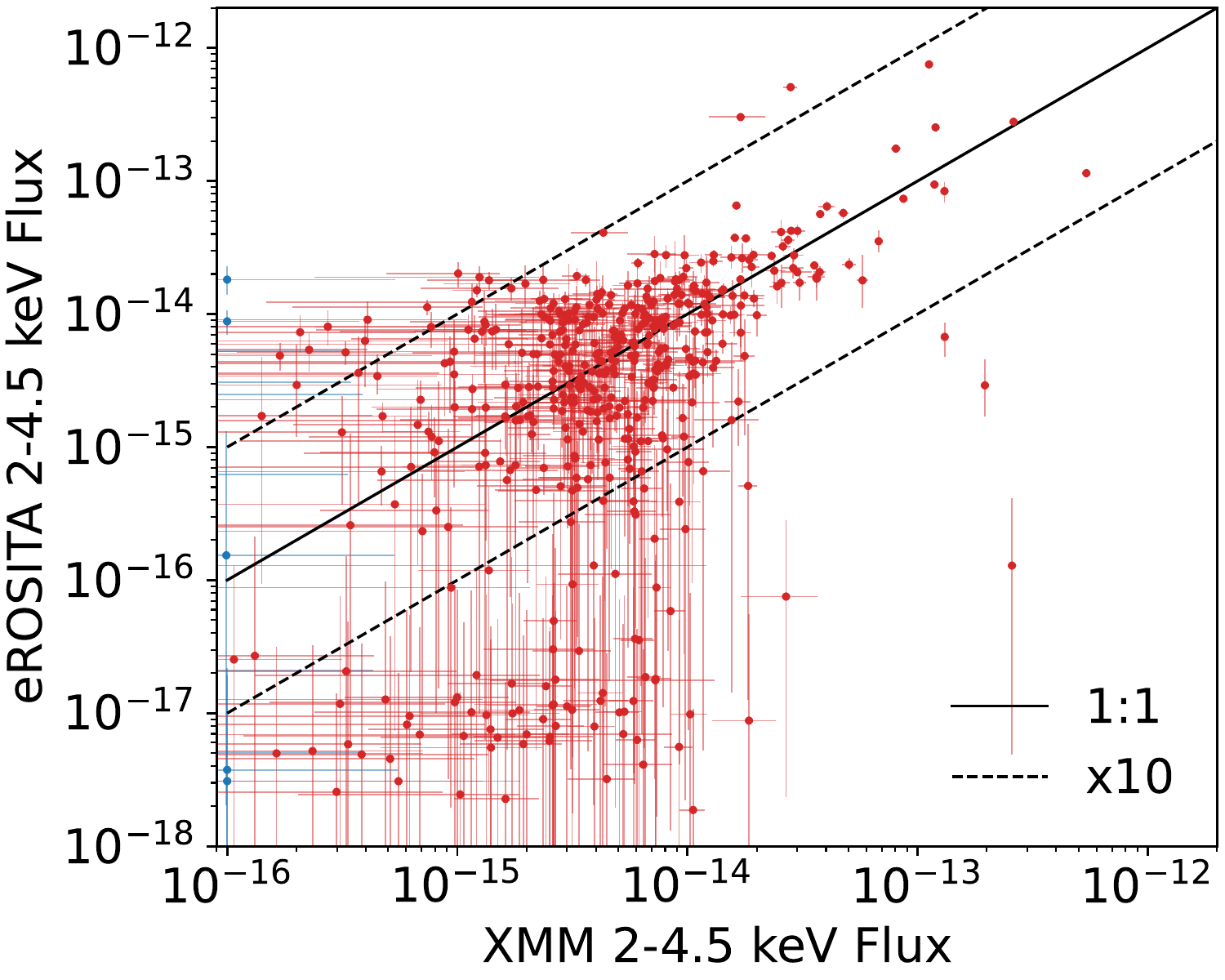}
\caption{Comparisons of the soft- (left) and hard-band (right) X-ray fluxes between \eROSITA and \Chandra measurements (upper) and between \eROSITA and \XMM measurements (lower).
  Sources with a \Chandra \Final{or} \XMM flux of 0 (even though detected) are plotted in blue and others are plotted in red.
  Black dashed lines indicate 10 times deviations in the \eROSITA fluxes from the 1:1 line.
  The \eROSITA 2--7~keV flux is converted from the 2--4.5~keV flux (by a factor of 1.657) assuming a power-law spectrum with a slope of 1.7.
\label{fig:eRO_CSC_XMM}}
\end{figure*}

\subsection{Source fluxes}
To convert source count rates to Galactic-absorption-corrected fluxes, we assume a power-law spectral model with a slope of $\Gamma=1.7$ with Galactic absorption.
Instead of a mean Galactic absorption column density \NH value for each field, we adopt the Galactic \NH at each position by creating a Galactic \NH map for each field as follows.
First, we obtain the HI4PI \citep{HI4PI2016} H$_\textrm{I}$ column density map and the SFD dust extinction map \citep{SFD1998}. Then we calculate the H$_2$ column density map from them adopting the empirical correlation presented by \citet{Willingale2013}. Finally we create the total \NH map as $N_\textrm{HI}+2N_\textrm{H2}$.
With the \NH map, we calculate the energy conversion factors (ECF) at each location and create an ECF map for each relevant band.
The ECF of each source in each band is read from the ECF maps and then used to convert \Final{the} source count rate to Galactic-absorption-corrected flux.

In addition to the Galactic-absorption-corrected flux measurement through model-dependent ECF, we also measure the observed fluxes for each source through spectral fitting.
We use the eSASS task \texttt{srctool} to calculate the source and background extraction regions and extract spectra from them.
The extraction regions are calculated using the same method as described in \citet{Liu2021_AGN} for eFEDS point sources, with the only difference being that we adopt here a ratio of 100 between the background and source extraction areas (instead of 200), because the observations of this work are much deeper than eFEDS.
We use the same Bayesian method as described in \citet{Liu2021_AGN} to analyze these spectra. Since redshift measurements are not yet available, we do not analyze the spectra with physical models in this work. As the aim is flux measurement, we adopt a phenomenological model composed of an absorbed power-law and an APEC plasma model (``wabs*powerlaw+apec'' in Xspec), balancing simplicity and flexibility.
We adopt a uniform prior between -1 and 5 for the power-law slope, a log-uniform prior between $4\times 10^{19}$ and $4\times 10^{24}$~cm$^{-2}$ for the absorbing column density, and a log-uniform prior between 0.04 and 4~keV for the plasma temperature.
We use this model to fit the data in the 0.2--3~keV band to measure the observed fluxes in the 0.2--0.5, 0.5--1, and 1--2~keV band, and fit the data in the 1.9$\sim$5.5~keV band to measure the observed fluxes in the 2.3--5, and 2--4.5~keV band.
Since our multicomponent model can broadly describe either thermal emission of hot plasma, inverse-Compton emission of AGN or X-ray binaries, or \Final{a} combination of them with various properties (e.g., temperature), the flux measurement based on this model is relatively model-independent.

We noticed that a large fraction of the field ES0102 (in SMC) was previously observed by both \XMM and \Chandra, so we use this field to test the \eROSITA flux measurement.
Adopting a search radius of 20\arcsec, we find the \XMM counterparts of our sources from the \XMM SMC point-source catalog \citep{Sturm2013} and the \Chandra counterparts from the Chandra Source Catalog Release 2.0 \citep[CSC2.0;][]{Evans2020}.
As displayed in Fig.~\ref{fig:eRO_CSC_XMM}, we find good agreements between the previous \XMM \Final{and} \Chandra measurements and the \eROSITA measurements in both the soft and hard bands.
Some sources are showing drastic variability between the \XMM \Final{or \Chandra observations} and the \eROSITA observations, which are likely HMXBs in the SMC which are characterized by large long-term flux changes \citep{Haberl2016}. A more detailed analysis of such sources is out of the scope of this paper and we refer to \citet{Haberl2021}, where an analysis of the \eROSITA CalPV data of HMXBs in the SMC is presented.

\begin{figure*}[!hptb]
\centering
\includegraphics[width=\textwidth]{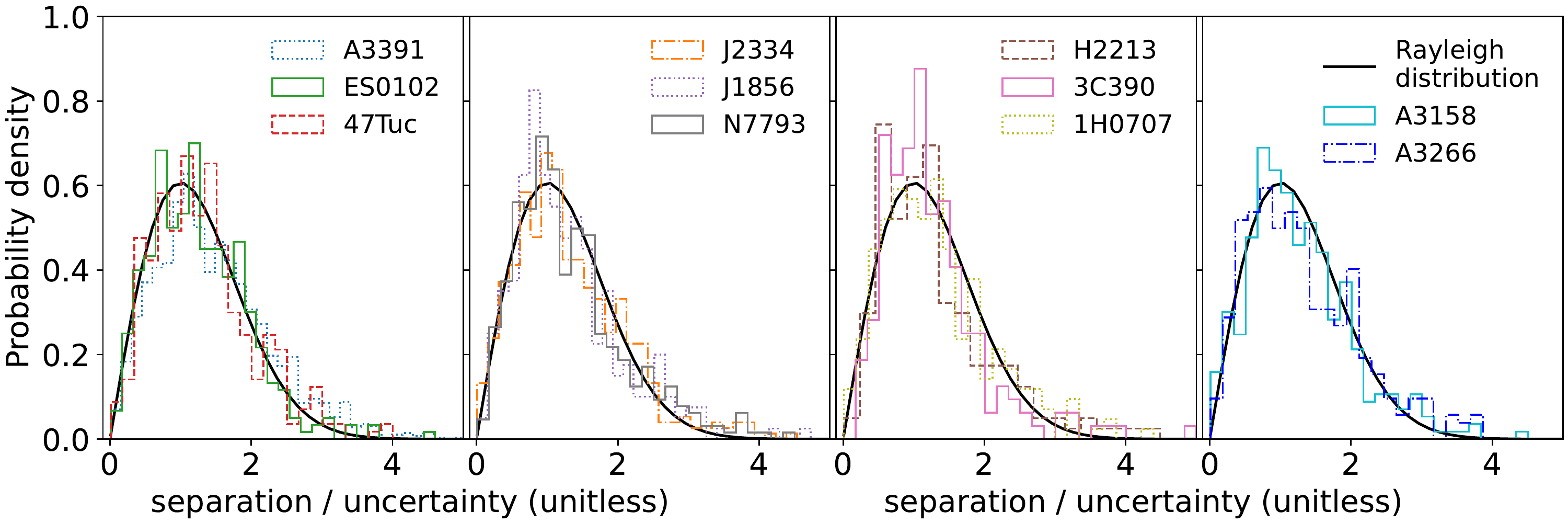}
\caption{Distributions of X-ray to CatWISE positional separations divided by positional uncertainties ($\sqrt{(\texttt{RADEC\_ERR}^2+\texttt{SYS\_ERR}^2)/2}$) for sources with highly reliable counterpart ($p_{any}>0.5$ and $p_i>0.5$) in the 11 fields. All of them are consistent with the Rayleigh distribution (\Edit{with $\sigma=1$;} black line).
\label{fig:Rayleigh}}
\end{figure*}

\subsection{Counterpart identification and astrometric correction}
\label{sec:ctp}

We identify the multiband counterparts of our X-ray sources following the counterpart identification of the eFEDS catalog \citep{Salvato2021}.
However, these CalPV fields are not as well covered by multiband surveys as the eFEDS field. So we only search the CatWISE2020 catalog \citep{Marocco2021}.
We run NWAY \citep{Salvato2018} on our catalog adopting the IR magnitude and color priors that were used for the eFEDS catalog \citep{Salvato2021}.
It could be helpful to perform further optimization on the priors and to include more multiband surveys in the counterpart identification for some of the fields when needed.
In this work, we only run the identification for all the fields systematically, and thus with not much elaboration.

We perform counterpart identification and astrometric corrections at the same time.
First, we use the raw X-ray coordinates and uncertainties to search for WISE counterparts.
Then we select the best counterparts with high reliability ($p_{any}>0.5$ and $p_i>0.5$) as a reference catalog and use the method described in Sect.\ref{sec:astrometric} to calculate the required astrometric corrections.
Since we have found that rotation corrections are negligible, we fix $\Delta\Theta$ at 0 and only measure a shift in $\Delta$RA and $\Delta$DEC. For the fields with a single pointing position, we can exclude the FOV border outside an off-axis angle of 25\arcmin and take the off-axis angle into account through the cost function (equation~\ref{equ1}).
For the four fields with multiple pointings or scanning observation, the off-axis selection is canceled and the off-axis item in the cost function is eliminated.
Besides calculating the required shift, we also add a systematic uncertainty to the positional uncertainty \texttt{RADEC\_ERR} in quadrature.
The systematic uncertainty \texttt{SYS\_ERR} is measured by comparing the sorted X-ray to IR normalized positional separations (separation$/\sqrt{0.5(\texttt{RADEC\_ERR}^2+\texttt{SYS\_ERR}^2)}$) with the inverse cumulative Rayleigh distribution function through a least square minimization in the range between 0.1 and 3.
Then we use the updated coordinates and positional uncertainties to rerun the counterpart identification.
Eventually, we find the WISE counterparts with $p_{any}>0.1$ for $7327$ (77\%) sources.
For $472$ out of them, we also find one \Final{or} more secondary counterparts.
Both the best and the secondary WISE counterparts are presented in the counterpart catalog (Appendix~\ref{sec:cat}).

Using the updated counterpart catalog, we repeat the calculation of corrections above and find that the required corrections are negligible compared with uncertainties, so no further iteration of astrometric correction is needed.
The astrometric \Final{shifts} required for the raw X-ray positions are listed in Table~\ref{tab:obs} and the additional systematic uncertainties required for the X-ray positional uncertainties are listed in Table~\ref{tab:fields}.
Figure~\ref{fig:Rayleigh} displays the distribution of X-ray to IR positional separations based on the corrected X-ray positions and positional uncertainties for the sources with highly reliable counterparts ($p_{any}>0.5$ and $p_i>0.5$) in each field.
All these distributions are consistent with the Rayleigh distribution.

We have identified WISE counterparts for 77\% of the X-ray sources.
For each WISE counterpart, we search for \Gaia EDR3 \citep{GAIAEDR3} counterpart within a maximum separation of 1\arcsec.
\TL{Such a small searching radius suppresses the possibility of finding multiple \Gaia counterparts for one WISE source. The fraction of \Final{noncounterpart} \Gaia sources within a separation of 1\arcsec is 2.5\% of the selected counterparts for the whole catalog. In the field ES0102, this fraction is relatively larger (6\%) because of the SMC.
Such additional \Gaia counterparts within 1\arcsec are often caused by spurious detection of saturated sources. We find \Gaia counterparts for 37\% of the WISE sources, and more detailed counterpart identifications are out of the scope of this work.
}
Based on the proper motion measurement of \Gaia (\texttt{pmra} and \texttt{pmdec} in the RA and DEC directions in units of mas/year and their uncertainties \texttt{pmra\_error}, \texttt{pmdec\_error}), we select the sources with at least 5-$\sigma$ significance of proper motion measurements ($\mathrm{(\texttt{pmra}/\texttt{pmra\_error})^2+(\texttt{pmdec}/\texttt{pmdec\_error})^2}>25$) as proper-motion confirmed stars.
This selection of stars is probably incomplete but relatively robust \citep{Salvato2021}.

\subsection{X-ray variability}
\label{sec:vary}

\begin{figure}[!ptb]
\centering
\includegraphics[width=\columnwidth]{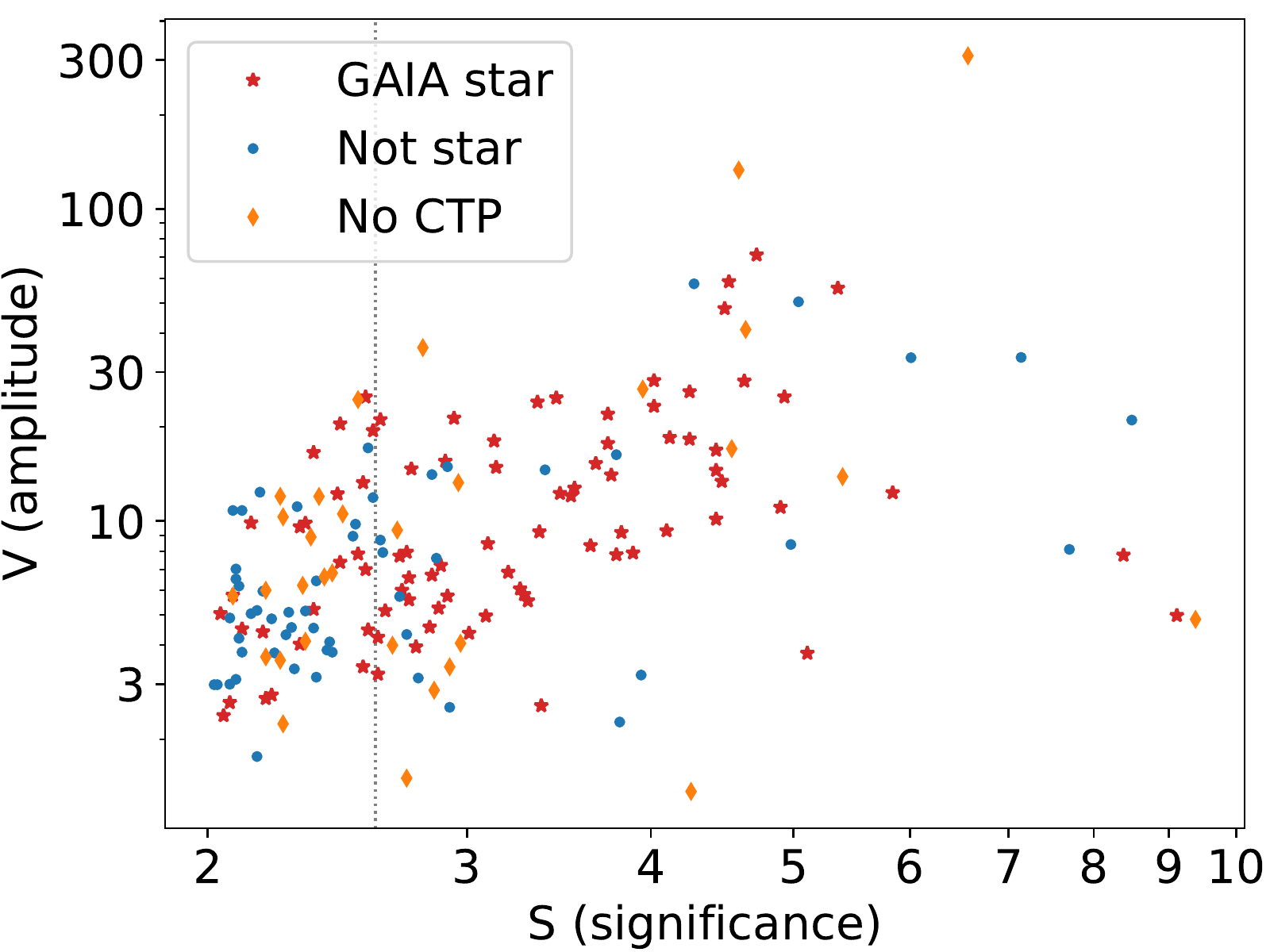}
\caption{\Final{Amplitude} and significance distribution of the variable sources.
  The red stars indicate sources that have a counterpart with proper motion measured by \Gaia above 5-$\sigma$ significance.
  The sources whose counterparts are not confirmed as stars through \Gaia proper-motion measurements are plotted as blue points.
  The ones without counterparts identified are plotted as orange diamonds.
  \Edit{The vertical line indicates a threshold of $2.6$.}
\label{fig:vary}}
\end{figure}

Using the source and background spectra extraction regions, we also extract light curves in the 0.2--5~keV band with a bin size of 100s and search for X-ray variability using the maximum amplitude method \citep{Boller2016, Buchner2021}.
From each light curve, we find the time bin with the lowest count rate upper limit ($\mathrm{Rate+\Delta Rate}$) and the bin with the highest count rate lower limit ($\mathrm{Rate-\Delta Rate}$, where $\Delta$Rate is the 1-$\sigma$ count rate uncertainty), and use these two bins to define a variability amplitude:
$$V=\mathrm{\frac{Rate_{high}}{max(Rate_{low},\Delta Rate_{low})}},$$
and a variability significance:
$$S=\mathrm{\frac{(Rate_{high}-\Delta Rate_{high})-(Rate_{low}+\Delta Rate_{low})}{\sqrt{\Delta Rate_{high}^2+\Delta Rate_{low}^2}}}.$$
The measurement of $V$ and $S$ is dependent on the time bin size. Narrow bin size helps identify short outbursts and wide bin size helps identify variabilities of small amplitude or low S/N.
Therefore, we rebin the light curves in two ways \Final{to have} at least $30$ total counts and at least $50$ total counts, respectively, in each bin. We select sources with $V>1$ and $S>2$, and for the cases selected with both the two binnings, we adopt the one that provides a larger $S$.
We find $173$ variable sources and present them in a catalog as described in Appendix~\ref{sec:cat}. Their $V$ and $S$ are displayed in Fig.~\ref{fig:vary}.
\Edit{Some low-$S$ cases might be due to random fluctuation.
  Following \citet{Buchner2021}, we adopt a threshold of $S>2.6$ to guarantee a 3-$\sigma$ purity. It results in $99$ variable sources.}
This method is sensitive in identifying large-amplitude flares but not small-amplitude variabilities \citep{Buchner2021}.
The counterparts of about half of these variable sources are classified as stars through \Final{\Gaia's} proper motion measurement; the other ones might also include not-yet-identified stars.
\TL{Both the two sources with $V>$100 have only a 1ks-scale flare and no emission outside the flare.}
A few largely variable sources might be HMXBs in the SMC \citep[][]{Haberl2021}.

\subsection{A subsurvey for point-sources}
\label{sec:pntsurvey}

\begin{figure*}[!hptb]
\centering
\includegraphics[width=0.35\textwidth]{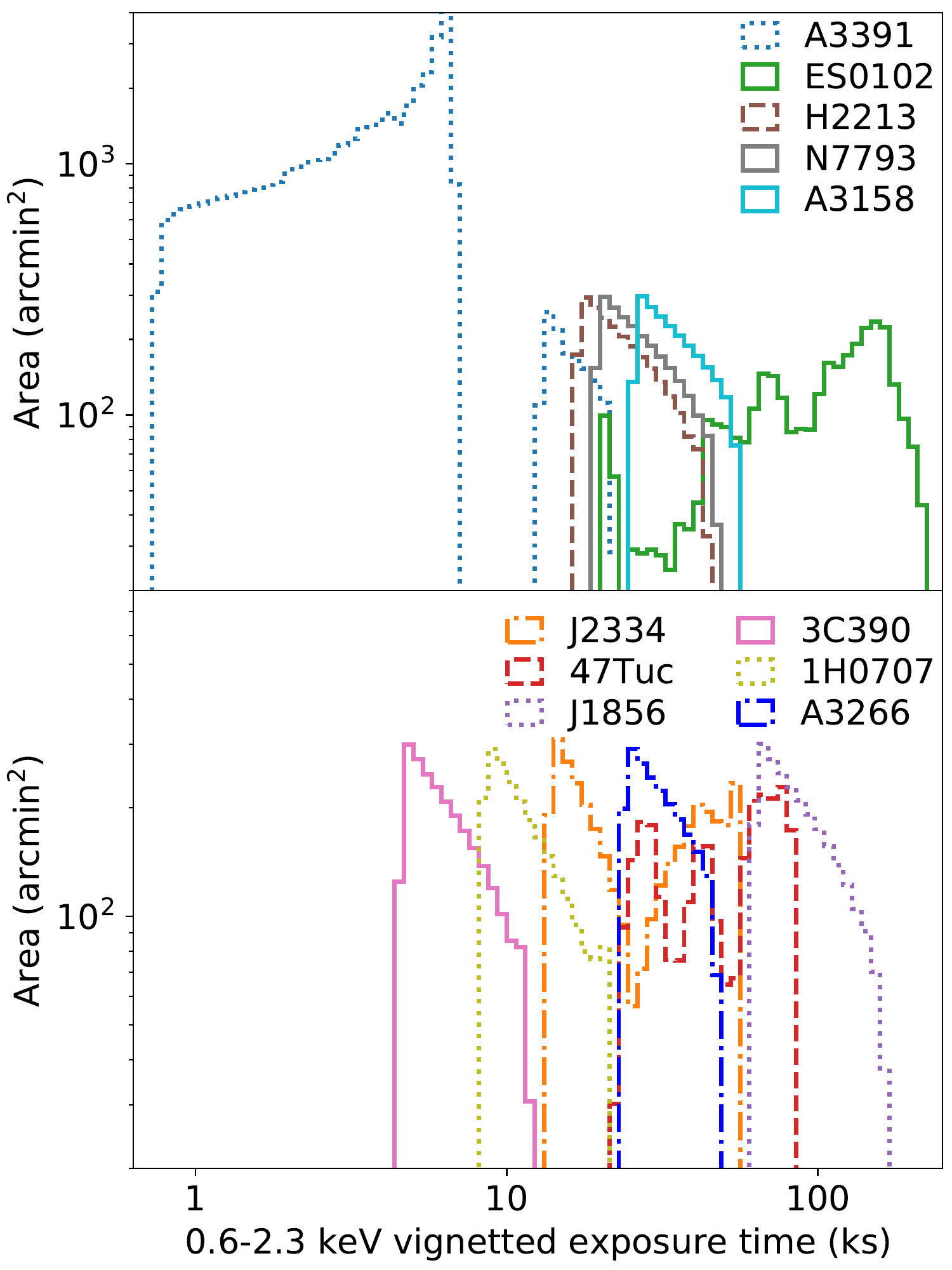}
\includegraphics[width=0.35\textwidth]{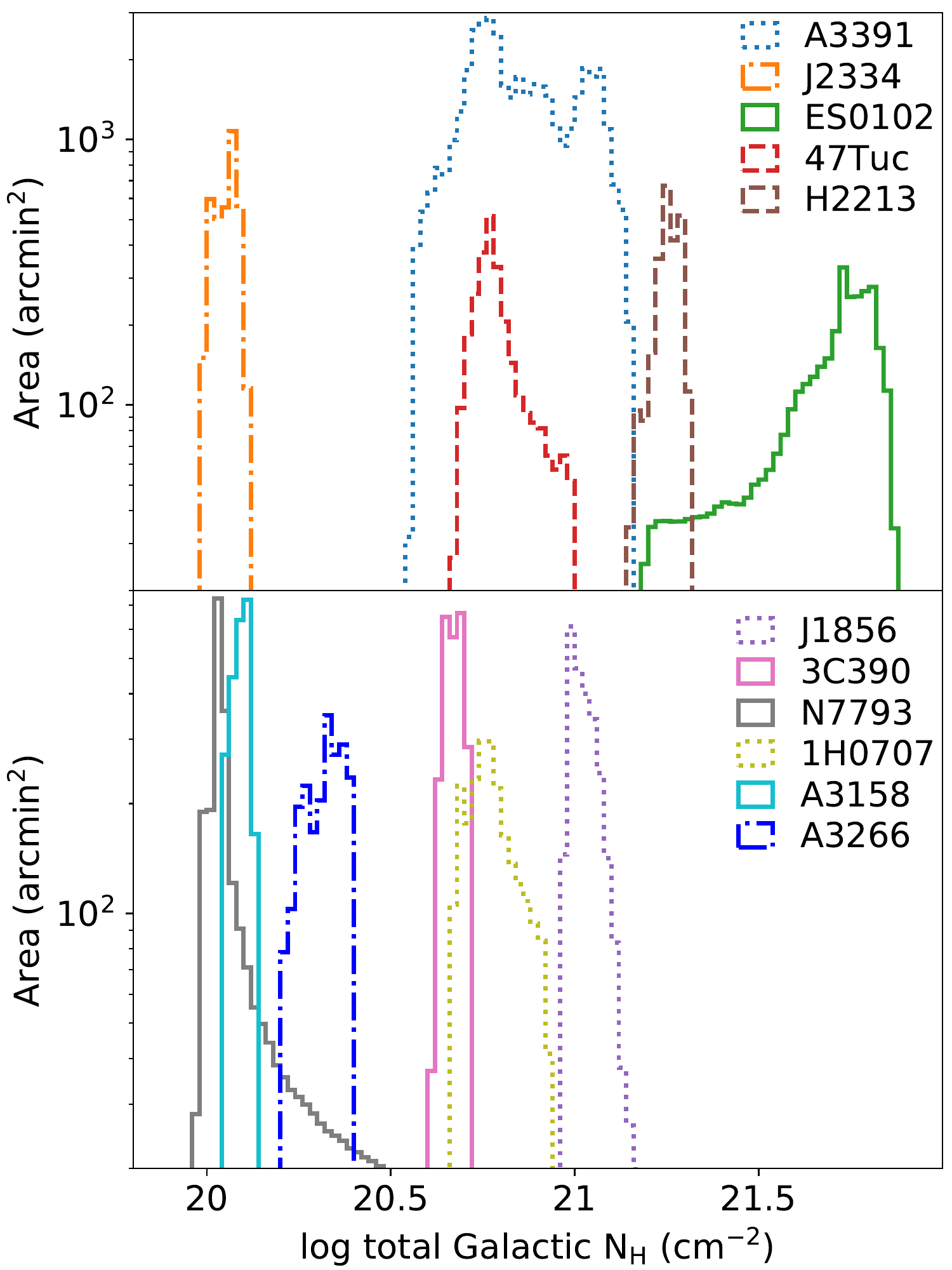}
\caption{Distributions of exposure depth and total Galactic \NH for each field in the point-source subsurvey region.
  The 11 fields are displayed separately in two panels for representation.
\label{fig:Exp_NH}}
\vspace*{\floatsep}
\includegraphics[width=0.35\textwidth]{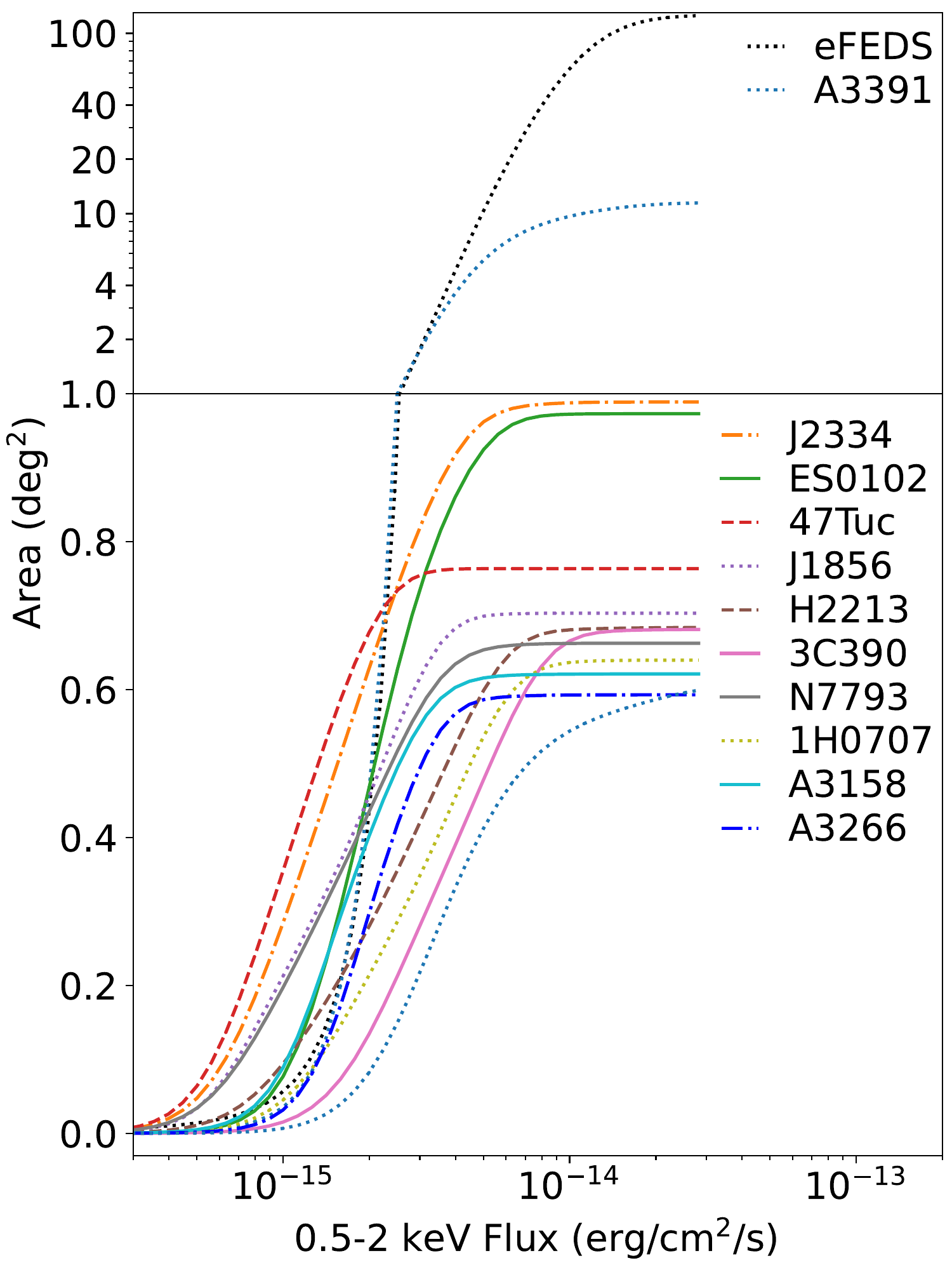}
\includegraphics[width=0.35\textwidth]{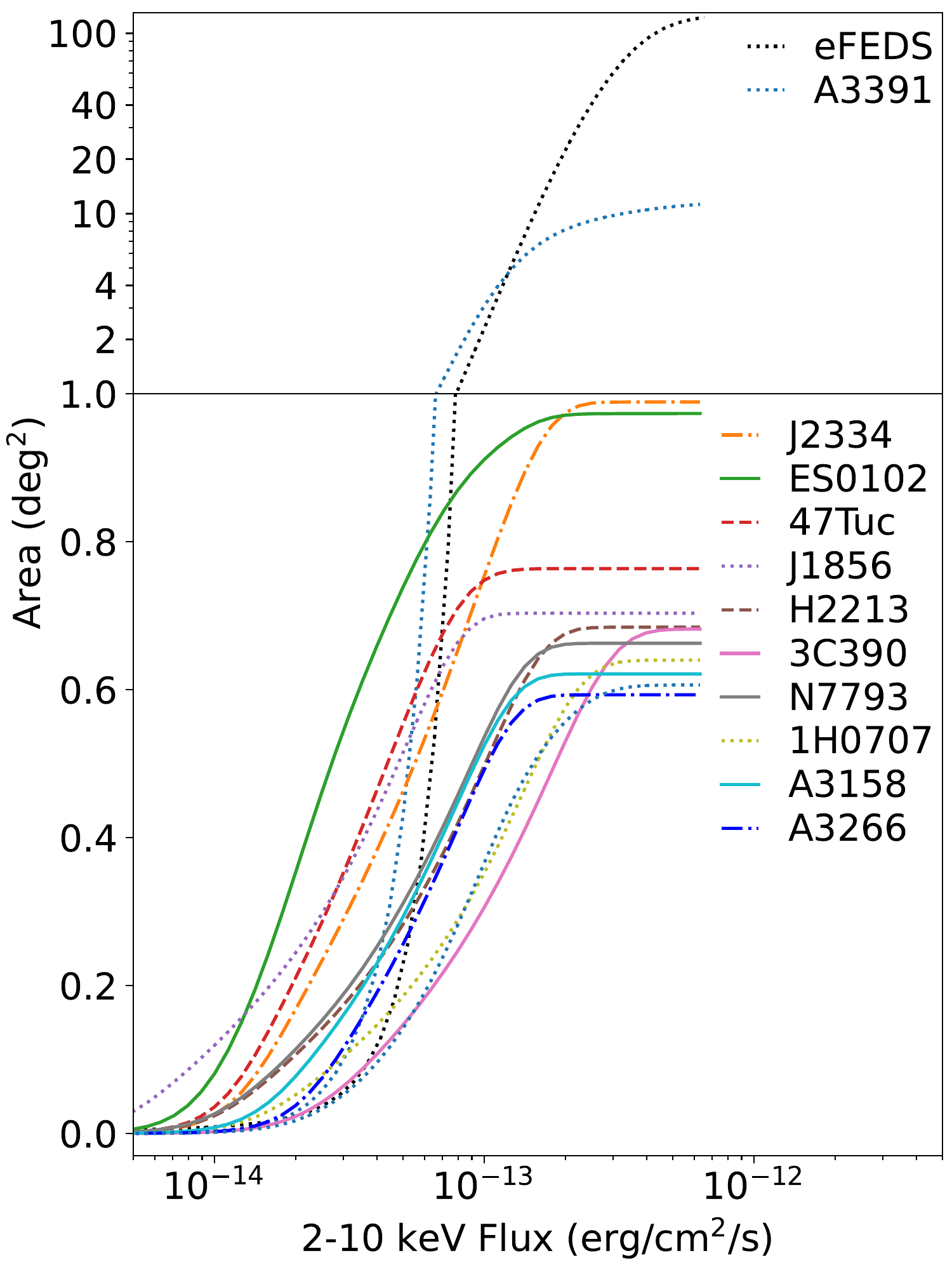}
\caption{\Final{Sky} coverage of each field as a function of 0.5--2 and 2--10~keV Galactic-absorption-corrected fluxes in the point-source subsurvey region.
  The eFEDS survey is also plotted with black dotted lines for comparison.
  All the sky coverage curves are calculated corresponding to an aperture-photometry likelihood threshold of 10.
  The same colors and line styles are adopted as in Fig.~\ref{fig:Exp_NH}.
  For representation, we divide each figure into two parts and adopt linear scale and log scale respectively.
  For the field A3391, besides the overall sky coverage, we also plot the sky coverage calculated for the pointing-mode observation (300014) within an off-axis angle of 28\arcmin.
\label{fig:skycov}}
\end{figure*}

The depth of a survey depends not only on exposure time but also on background and spatial resolution. To make a relatively uniform survey of point sources from our data, we define a point-source survey region (green footprint in Fig.~\ref{fig:images1}, \ref{fig:images2}) in each field, excluding the field border where the depth and resolution are significantly worse and the regions that are dominated by bright, particular sources.
For the A3391 field, covered by scanning, we select the region where the 2.3--5~keV vignetted exposure time is above 0.5~ks, which comprises 80\% area of the whole field.
For the other pointing-mode fields, we select circular regions within a radius of 28\arcmin from FOV centers.
For 47Tuc and ES0102, we also manually select circles with a radius of 30\arcmin and 35\arcmin, respectively, at the deep overlapping regions, excluding the field border where the exposure depth is much lower than the overlapping regions.

When creating the final background map (Sect.\ref{sec:bkg5}), we have created a smoothed count rate map (image divided by exposure map). In this map, after excluding the cores of extended sources as a point source, the diffuse outskirt of extended sources is included as a background component. We make a cut at twice the mean value in the 0.6--2.3~keV map to exclude the regions containing strong diffuse emissions. In this way, we exclude the central regions of the A3266, A3158, A3408, A3391, A3395 clusters, and also a few other extended sources in the 47Tuc and A3391 field (see Fig.~\ref{fig:images1}, \ref{fig:images2}).
\Final{We note} that only the brightest part, where the diffuse emission is too high as background, is excluded, and a significant outer part of the big clusters is still included.
Particularly for the ES0102 field, which contains several \Final{SNRs} in SMC, we manually draw a few circles around SNR 0103-72.6, SNR 0057-72.2, and a few other SNRs around the target 1ES~0102-72.2, to exclude such bright regions (Fig.~\ref{fig:images2}), in which the diffuse emission failed to be detected as extended because of its irregular shape.

In addition to such regions dominated by diffuse emission, we also manually remove the regions around a few particularly selected observation targets.
In the field N7793, we exclude a 5\arcmin-radius circle around the bright nearby galaxy NGC~7793, which harbors a significant population of XRB.
In the field 47Tuc, we exclude the region within the half-mass radius \citep[2.8\arcmin;][]{Harris1996} of the 47~Tuc star cluster, in which hundreds of X-ray sources detected by Chandra \citep{Heinke2005} blend together and appear as diffuse emission under the spatial resolution of \eROSITA.
In the fields J1856 and 3C390, the targets (neutron star J185635.1-375433 and AGN 3C390.3) are ultra-bright point sources. Such bright sources are rare in the sky and did not fall in the region by chance. We draw circles with a 2\arcmin radius to mask them out.

The selected regions compose a random, extragalactic survey with a relatively uniform depth in each field.
\TL{The sources in this subsurvey region are flagged in the catalog (\texttt{inMskPnt}=True).}
We plot the distributions of 0.6--2.3~keV vignetted exposure time and total Galactic \NH for each field in Fig.~\ref{fig:Exp_NH}.
The majority of the A3391 field, covered by only scanning-mode observations, has a much shallower exposure depth ($<$10 ks) than the other pointing-mode observations.
The 3C390 and 1H0707 fields have relatively shallower exposure depth than the others because only one TM was active for 3C390 and only three TMs for 1H0707.
The deepest exposure occurs in the J1856 field and the FOV-overlapping region of the ES0102 field.
Most of the fields have a typical total Galactic \NH $\leqslant 10^{21}$ cm$^{-2}$, except H2213, which targets a dark cloud, and ES0102, which is located in the SMC where the total absorption includes both the Galactic and the SMC components.
This is the reason why, compared with other fields, these two fields have relatively low 0.2--0.6~keV count rate distributions in comparison to the 0.6--2.3 and 2.3--5~keV count rate distributions, as displayed in Fig.~\ref{fig:ratehist}.

In the point-source subsurvey regions, we calculate the number counts of point sources based on the aperture photometry and sensitivity map built using \texttt{apetool} in the 0.5--2 and 2.3--5~keV band.
We assume a power-law spectral model with a slope of $\Gamma=1.7$ and local total Galactic absorption to convert the 0.5--2 and 2.3--5~keV count rates into the 0.5--2 and 2--10~keV Galactic-absorption-corrected fluxes.
Figure~\ref{fig:skycov} displays the sky coverage of each field as a function of soft- and hard-band fluxes.
Since the Galactic-absorption-corrected fluxes are adopted, the sky coverage shows the survey depth for extragalactic sources.
The ES0102, 47Tuc, J2334, and J1856 fields comprise the deepest extragalactic survey based on these data.
The ES0102 field is deep in the hard band but shallow in the soft band because of \Final{the} high total \NH combining the Galactic and the SMC absorption.
The A3391 field is the largest one, with a median exposure depth of 6.9 ks. The eFEDS survey is about 10 times larger than A3391 with a shallower exposure depth (2.2 ks).
Thanks to the deeper exposure, the A3391 field is deeper than eFEDS in the hard band. But it is not deeper than eFEDS in the soft band, because of the high Galactic absorption in this field.
Covered by a pointing-mode observation (300014), the central region of A3391 is much deeper than the rest of the field (Fig.~\ref{fig:Exp_NH}).
This small region is of particular interest because it targets the filament between the galaxy \Final{clusters} Abell~3391 and Abell~3395.
So we also calculate the sky coverage (Fig.~\ref{fig:skycov}) and number counts using only the pointing-mode observation within an off-axis angle of 28\arcmin and masking out the clusters (Fig.~\ref{fig:images2}).
The number counts of all the fields are \Edit{presented in Appendix~\ref{sec:cat}} and discussed in Sect.\ref{sec:popu}.

\section{Discussion}
\label{sec:discussion}
\subsection{The population of X-ray sources}
\label{sec:popu}
\begin{figure*}[!hptb]
\centering
\includegraphics[width=\textwidth]{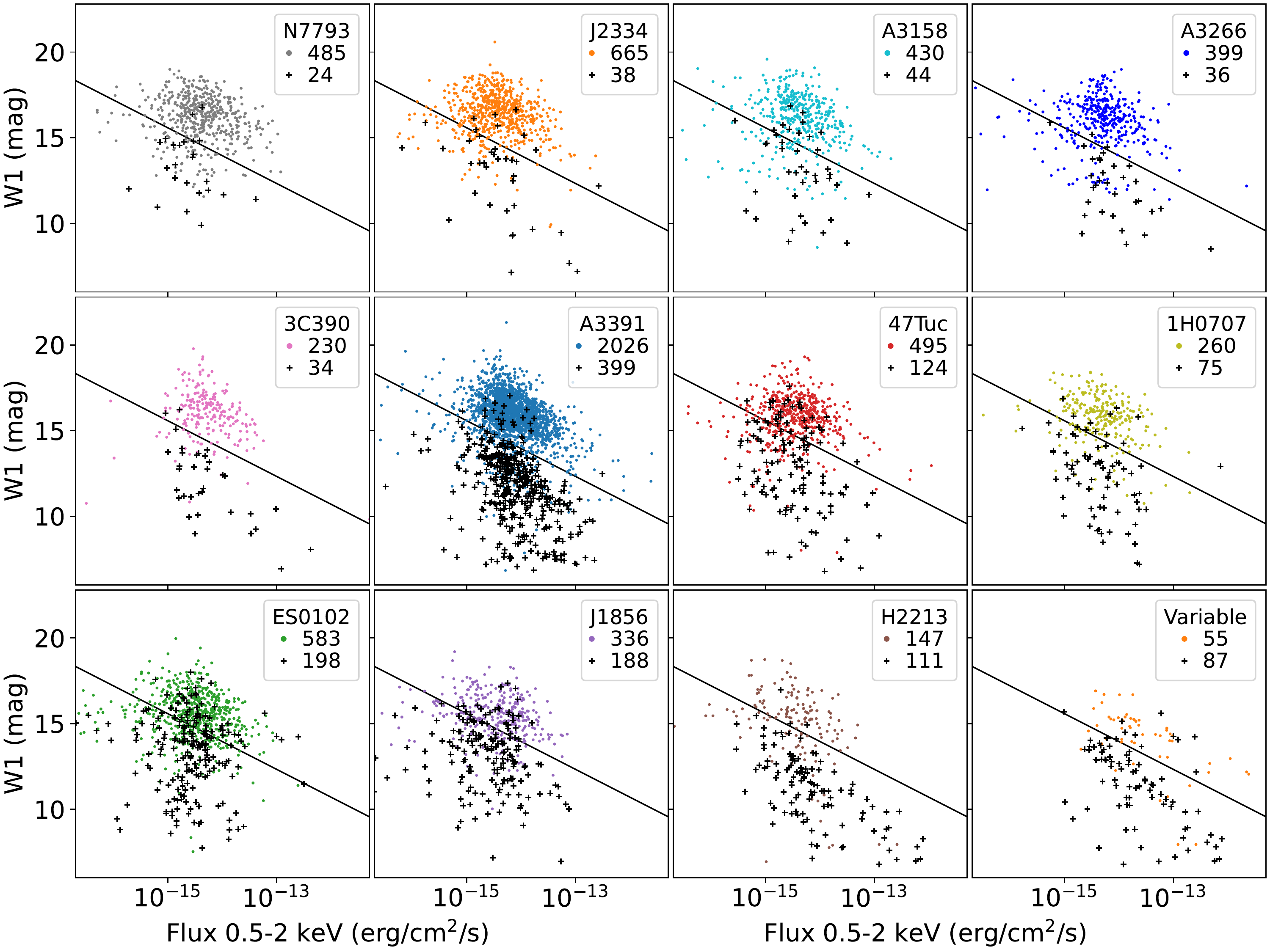}
\caption{Scatter plots in the 0.5--2 keV X-ray flux ($F_x$; measured through spectra fitting) and WISE W1-band Vega magnitude ($W1$) space for catalogs in the 11 fields and for the X-ray variable sources (lower right panel).
  The black line corresponds to $W1=-1.625\times\log F_x-8.8$, which is an empirical boundary between AGN and stars \citep{Salvato2018}.
  The sources with significantly measured proper motions by \Gaia are plotted in black crosses, and the others are plotted as points.
  The numbers of sources are printed in each panel.
\label{fig:X_W1}}
\vspace*{\floatsep}
\includegraphics[width=0.7\textwidth]{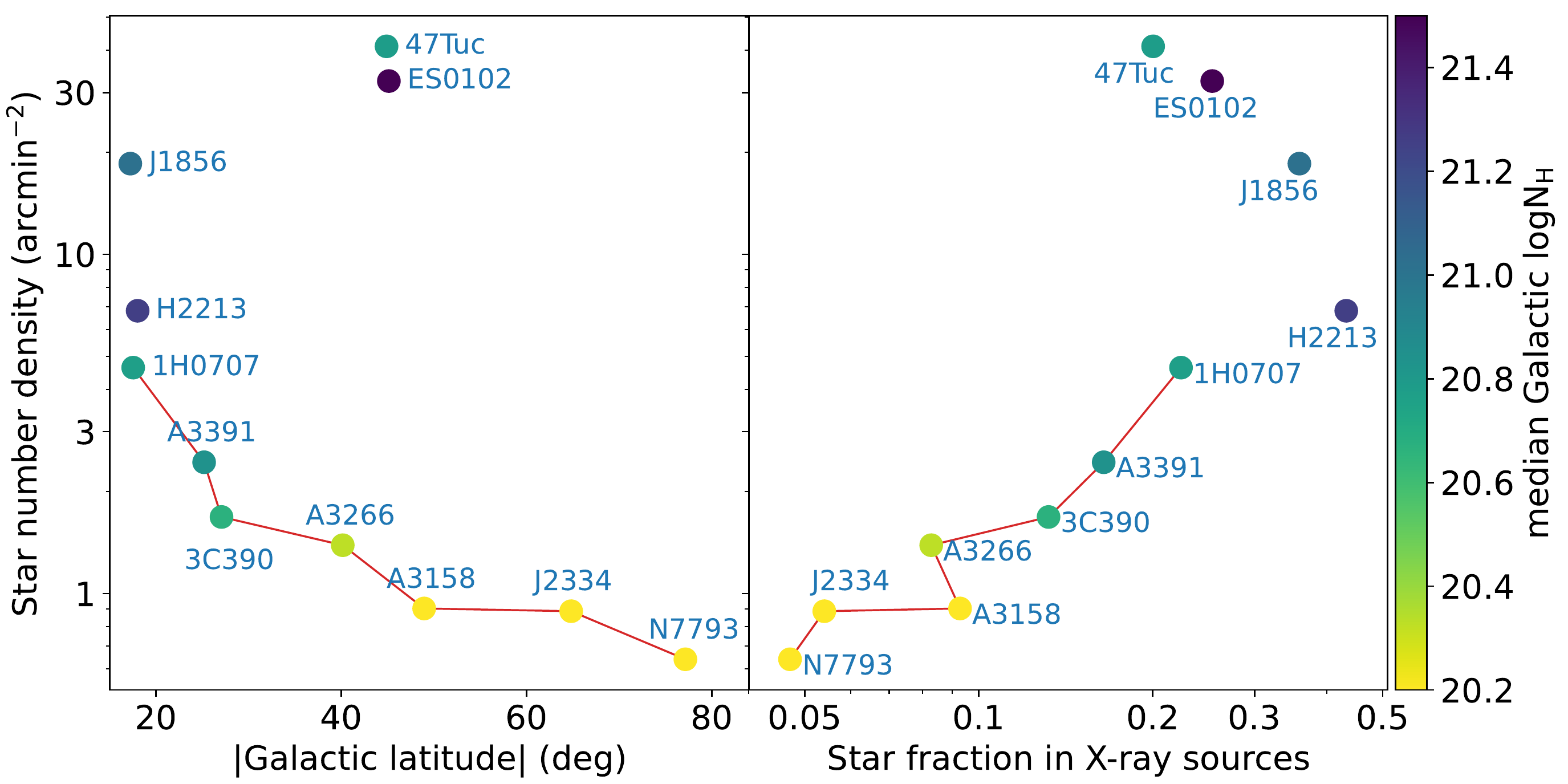}
\caption{Scatter plot of the 11 fields in the space of the number density of proper-motion-confirmed stars versus the absolute Galactic latitude (left) and the fraction of proper-motion confirmed stars among the best counterparts (right).
  All the fields except 3C390 are located in the Southern Galactic sky.
  The color indicates the median value of total Galactic \NH in the field.
\label{fig:starfrac}}
\end{figure*}

  In the extragalactic sky, the main X-ray sources are AGN and galaxy clusters.
  In this catalog, a small fraction of the sources are extended ($225$ sources) and thus likely galaxy clusters, and the other point sources should be AGN or stellar objects.
  In this work, we only performed a rough counterpart identification for the sources.
  To study the different populations of astronomical objects, which is out of the scope of this paper, we recommend more detailed counterpart identification and analysis with more multiband data.
  Here, we only test whether these fields are representative of the extragalactic sky using the point-source number counts and the WISE and \Gaia counterparts.

\Edit{We compare the point-source number counts of each field (see more details in Appendix~\ref{sec:cat}) with that of} a few typical extragalactic surveys, including the eFEDS survey \citep[][]{Brunner2021}, the XMM-COSMOS survey \citep[][]{Cappelluti2009}, and a combination of a few \Chandra surveys presented by \citet{Georgakakis2008}.
These surveys are selected at extragalactic positions avoiding high star density and absorbing column density in the Galactic plane and any ultra-bright sources in the X-ray sky. Such surveys are perfect for the characterization of extragalactic populations, especially AGN and galaxy clusters, if we put aside the residual potential cosmic variance.
On the other hand, the fields used in this work were \Final{observed} targeting particular sources for various purposes (Table~\ref{tab:fields}) of the \eROSITA CalPV program.
By manually masking out ultra-bright targets and regions with bright diffuse emission (Sect.\ref{sec:pntsurvey}), we managed to turn this serendipitous survey into a more stochastic sampling of the X-ray point source population, with an acceptable level of background contamination (diffuse emission included).
We find that most of the fields have consistent point-source number counts as the more widely studied extragalactic surveys.
However, a few fields show some peculiarities, indicative of different populations of X-ray objects, as discussed below.

In Fig.~\ref{fig:X_W1}, we plot the distributions of best IR counterparts (\texttt{match\_flag}==1 and $p_{any}>0.1$) in the space of X-ray 0.5--2~keV observed flux ($F_x$) and IR W1-band magnitude ($W1$) for the 11 fields.
The number of sources is also printed in the figure.
Most of the proper-motion confirmed stars are located below the empirical line $W1=-1.625\times\log F_x-8.8$ suggested by \citet{Salvato2018} to separate extragalactic (with high X-ray to IR flux ratios) from galactic (with low X-ray to IR flux ratios) sources.

In Fig.~\ref{fig:starfrac}, we plot the number density of proper-motion confirmed stars in the \Gaia catalog within a 30\arcmin-radius circular region at the position of each field.
For the relatively larger fields J2334, ES0102, and A3391, we adopt larger radii of 40\arcmin, 40\arcmin, and 100\arcmin, respectively.
The number density is plotted as a function of the Galactic latitude and the fraction of stars among the selected counterparts of the X-ray sources.
We find that seven fields (connected by lines in Fig.~\ref{fig:starfrac}) show a relatively smooth anticorrelation (correlation) between the star number density and the Galactic latitude (the star fractions).
All these seven fields show similar point-source number counts as typical extragalactic fields.
For A3158 and A3266, the selected regions (Fig.~\ref{fig:images1}) are inside galaxy clusters but with the central region ($8\sim 10$\arcmin from center) excluded. These two regions show almost identical point-source number counts as typical extragalactic fields in both the soft and hard bands, because most of the X-ray sources are background AGN, on which the clusters have no impact other than providing enhanced diffuse background.
The fields J2334 and N7793 are the two fields with the lowest Galactic absorption, which are ideal regions for extragalactic surveys, after excluding the particular target NGC~7793.
The 3C390 field is also a representative extragalactic region after excluding the target, but it is the shallowest one and thus has a small number of sources.
Small bumps can be seen in the soft-band number counts around $10^{-14}$ erg cm$^{-2}$ s$^{-1}$ for 1H0707 and N7793, possibly because of cosmic variance, or uncertainty in Galactic absorbing column density, or relatively larger number of X-ray stars (as for 1H0707).

The target of the H2213 field is the Corona Australis (CrA) Molecular Cloud \citep{Dobashi2005, Peterson2011}, which is one of the nearest star-forming regions to the Sun. The cloud gives birth to a crowd of young stellar objects \citep{Peterson2011} and obscures distant AGN.
As shown in Fig.~\ref{fig:X_W1} and Fig.~\ref{fig:starfrac}, this field has an extremely high fraction of stars and many of these stars are abnormally bright in X-ray.
As a result, the point-source number counts in this region appear completely different from typical extragalactic fields (Fig.~\ref{fig:logNlogS}).
Assuming all the X-ray sources are extragalactic and falsely correcting the fluxes of stars, which are already abnormally high, with the column density of the dark cloud, we get an extremely high density of bright sources.

Located only 1.4\arcdeg away from H2213, the field J1856 is also close to CrA, but not in the main region. It is located at the far end of the ``streamer'' of CrA \citep{Peterson2011}.
As shown in Fig.~\ref{fig:X_W1} and Fig.~\ref{fig:starfrac}, this field also has a high fraction of stars, but lacks extremely bright ones as in H2213.
As discussed in \citet{Peterson2011}, the $116$ candidate young stellar objects of CrA strongly cluster around the center (H2213) and the J1856 region does not harbor any \Final{of them}.

As displayed in Fig.~\ref{fig:starfrac}, the fields J1856, 47Tuc and ES0102 have extremely high number densities of stars, which might lead to high number counts of point sources in these fields.
Both J1856 and ES0102 have high \NH of absorption, which applies to the background AGN but not to the stars in the Galaxy or SMC.
However, we applied the same absorption correction to all the sources.
The over-correction to stars causes overestimated soft-band fluxes and thus higher number counts.
In the hard band, where the fluxes are less affected by absorption correction, their number counts only show slight excess over that of typical extragalactic fields (Fig.~\ref{fig:logNlogS}), possibly caused by a small number of HMXBs.

\TL{
There are $198$ proper-motion confirmed stars among the best counterparts of the ES0102 sources.
Out of them, $92$ concentrate in a narrow region in the \texttt{pmra}--\texttt{pmdec} space (0.4<\texttt{pmra}<1.2 and -1.7<\texttt{pmdec}<-0.8), which corresponds to the typical proper motion of SMC members \citep{Gaia2018}.
We use this region to roughly select SMC members and exclude them from the number counts calculation.
As displayed by the dashed lines in Fig.~\ref{fig:logNlogS}, the soft-band number counts in the ES0102 field are reduced.
However, this selection of SMC members is far from complete, and there is still a significant excess of soft-band number counts over that of typical extragalactic fields.
}

Among the three high-star-density fields, 47Tuc has the largest number density of stars but the lowest fraction of stars among X-ray sources (Fig.~\ref{fig:starfrac}). This is because the globular cluster 47~Tuc is extremely old \citep[13 Gyr;][]{Forbes2010} and lacks young (coronally emitting) stars. On the other hand, the dense core of 47~Tuc (within the 2.8\arcmin half-mass radius) has a large number of binary systems and thus a significant population of X-ray emitters \citep{Heinke2005}. However, this core region has been excluded because all the sources are blended under the spatial resolution of \eROSITA.
As a result, the 47~Tuc field only shows slight excess in the soft- and hard-band number counts over that of typical extragalactic fields (Fig.~\ref{fig:logNlogS}).

For the A3391 field, the point source number counts in the large outer region are similar to that of typical extragalactic surveys.
In the small inner region, the number counts are lower than that of typical extragalactic surveys in the soft band and similar to that in the hard band.
One possible reason is underestimated absorption to the background AGN, which causes underestimation of the absorption-corrected soft-band fluxes.
This region targets the filament between Abell~3391 and Abell~3395, in which \eROSITA has found prominent hot gas and potential warm gas \citep{Reiprich2021}.
If neutral gas also exists in the direction of such filaments, it could reduce the soft-X-ray number counts of the background AGN.

\subsection{100ks pointing-mode simulation}

Based on detailed simulations of the eFEDS survey, \citet{Liu2021_sim} discussed the source detection strategy in survey mode and measured the selection function and sample purity of the eFEDS catalogs.
In this work, we simulate a pointing-mode observation of an extragalactic field using the same method.
The 11 fields have different observing modes and exposure depths.
We do not simulate each field with the same exposure map as the real data.
Instead, we simulate a 100 ks observation with a single pointing direction.
We create an observing attitude file (pointing attitude at an array of time) by truncating the 166 ks stacked observation of J1856 to 100 ks and shifting the pointing directions to around RA 100\arcdeg, DEC 0\arcdeg.
Assuming an extragalactic case and ignoring stars, we input mock catalogs of AGN and galaxy clusters created by \citet{Comparat2019,Comparat2020} to SIXTE \citep[][provided by ECAP/Remeis observatory\footnote{\texttt{https://www.sternwarte.uni-erlangen.de/research/sixte/}}]{Dauser2019} to create mock signal events. More details are described in \citet{Liu2021_sim}.
We also use the method of \citet{Liu2021_sim} to create the input background model from the real data.
Two background spectra are extracted from source-free regions of the N7793 field within an off-axis angle of 23\arcmin and between an off-axis angle of 23 and 30\arcmin, respectively. We model them simultaneously to decompose the background into vignetted (cosmic X-ray background and Galactic diffuse emission) and unvignetted (mainly particle background) components. We use the phenomenological model used in \citet{Liu2021_sim} to model the vignetted X-ray emission, and use the spectral shape of the Filter-wheel-closed data and an additional power-law to model the unvignetted component.
Fitting the inner- and outer-region spectra, which have different vignetting, with the same model simultaneously, we decompose the background spectrum into vignetted and unvignetted components. Then we simulate them separately and finally merge them with the mock signals.
The simulation is repeated 100 times and the mock data is analyzed using eSASS as done for the real data.

\begin{figure*}[!hptb]
\centering
  \includegraphics[width=0.245\textwidth]{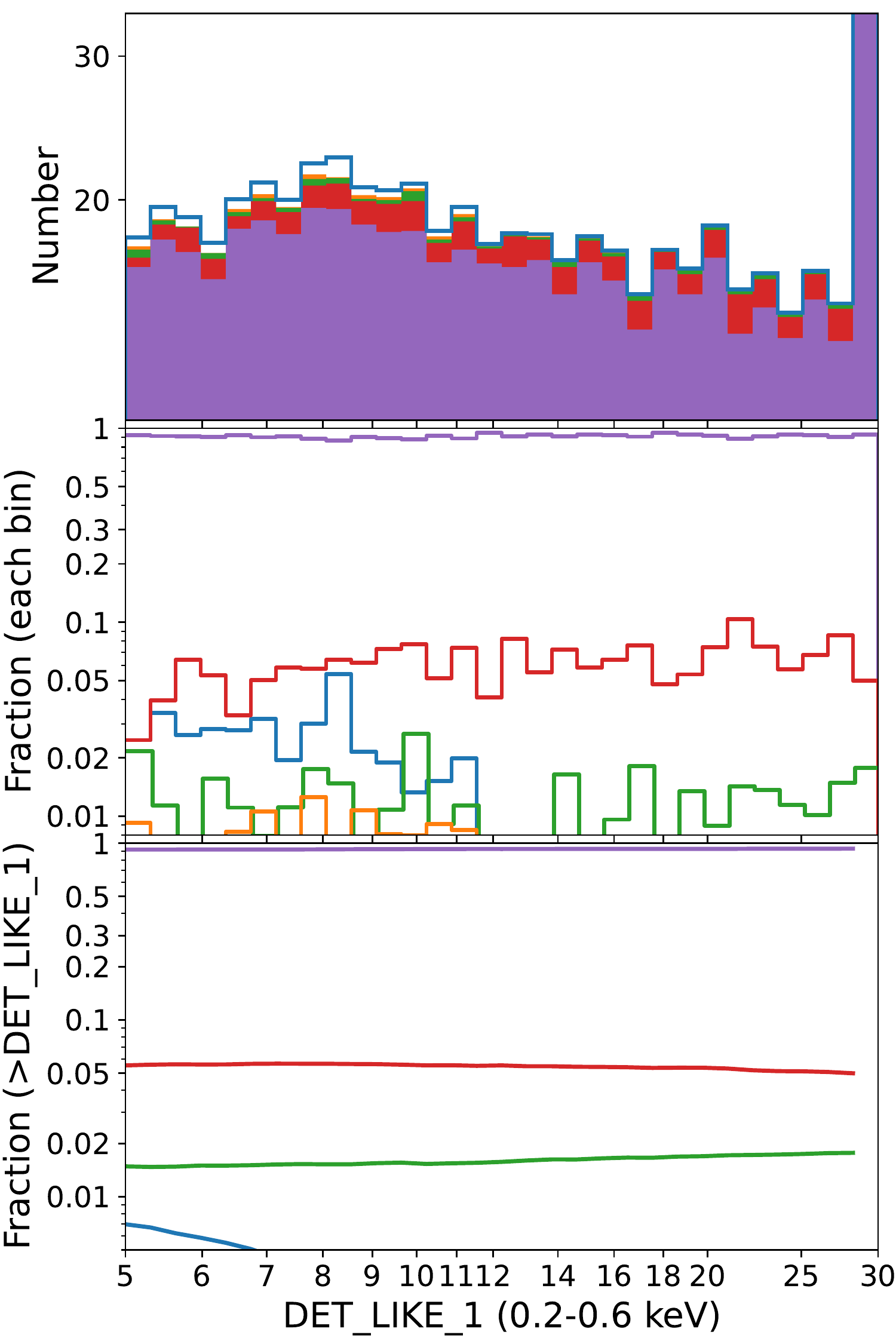}
  \includegraphics[width=0.245\textwidth]{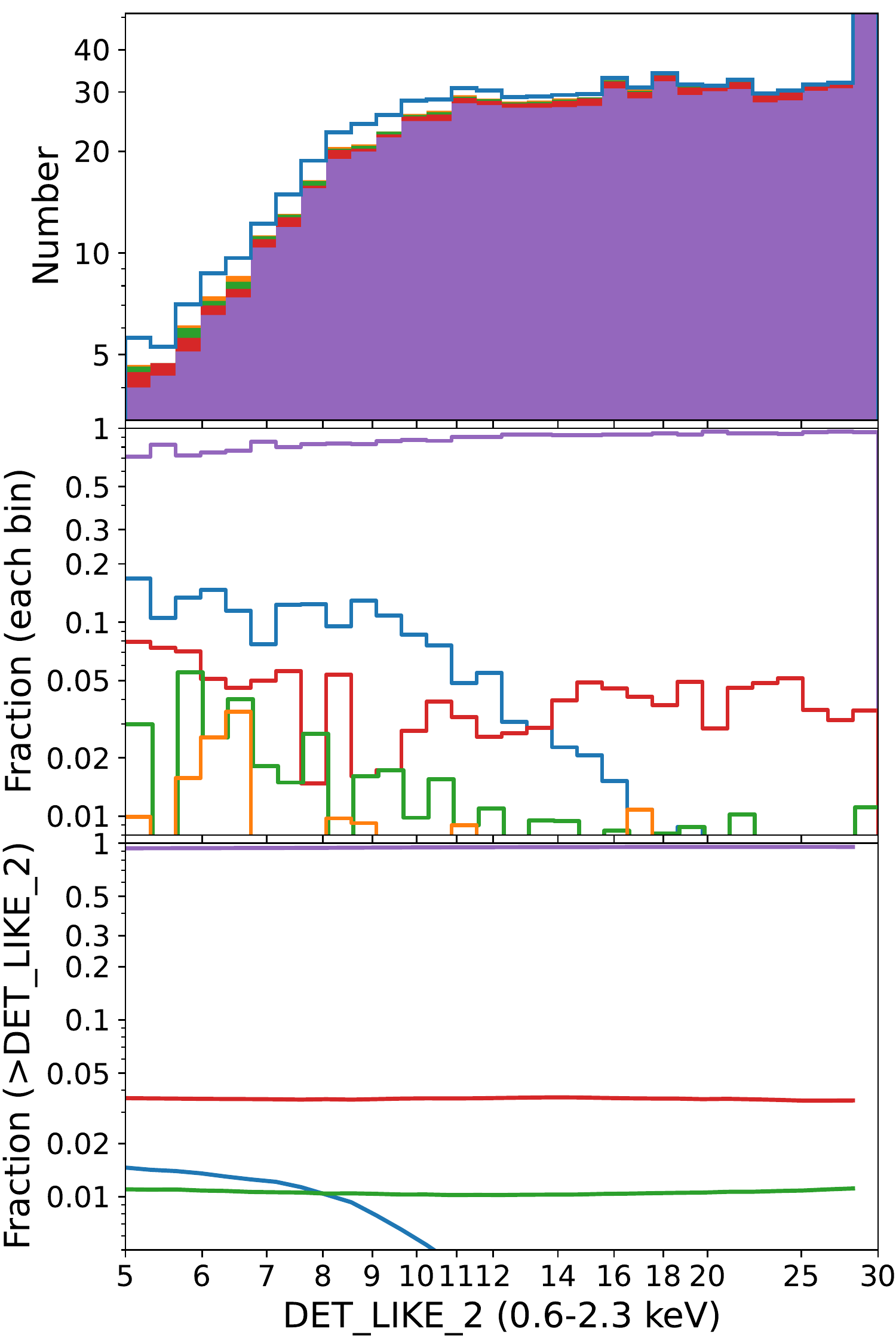}
  \includegraphics[width=0.245\textwidth]{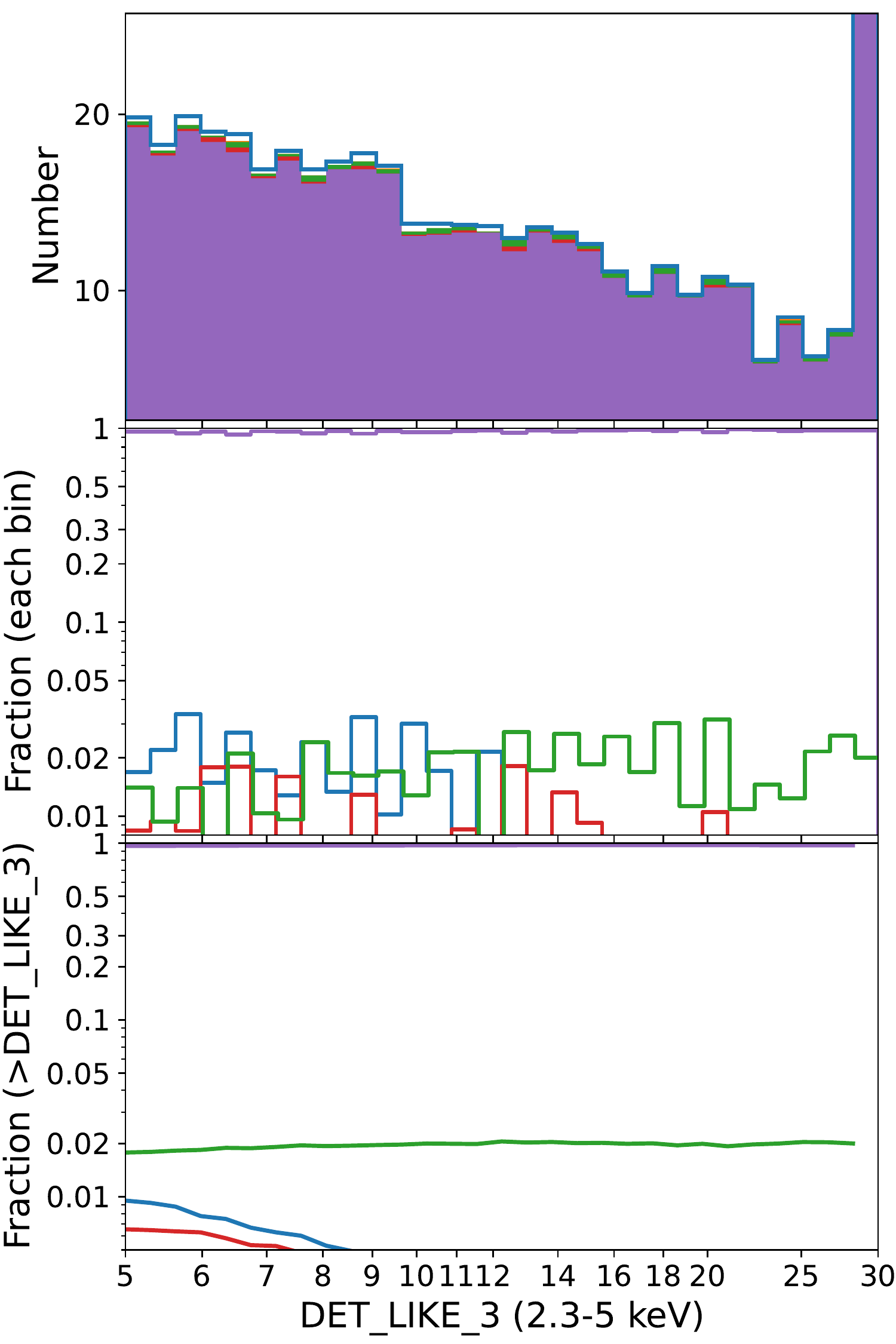}
  \includegraphics[width=0.245\textwidth]{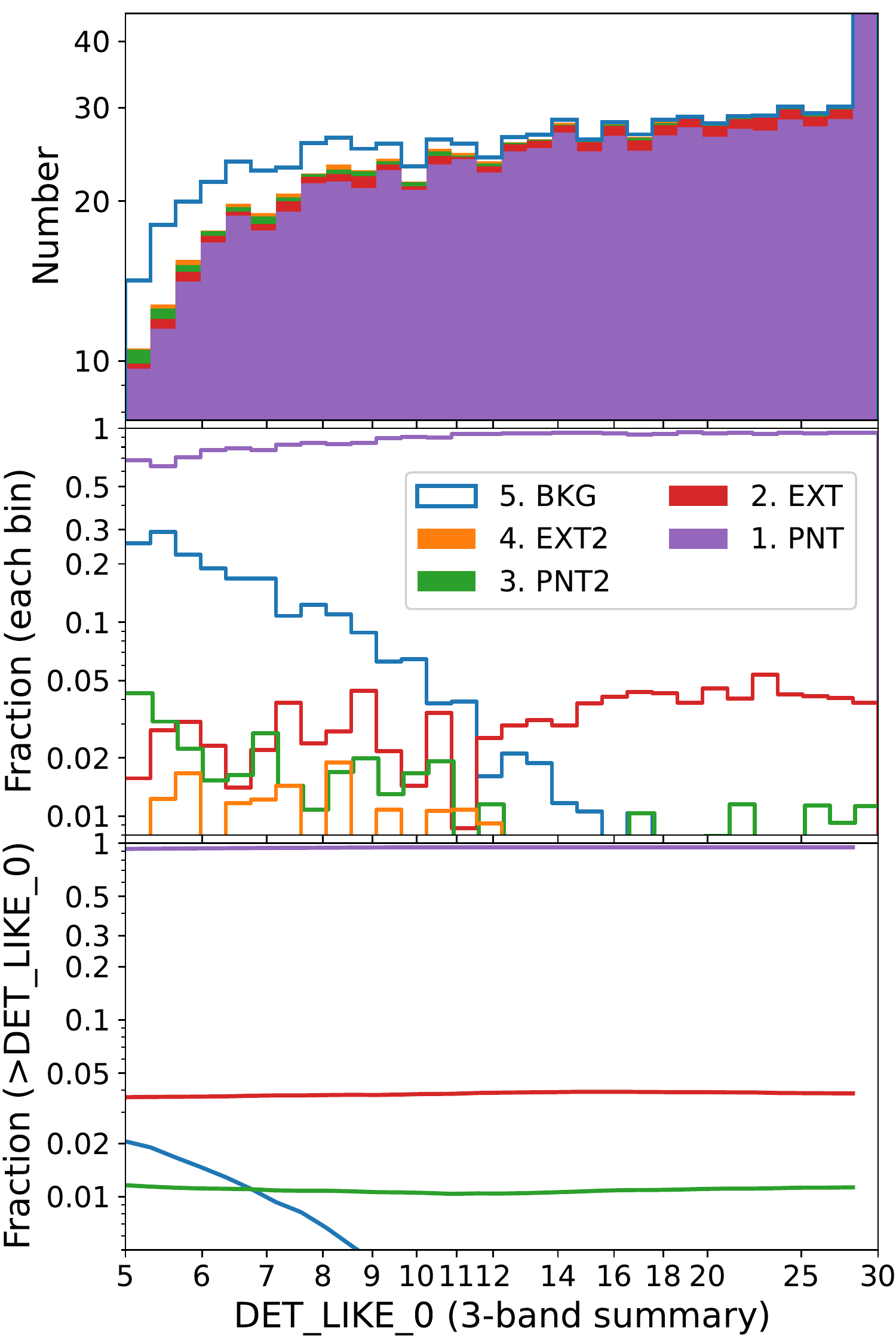}
\caption{Distributions of the five classes of simulated sources detected within an off-axis angle of 28\arcmin and with \texttt{EXT\_LIKE}=0, plotted as a function of the detection likelihood \texttt{DET\_LIKE\_$n$}, where $n$ indicates the energy band 1 (0.2--0.6~keV), 2 (0.6--2.3~keV), 3 (2.3--5~keV), or 0 (summary of the three bands).
  The upper, middle, and lower panels display the source numbers, the fraction in each bin, and the fraction above a given likelihood, respectively.
\label{fig:DET_dis}
  }
\vspace*{\floatsep}
\includegraphics[width=0.32\textwidth]{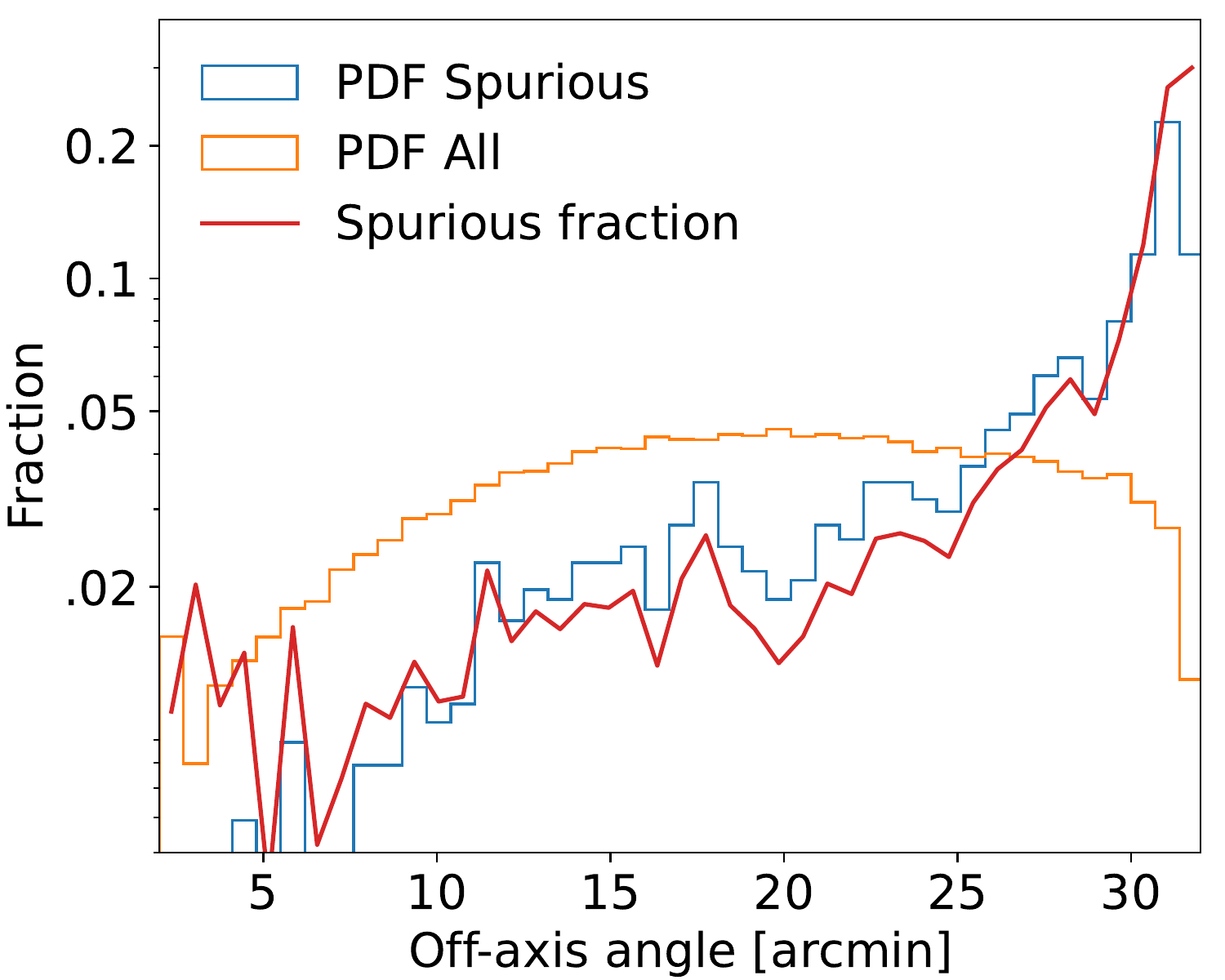}
\includegraphics[width=0.32\textwidth]{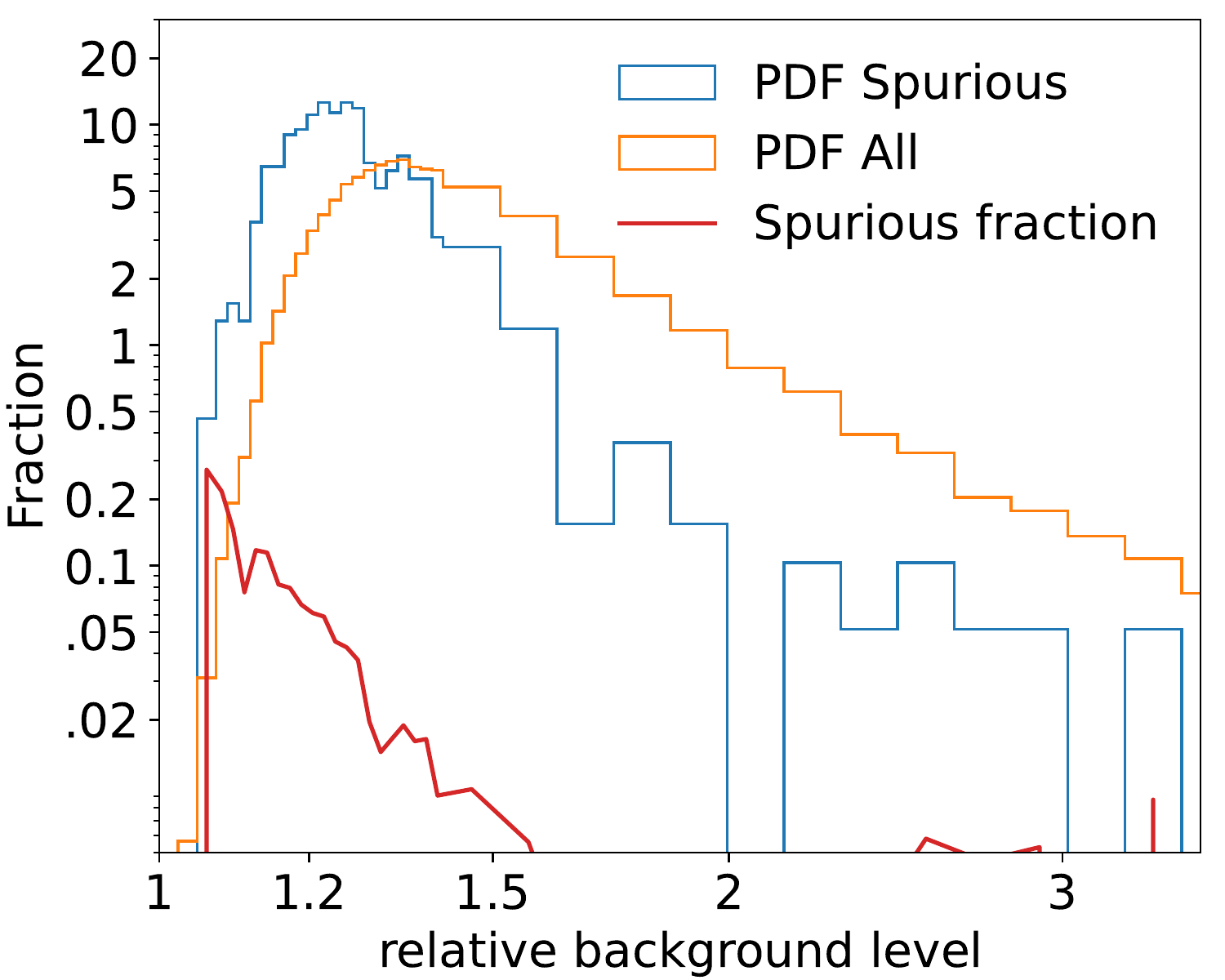}
\includegraphics[width=0.32\textwidth]{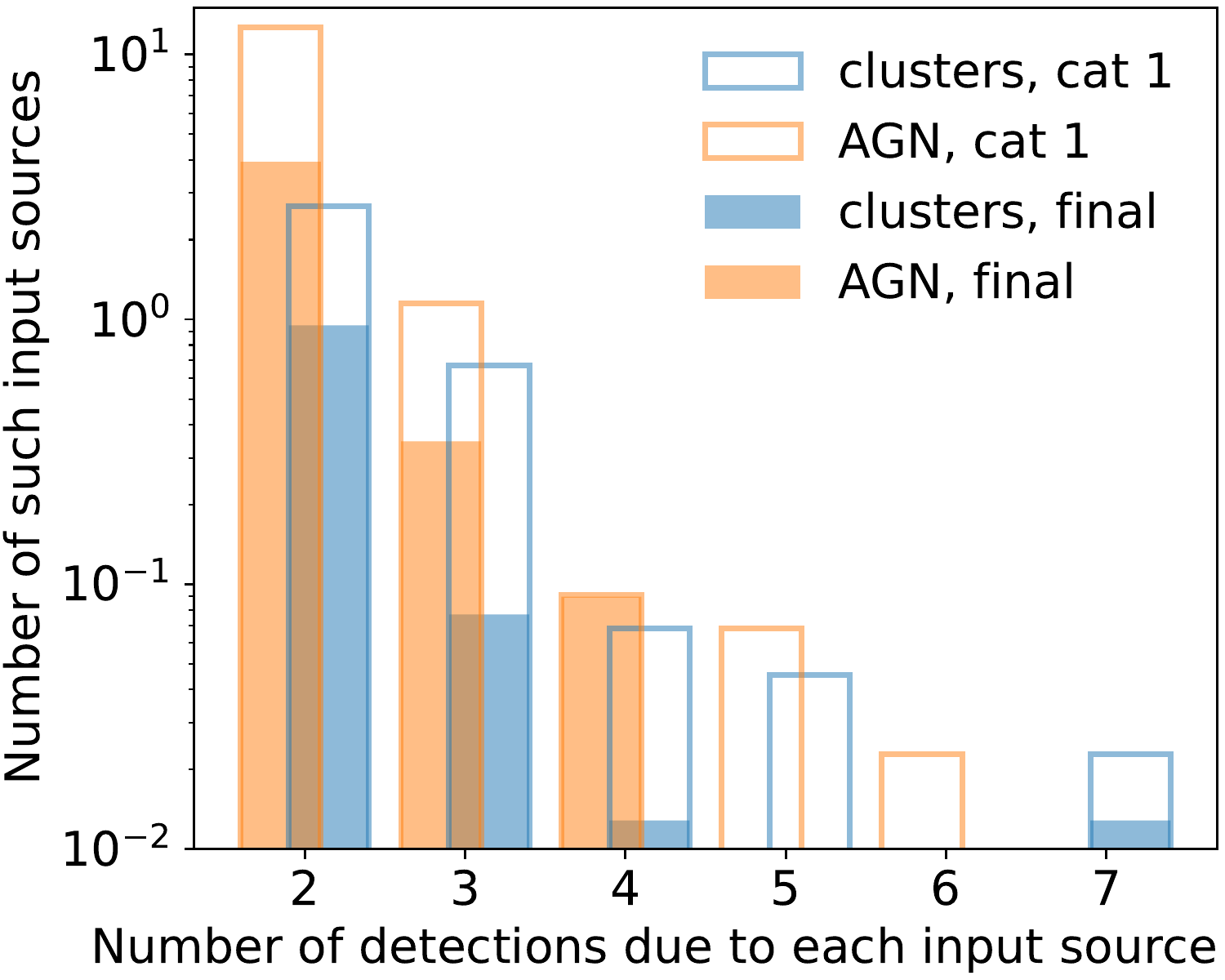}
\caption{The left and middle panels display the probability distributions of spurious sources (\texttt{ID\_Any}$<$0; blue) and all sources (orange) for the 100ks simulations as a function of off-axis angle (left) and relative background level (middle), respectively.
  The right panel displays the numbers of AGN (orange) and galaxy clusters (blue) detected as multiple sources in one field for the ``PSF-fitting catalog 1'' (empty) and the final catalog (filled), respectively.
  Only sources with an off-axis angle $<$26\arcmin are plotted in the middle and right panels.
\label{fig:spuri}}
\end{figure*}

As described in \citet{Liu2021_sim}, we associate the detected sources to the input ones based on the input-source ID flag on each photon, and classify detected sources into five classes:
\Final{
(1) primary counterpart of a point source                 (``PNT''),
(2) primary counterpart of an extended source             (``EXT''),
(3) secondary counterpart of a point source              (``PNT2''),
(4) secondary counterpart of an extended source          (``EXT2''), and
(5) background fluctuation                                (``BKG'').
}
The distributions of the five classes of sources detected within an off-axis angle of 28\arcmin are displayed in Fig.~\ref{fig:DET_dis} as a function of detection likelihoods. Compared with eFEDS \citep[][Fig.5]{Liu2021_sim}, the fraction of spurious sources in this pointing-mode survey is much lower, because of both the deeper exposure and better spatial resolution and the improved source detection method.

In Fig.~\ref{fig:spuri}, we plot the spurious fraction as a function of the off-axis angle.
It increases sharply at off-axis angles $\gtrsim26\arcmin$, mainly because the spatial resolution is much worse at the FOV border than in the center. We thus suggest, for future analysis of \eROSITA pointed observations, excluding the FOV border if a high-purity catalog is needed.
Excluding the border outside an off-axis angle of $26\arcmin$, we also plot the distribution of relative background \Final{levels} at source positions in Fig.~\ref{fig:spuri}. 
Knowing the origin of each photon, we create a pure-background image and smooth it to reduce fluctuations. The relative background is calculated as the ratio between the measured background map of each field and this pure-background image.
The measured background is higher than the value in the pure-background image because the undetectable sources and the outer wing of the detectable sources contribute an additional background component.
As shown in the figure, the spurious sources have lower background than the others, indicating that a main cause of spurious sources is local background underestimation.
As discussed in \citet{Liu2021_sim}, the definition of source detection likelihood, either measured through PSF-fitting or aperture photometry, always underestimates the fraction of spurious sources, because practically there are additional uncertainties in the data. One major origin of the uncertainties \Final{consists in} the background map created by smoothing.
For smoothing, we have to compromise between final S/N and spatial variability, and thus with a limited smoothing scale, fluctuations inevitably result in some regions with local underestimation.

In addition to background fluctuations, there are also false detections caused by real source \Final{signals}, \Final{that is}, one bright source detected as multiple sources (``PNT2'' and ``EXT2'' in Fig.~\ref{fig:DET_dis}).
In Fig.~\ref{fig:spuri}, We also display the number of input sources detected as multiple ones in one field within an off-axis angle of $26\arcmin$.
Compared with the ``PSF-fitting catalog 1'', which is based on the background maps created using the eFEDS method, the improved background maps introduced in this work suppress such cases in our final catalog.
To eliminate such cases clearly, a more-detailed image modeling of bright sources with improved models and within a larger region is needed.
Alternatively, a practical solution for the real-data catalog is to identify such spurious sources through visual inspection.

\begin{figure}[!hptb]
\centering
\includegraphics[width=0.8\columnwidth]{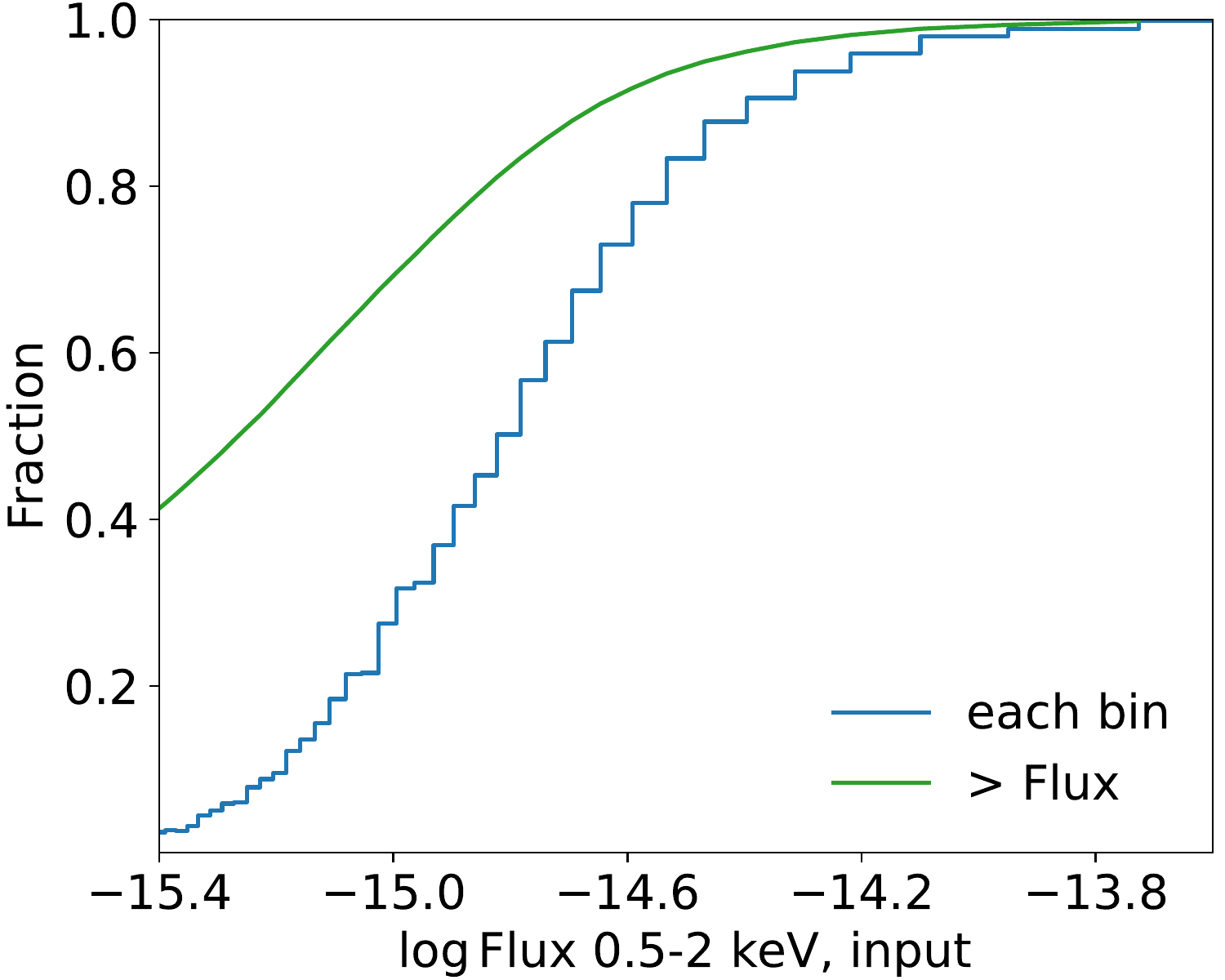}
\caption{\Final{Detected} fraction of AGN for the 100ks pointing-mode simulation within an off-axis angle of 28\arcmin as a function of the input 0.5--2~keV flux in each bin (blue) and above a given flux (green), respectively.
\label{fig:comp}}
\end{figure}

Based on the input-output association, we also plot the detection completeness of AGN as a function of fluxes in Fig.~\ref{fig:comp} for sources within an off-axis angle of 28\arcmin. For such a 100ks pointing-mode observation, we obtain a 90\% completeness of AGN down to a 0.5--2~keV flux of $2.3\times 10^{-15}$ \egs.
This flux limit is dependent on the exposure depth, spatial resolution, and background level, which are different among the 11 fields. This simulation only presents a typical situation of 100ks exposure.
If an accurate measurement of the AGN selection function is needed, the simulation needs to be rerun with the required exposure.


We also compared the photon counts measured through PSF-fitting with the input counts, and find an underestimation by a few percent, the same as found in the eFEDS simulation \citep[][Appendix. A]{Liu2021_sim}. It is caused by calibration uncertainties in the current \eROSITA PSF library and will be improved in the future.

\section{Conclusions}
\label{sec:conclusion}
In this work, we present a serendipitous catalog of X-ray sources detected within 11 extragalactic fields observed during the Calibration and Performance Verification phase of \eROSITA. 
We develop a source detection method optimized for point-like X-ray sources by including extended X-ray emission in the background measurement.

We release to the public a catalog of $9515$ X-ray sources, whose X-ray fluxes are measured through spectral fitting. CatWISE counterparts are presented for 77\% of these sources. 
A subsample of significantly variable sources \Final{is} also selected and presented with this paper.

\TL{
By comparing the point-source number counts obtained in each of the observed fields, we unveil variations of large-scale properties which determine the characteristics of the X-ray sky (X-ray stellar density, Galactic gas density).
Excluding four particular fields (47Tuc, ES0102, H2213, J1856), the other seven show similar populations of X-ray sources dominated by AGN.
}


With the help of detailed \texttt{SIXTE} simulations, we provide a characterization of the \eROSITA performance (detection sensitivity, catalog completeness and purity) in deep pointed observation mode, complementing the study published earlier for the large scanning-mode observations of the eFEDS field \citep{Liu2021_sim}. This will aid the planning and analysis of \eROSITA observations in the second phase of the mission, starting in 2024, after the completion of the four-year all-sky survey.

\begin{acknowledgement}
This work is based on data from eROSITA, the soft X-ray instrument aboard SRG, a joint Russian-German science mission supported by the Russian Space Agency (Roskosmos), in the interests of the Russian Academy of Sciences represented by its Space Research Institute (IKI), and the Deutsches Zentrum f\"ur Luft- und Raumfahrt (DLR). The SRG spacecraft was built by Lavochkin Association (NPOL) and its subcontractors, and is operated by NPOL with support from the Max Planck Institute for Extraterrestrial Physics (MPE). The development and construction of the eROSITA X-ray instrument was led by MPE, with contributions from the Dr. Karl Remeis Observatory Bamberg \& ECAP (FAU Erlangen-Nuernberg), the University of Hamburg Observatory, the Leibniz Institute for Astrophysics Potsdam (AIP), and the Institute for Astronomy and Astrophysics of the University of T\"ubingen, with the support of DLR and the Max Planck Society. The Argelander Institute for Astronomy of the University of Bonn and the Ludwig Maximilians Universit\"at Munich also participated in the science preparation for eROSITA.

The eROSITA data shown here were processed using the eSASS software system developed by the German eROSITA consortium.

MB is supported by the European Innovative Training Network (ITN) "BiD4BEST" funded by the Marie Sklodowska-Curie Actions in Horizon 2020 (GA 860744).
\end{acknowledgement}

\bibliography{calpv}
\begin{appendix}

\section{The catalogs}
\label{sec:cat}

\begin{table*}[!hptb]
  \centering
  \caption{Descriptions of the X-ray source catalog columns}
  \begin{tabular}{p{0.24\textwidth} p{0.08\textwidth} p{0.67\textwidth}}

   \hline
Column         &Units & Description \\
   \hline
    \multicolumn{3}{l}{1. Source properties and PSF-fitting results}\\
   \hline
    Name           & & Source name \\ 
    UID            & & Unique source ID\\
    ID\_SRC        & & Source ID in each field \\
    Field          & & Field name \\
    inMskPnt       &bool & Whether in the point-source subsurvey region\\
    RA\_corr        	&deg	& Right ascension, corrected using CatWISE2020\\
    DEC\_corr       	&deg	& Declination, corrected using CatWISE2020\\
    RADEC\_ERR\_corr 	&arcsec	& Combined positional error, corrected\\
    RA             &deg & Right ascension (J2000), uncorrected\\
    RA\_LOWERR     &arcsec & 1-$\sigma$ RA lower error \\
    RA\_UPERR      &arcsec & 1-$\sigma$ RA upper error \\
    DEC            &deg & Declination (J2000), uncorrected \\
    DEC\_LOWERR    &arcsec & 1-$\sigma$ DEC lower error \\
    DEC\_UPERR     &arcsec & 1-$\sigma$ DEC upper error \\
    RADEC\_ERR     &arcsec & Combined positional error, uncorrected \\
    DET\_LIKE\_$n$       	&	& Detection likelihood for each detection band ($n$=1,2,3) or combining them ($n$=0)\\
    EXT            &arcsec	& Source extent parameter\\
    EXT\_{\sl err}       &arcsec	& 1-$\sigma$ Extent error\\
    EXT\_LIKE       	&	& Extent likelihood\\
    Exp            &s      & Unvignetted exposure time\\
    galNHI                          &cm$^{-2}$ & Galactic HI column density\\
    galNH                           &cm$^{-2}$ & Galactic total column density\\
    ML\_RATE\_{\sl band}        	&cts/s	& Source count rate measured by PSF-fitting\\
    ML\_RATE\_{\sl err}\_{\sl band}    	&cts/s	& 1-$\sigma$ count rate error\\
    ML\_CTS\_{\sl band}         	&cts	& Source net counts measured from count rate\\
    ML\_CTS\_{\sl err}\_{\sl band}     	&cts	& 1-$\sigma$ source counts error\\
    ML\_EXP\_{\sl band}         	&s	& Vignetted exposure time at the source position\\
    ML\_BKG\_{\sl band}         	&cts/arcmin$^2$	& Background at the source position\\
    ML\_EEF\_{\sl band}         	&s	& PSF enclosed energy fraction in the fitting aperture\\
    ECF\_2d3\_5\_h       &cm$^{2}$/erg &Energy conversion factor between ML\_FLUX\_h and ML\_RATE\_t\\
    ECF\_0d5\_2\_s       &cm$^{2}$/erg &Energy conversion factor between ML\_FLUX\_s and ML\_RATE\_s\\
    ECF\_2d3\_5          &cm$^{2}$/erg &Energy conversion factor between ML\_FLUX\_3 and ML\_RATE\_3\\
    ECF\_0d6\_2d3        &cm$^{2}$/erg &Energy conversion factor between ML\_FLUX\_2 and ML\_RATE\_2\\
    ECF\_0d2\_0d6        &cm$^{2}$/erg &Energy conversion factor between ML\_FLUX\_1 and ML\_RATE\_1\\
    ML\_FLUX\_{\sl band}        	&erg/cm$^2$/s	& Source flux corrected for Galactic absorption \\
    ML\_FLUX\_{\sl err}\_{\sl band}    	&erg/cm$^2$/s	& 1-$\sigma$ source flux error\\
\hline
    \multicolumn{3}{l}{2. Aperture photometry results in two {\sl Band}s: s and t}\\
    \hline
APE\_CTS\_{\sl\small Band}    	&cts	& Total counts extracted within the aperture\\
APE\_EXP\_{\sl\small Band}    	&s	& Vignetted exposure time at the given position\\
APE\_BKG\_{\sl\small Band}    	&cts	& Background counts extracted within the aperture\\
APE\_RADIUS\_{\sl\small Band} 	&pixels	& Extraction radius\\
APE\_POIS\_{\sl\small Band}   	&	& Poisson probability of the source being background fluctuation\\
\hline
    \multicolumn{3}{l}{3. Observed fluxes measured from spectra in five {\sl\small Band}s: b1, b2, b3, b4, t}\\
    \hline
    FluxObsv\_Med\_{\sl\small Band} &erg/cm$^2$/s &Observed fluxes, median \\
    FluxObsv\_Lo1\_{\sl\small Band} &erg/cm$^2$/s &Observed fluxes, 1-$\sigma$ lower limit \\
    FluxObsv\_Lo2\_{\sl\small Band} &erg/cm$^2$/s &Observed fluxes, 2-$\sigma$ lower limit  \\
    FluxObsv\_Up1\_{\sl\small Band} &erg/cm$^2$/s &Observed fluxes, 1-$\sigma$ upper limit  \\
    FluxObsv\_Up2\_{\sl\small Band} &erg/cm$^2$/s &Observed fluxes, 2-$\sigma$ upper limit  \\
\hline	
  \end{tabular}
    \tablefoot{
      \Edit{
      The involved energy bands are listed in Table.~\ref{table:bands}.
      In section 1, the PSF-fitting energy bands ({\sl band} suffix) include three source-detection bands (1, 2, 3) and two forced-photometry bands (s, t). The count rate in the bands 1, 2, 3, s, and t are converted to Galactic-absorption-corrected flux in the bands 1, 2, 3, s, and h.
      }
      The {\sl err} suffix can be ``LOWERR'', ``UPERR'', or ``ERR'', which correspond to 1-$\sigma$ lower error, upper error, and combined error, respectively.
      The current version of the eSASS PSF-fitting task \texttt{ermldet} fails in estimating errors in a small fraction of cases, leaving an undefined value (NULL).
      ML\_EXP\_3 is not given, since it equals ML\_EXP\_t.
      \Edit{
      In section 2, the aperture photometry is done in bands s and t.
      In section 3, the fluxes are measured from spectra in the band t and four bands b1--b4 used in the 4XMM catalog \citep{Webb2020}.
      }
      \label{table:columns}}
\end{table*}

\begin{table}[!hptb]
  \centering
  \caption{Energy bands used in this work}
  \begin{tabular}{p{0.2\columnwidth} p{0.3\columnwidth}}
   \hline
{\sl band} mark & energy range\\
    & keV\\
   \hline
    \multicolumn{2}{l}{source-detection bands}\\
   \hline
    1 & 0.2--0.6\\
    2 & 0.6--2.3\\
    3 & 2.3--5 \\
   \hline
    \multicolumn{2}{l}{forced-photometry bands}\\
   \hline
    s & 0.5--2\\
    t & 2.3--5\\
    h & 2--10 \\
    \hline
    \multicolumn{2}{l}{spectral flux measurement bands}\\
   \hline
    b1 & 0.2--0.5\\
    b2 & 0.5--1\\
    b3 & 1--2\\
    b4 & 2--4.5\\
    t & 2.3--5\\
   \hline
  \end{tabular}
    \tablefoot{Forced photometry is done in the band s and t. The band h is used only in one case, that is, converting the t band count rate to h band flux.
      \label{table:bands}}
\end{table}

\begin{table}[!hptb]
  \centering
  \caption{Column descriptions for the variable source catalog}
  \begin{tabular}{p{0.14\columnwidth}p{0.08\columnwidth} p{0.68\columnwidth}}
   \hline
Column & Units  &        Description \\
   \hline
UID     &   & Unique source ID\\
T\_lo    &s  & Time of the low-rate bin\\
T\_hi    &s  & Time of the high-rate bin\\
Rate\_lo &cts/s  & 0.2-5 keV count rate of the low-rate bin\\
Rate\_hi &cts/s  & 0.2-5 keV count rate of the high-rate bin\\
V       &  & Variability amplitude\\
S       &  & Variability significance\\
mCts    &cts  & Adopted minimum counts in each bin (30 or 50)\\
   \hline
  \end{tabular}
    \tablefoot{The reference time of the light curves is MJD 51543.875.
      \label{table:vary}}
\end{table}

\begin{table*}[!hptb]
  \centering
  \caption{Descriptions of the counterpart catalog columns}
  \begin{tabular}{p{0.20\textwidth} p{0.08\textwidth} p{0.71\textwidth}}

   \hline
Column         &Units & Description \\
   \hline
Field              &      & Field name\\
UID                &      & X-ray unique source ID\\
ID\_SRC             &      & X-ray source ID in each field\\
RA\_corr            &deg      & X-ray Right ascension, corrected using CatWISE2020\\
DEC\_corr           &deg      & X-ray Declination, corrected using CatWISE2020\\
RADEC\_ERR\_corr     &arcsec      & X-ray combined positional error, corrected\\
CW2\_source\_name    &      & CatWISE source name\\
CW2\_source\_id      &      & CatWISE source ID\\
CW2\_ra             &deg      & CatWISE RA (ICRS) at reference epoch 2015\\
CW2\_dec            &deg      & CatWISE DEC (ICRS) at reference epoch 2015\\
CW2\_sigra          &arcsec      & CatWISE RA error\\
CW2\_sigdec         &arcsec      & CatWISE DEC error\\
CW2\_w1mag          &mag      & CatWISE W1 magnitude\\
CW2\_w1sigm         &mag      & CatWISE W1 magnitude error\\
CW2\_w2mag          &mag      & CatWISE W2 magnitude error\\
CW2\_w2sigm         &mag      & CatWISE W2 magnitude error\\
dist\_post          &      & distance posterior probability comparing this association versus no association\\
p\_any              &      & the probability that any of the associations is the correct one\\
p\_i                &      & relative probability of the match\\
p\_X                &      & \Edit{the probability of being X-ray source measured by machine learning based only on the WISE photometry} \\
match\_flag         &      & 1 for the most probable match, 2 for \Edit{secondary matching solution} with $p_i/p_i^{best}>0.5$\\
Separation         &arcsec      & Separation between the CatWISE and X-ray positions\\
GA\_source\_id       &      & Gaia EDR3 unique source ID\\
GA\_ra              &deg      & Gaia RA (ICRS) at reference epoch 2016\\
GA\_ra\_error        &mas      & Gaia RA error \\
GA\_dec             &deg      & Gaia DEC (deg;ICRS) at reference epoch 2016\\
GA\_dec\_error       &mas      & Gaia DEC error\\
GA\_parallax        &mas      & Gaia absolute stellar parallax\\
GA\_parallax\_error  &mas      & Gaia parallax error\\
GA\_pmra            &mas/year & Gaia proper motion in RA \\
GA\_pmra\_error      &mas/year & Gaia RA proper motion error \\
GA\_pmdec           &mas/year & Gaia proper motion in DEC \\
GA\_pmdec\_error     &mas/year & Gaia DEC proper motion error\\
    \hline
  \end{tabular}
    \tablefoot{When a \Gaia counterpart is not found, the relevant values are left undefined (NULL).}
    \label{table:ctp}
\end{table*}

In this work, we present the X-ray source catalog of this survey, the catalog of variable sources selected in Sect.~\ref{sec:vary}, and the catalog of CatWISE counterparts identified in Sect.~\ref{sec:ctp}.
The columns of these three catalogs are described in Tables~\ref{table:columns}, \ref{table:vary}, and \ref{table:ctp}, respectively.
The catalogs are available on the eROSITA Early Data Release (EDR) website\footnote{\texttt{https://erosita.mpe.mpg.de/edr/eROSITAObservations/Catalogues/}, \Edit{where the X-ray catalog (``CalPVexG\_V3.3.fits''), the counterpart catalog (``CalPVexG\_CTP\_V3.3.fits''), and the list of variable sources (``CalPVexG\_vary\_V3.3.fits'') are available in FITS format.}} and at the CDS.

\begin{figure*}[!hptb]
\centering
\includegraphics[width=0.24\textwidth]{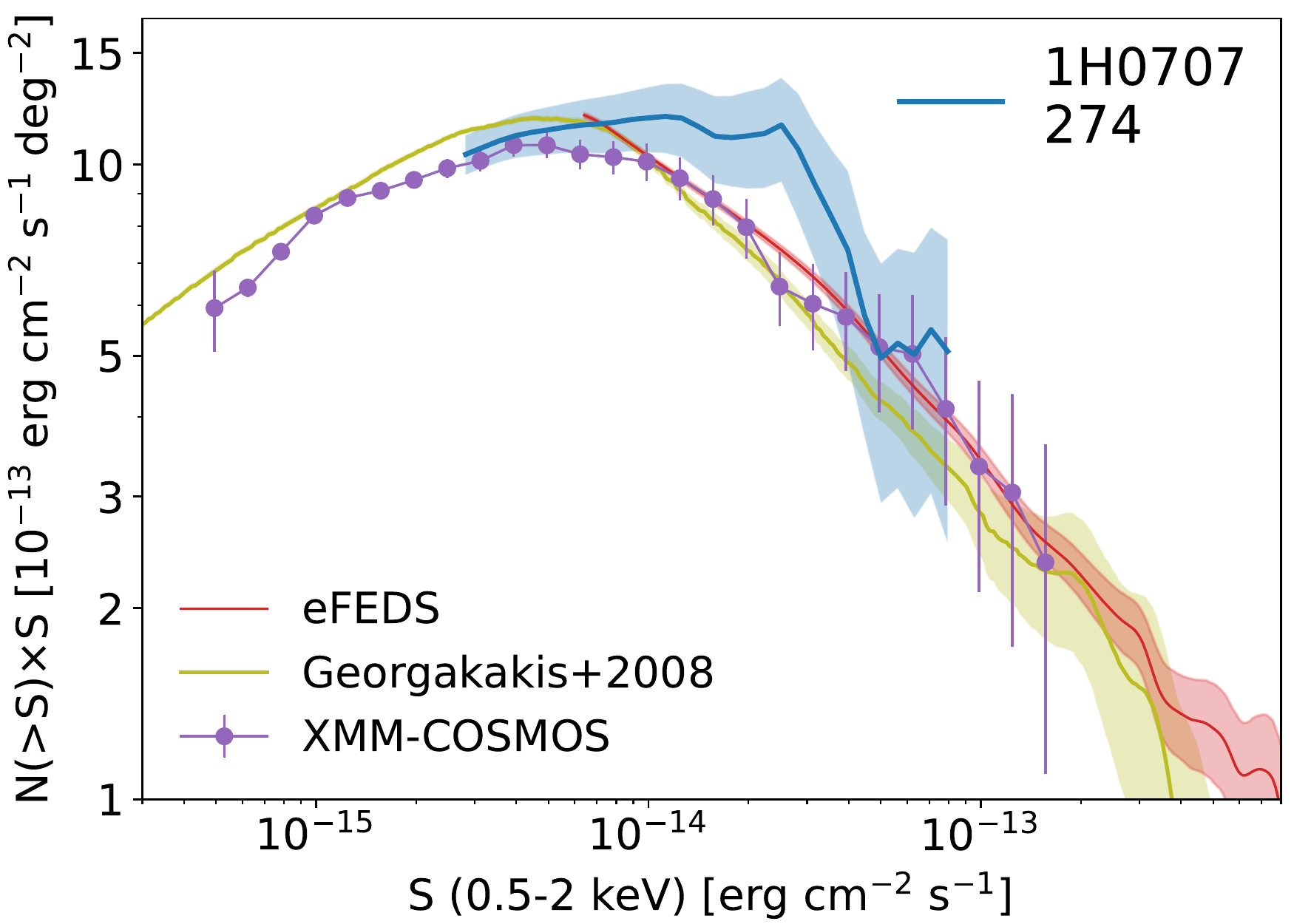}
\includegraphics[width=0.24\textwidth]{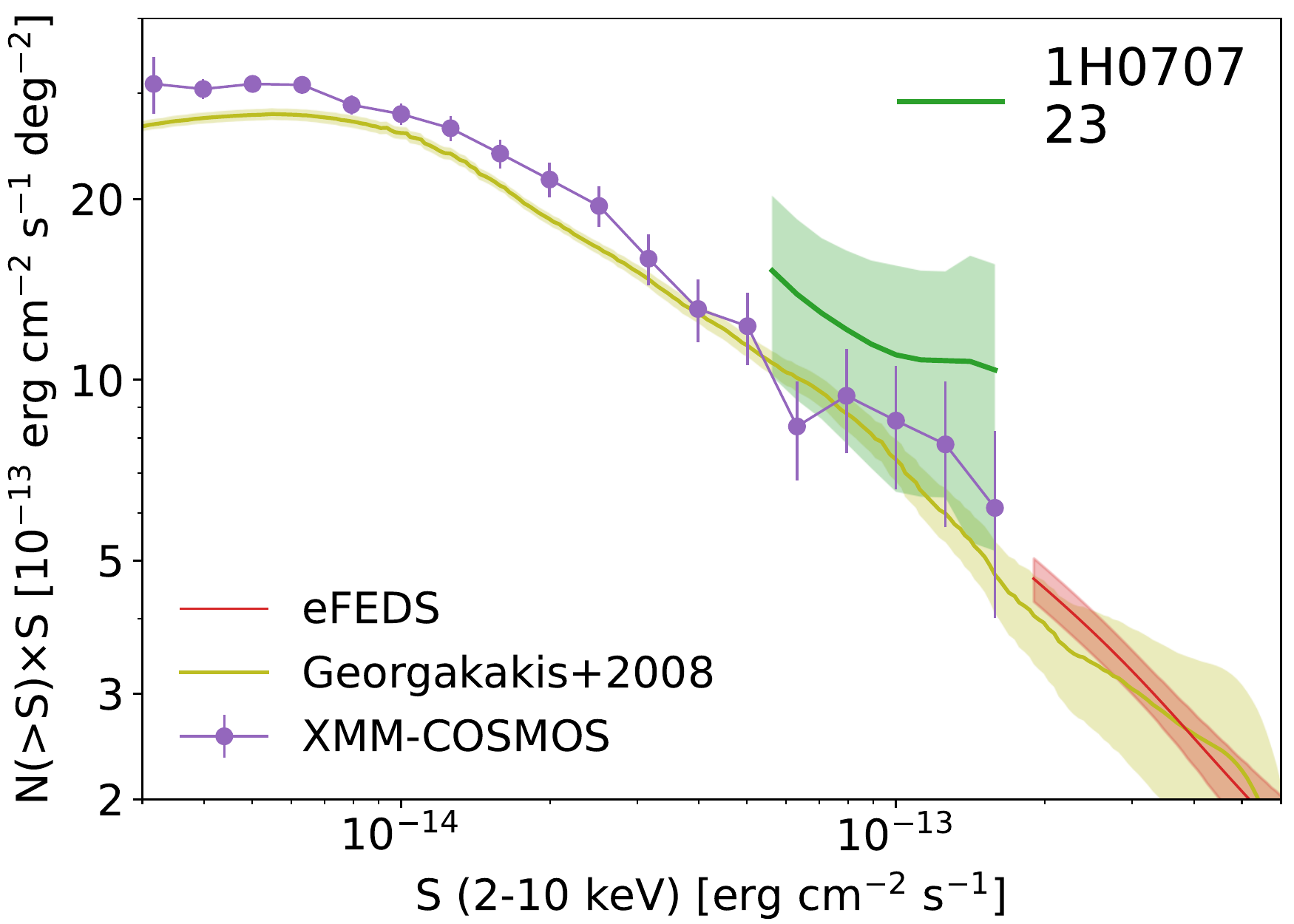}
\includegraphics[width=0.24\textwidth]{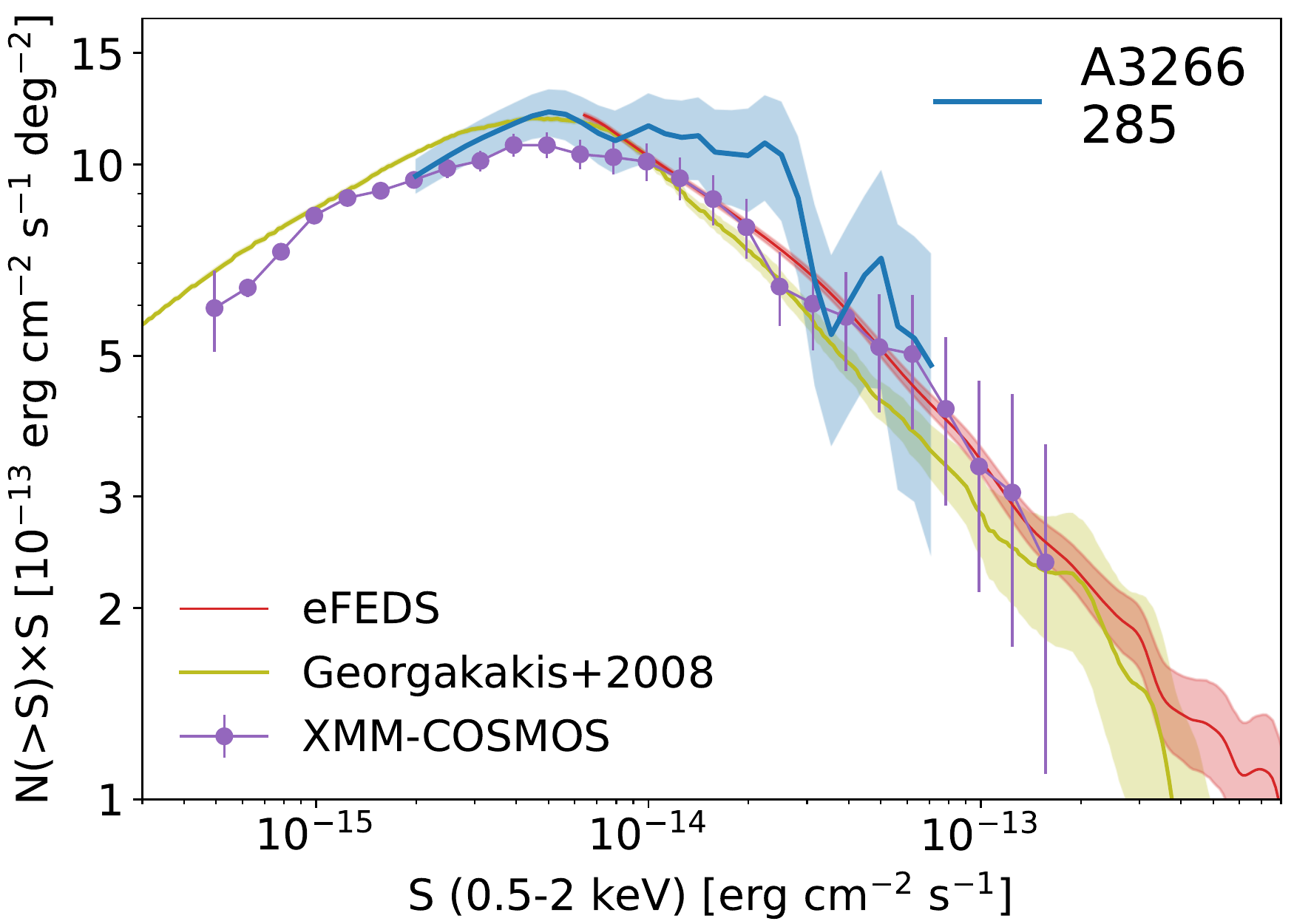}
\includegraphics[width=0.24\textwidth]{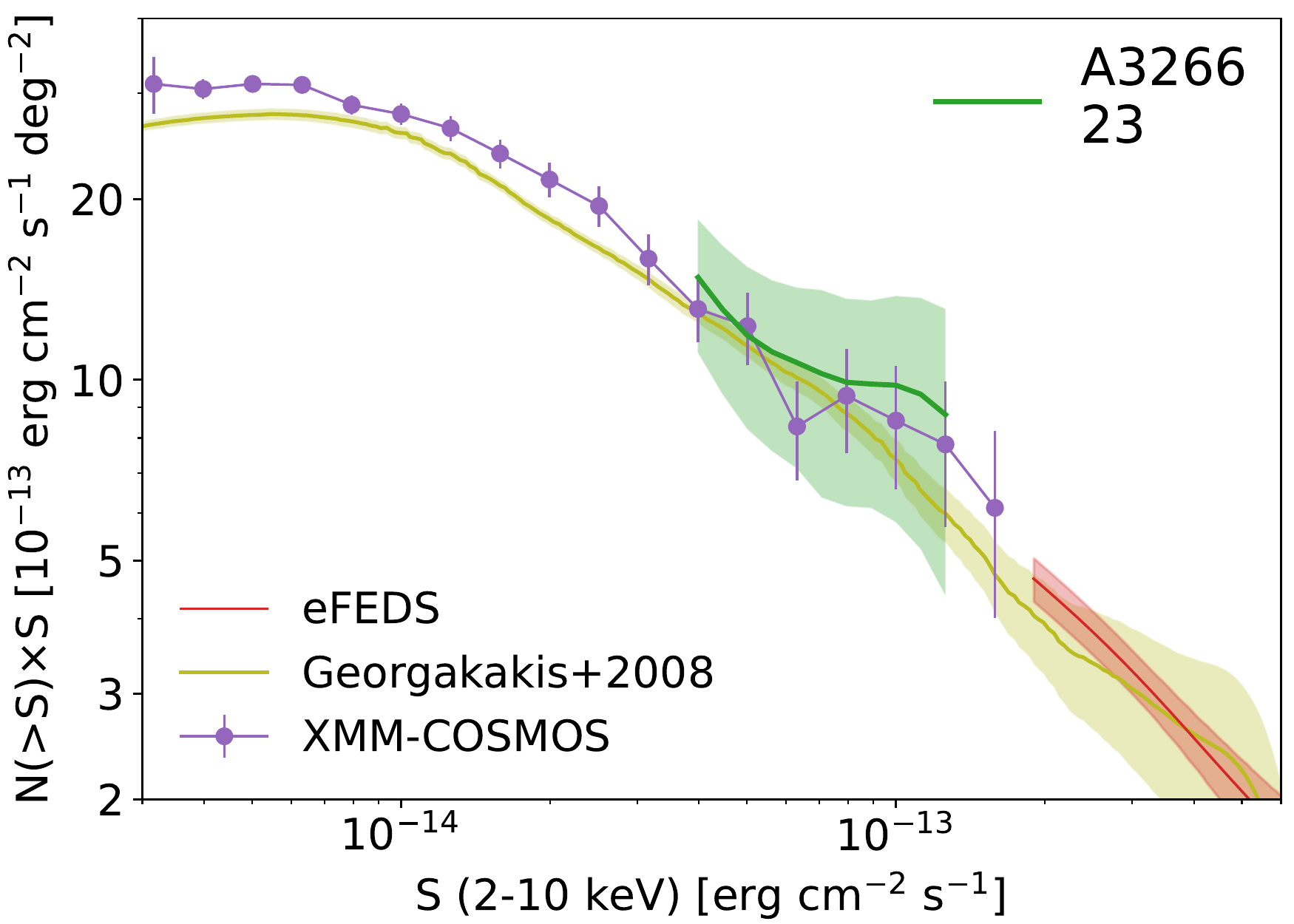}
\includegraphics[width=0.24\textwidth]{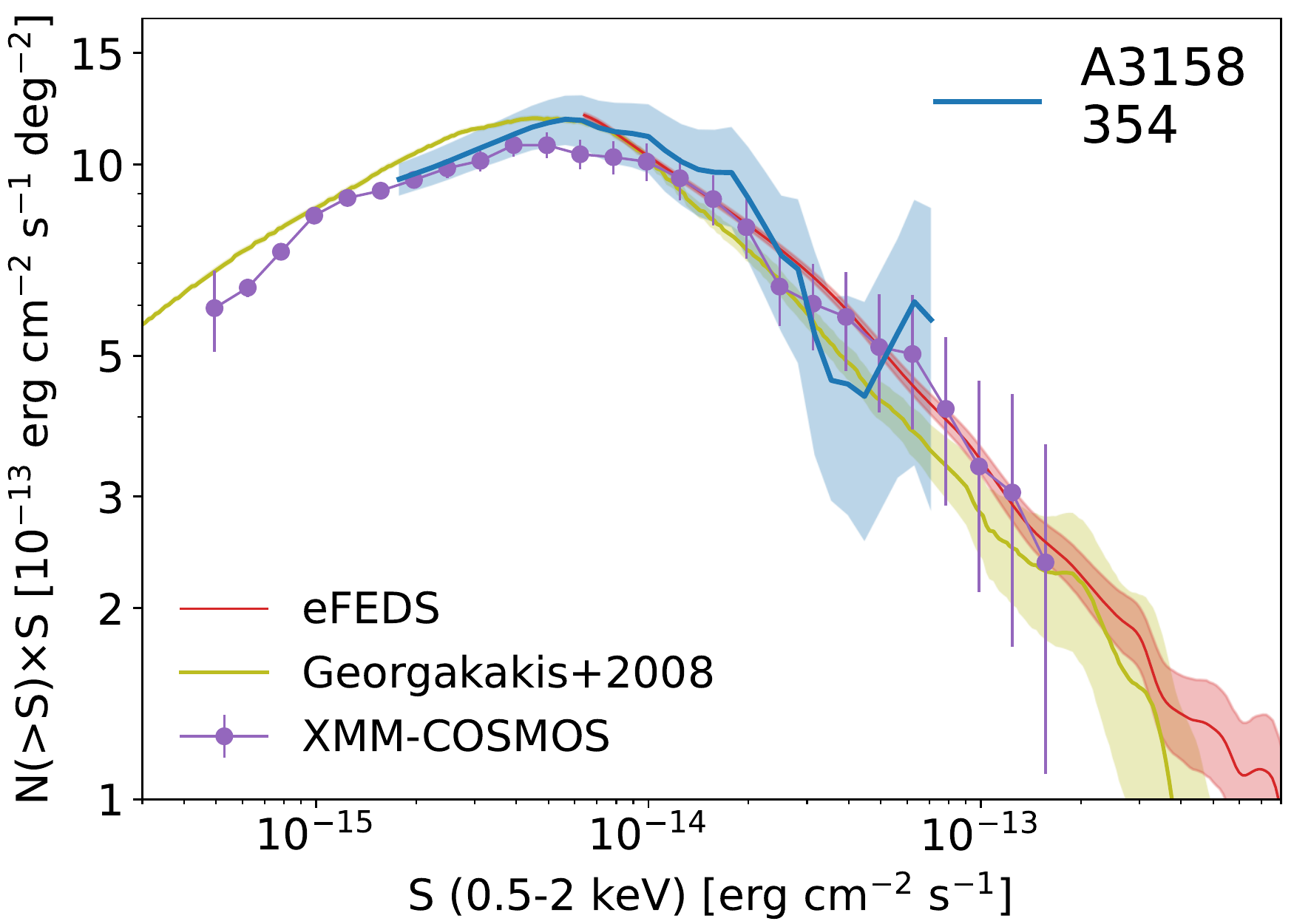}
\includegraphics[width=0.24\textwidth]{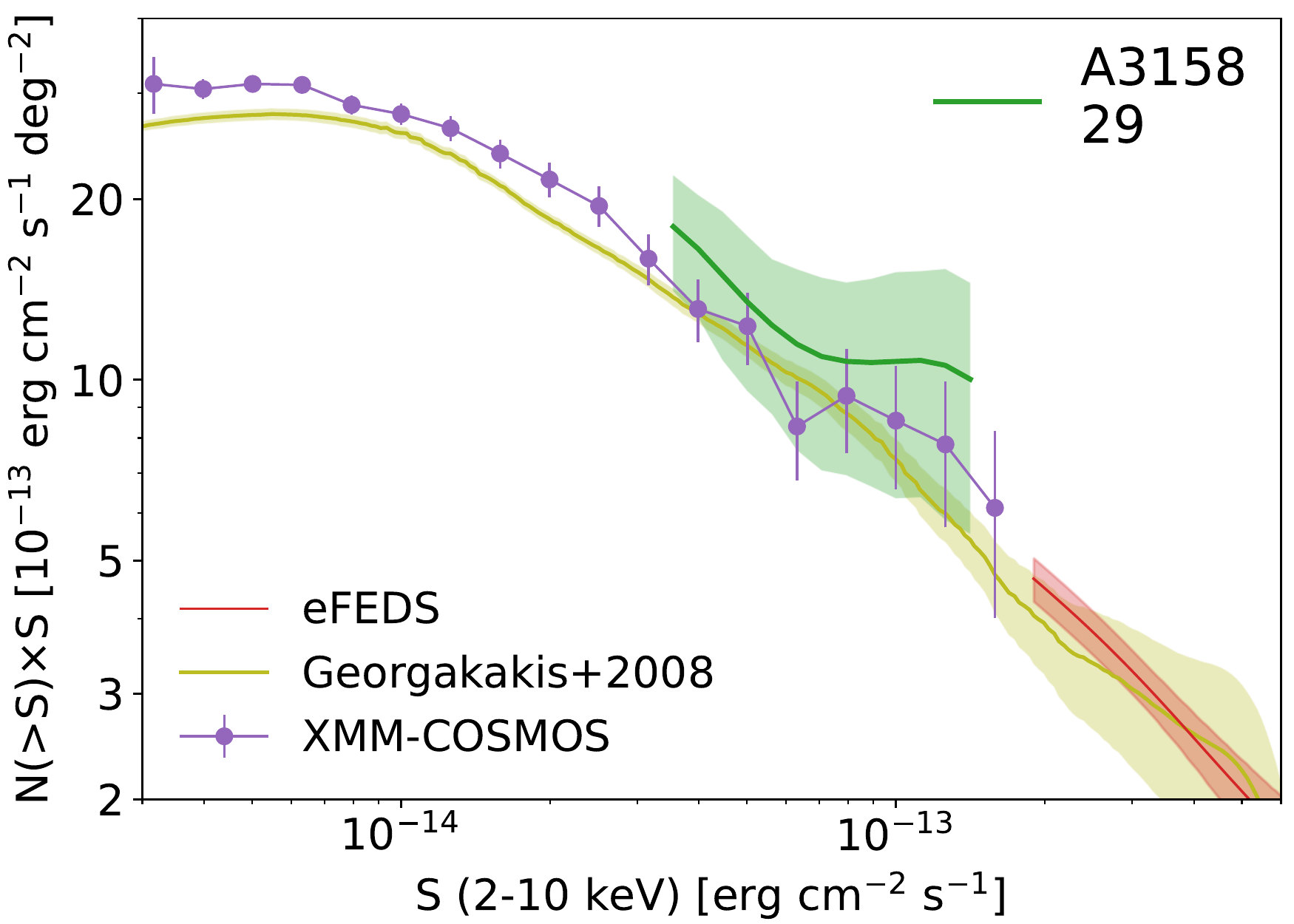}
\includegraphics[width=0.24\textwidth]{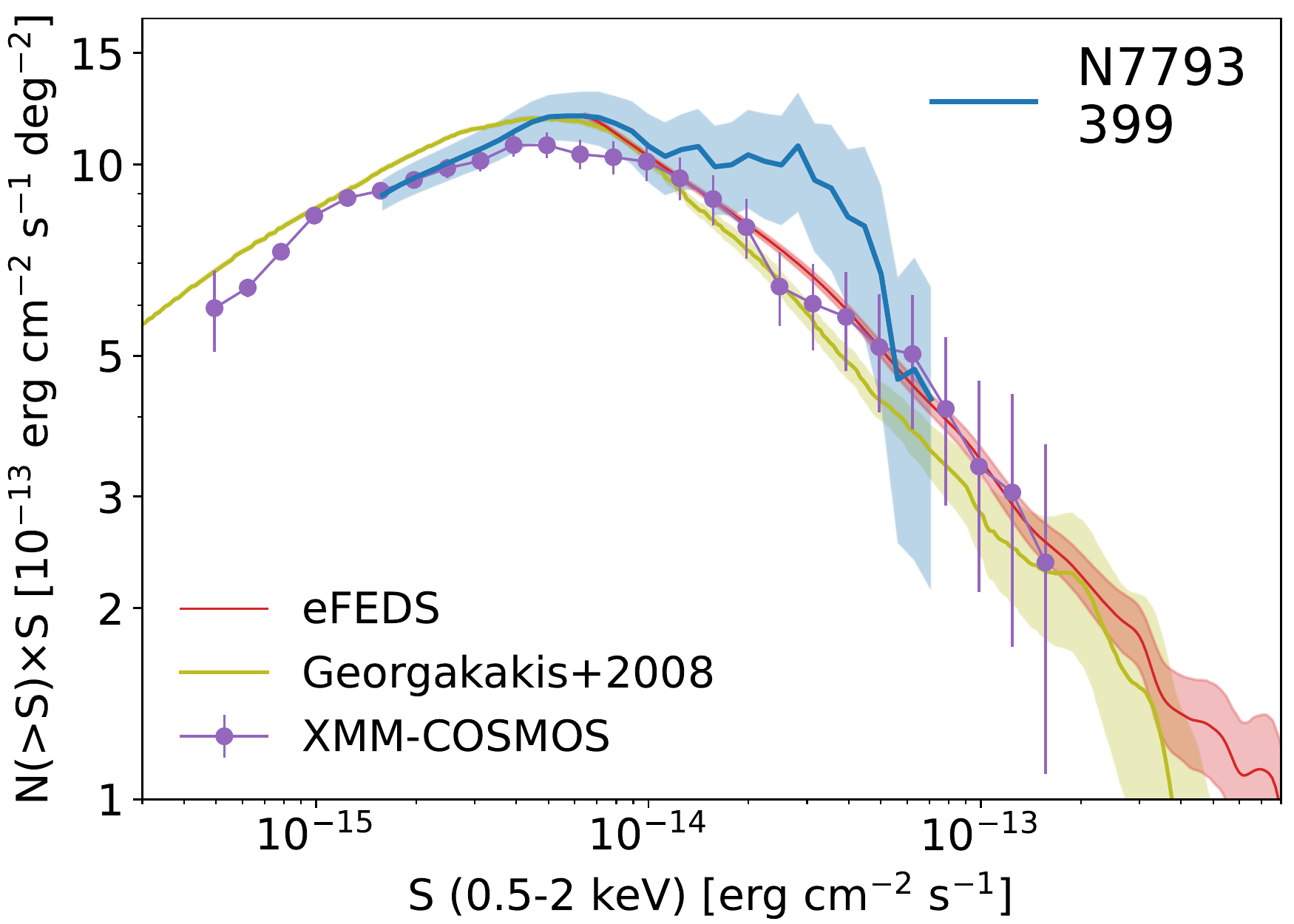}
\includegraphics[width=0.24\textwidth]{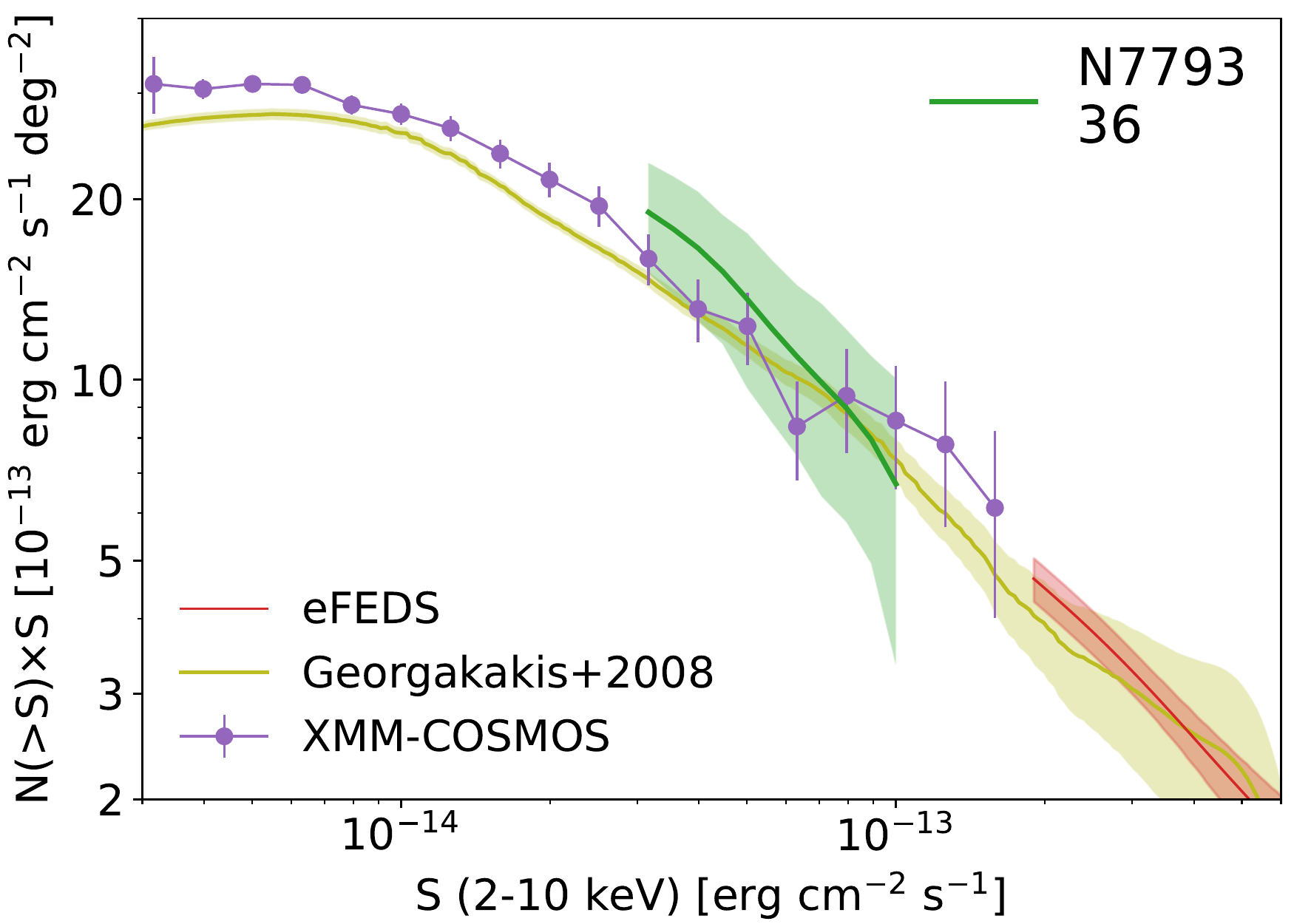}
\includegraphics[width=0.24\textwidth]{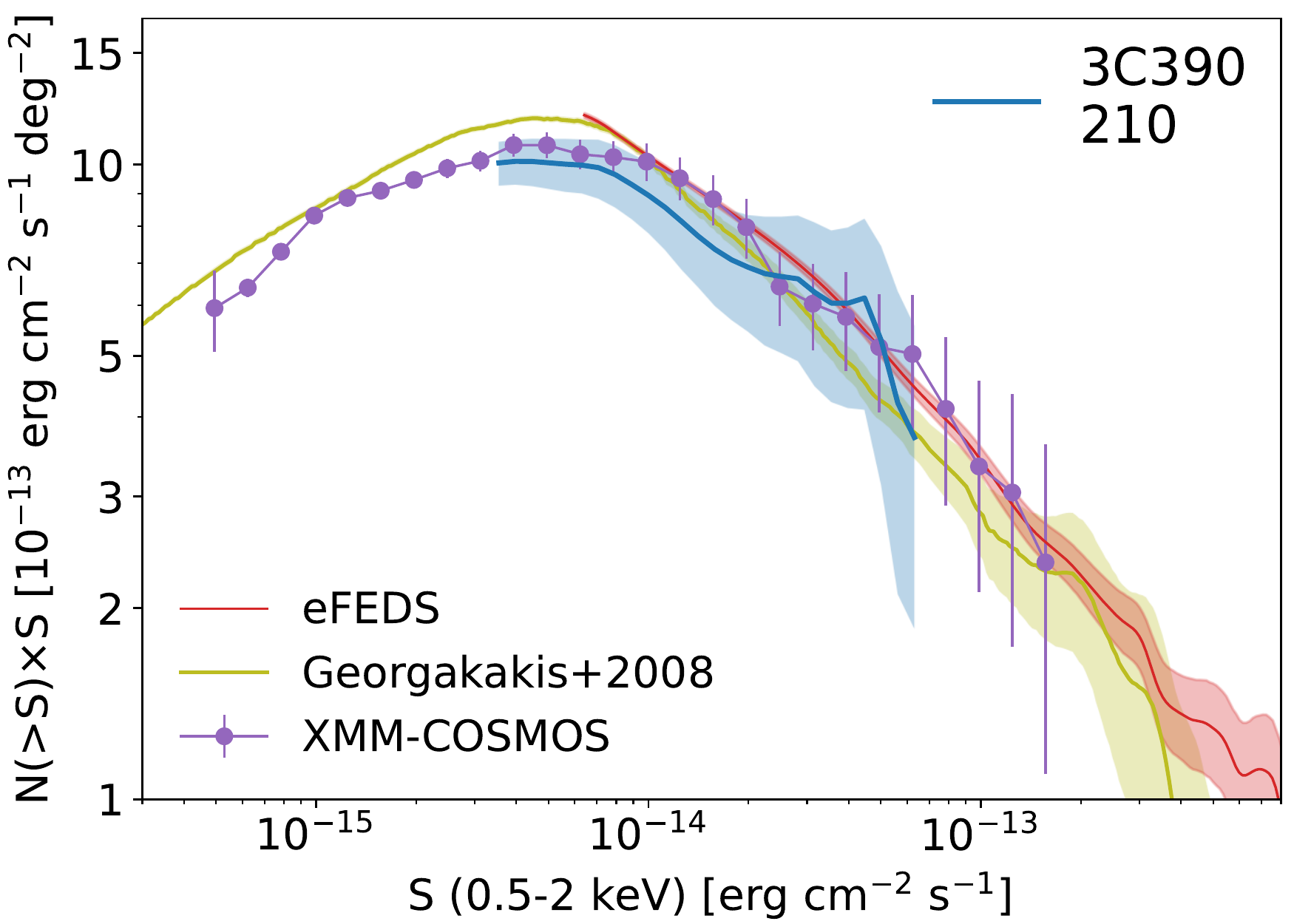}
\includegraphics[width=0.24\textwidth]{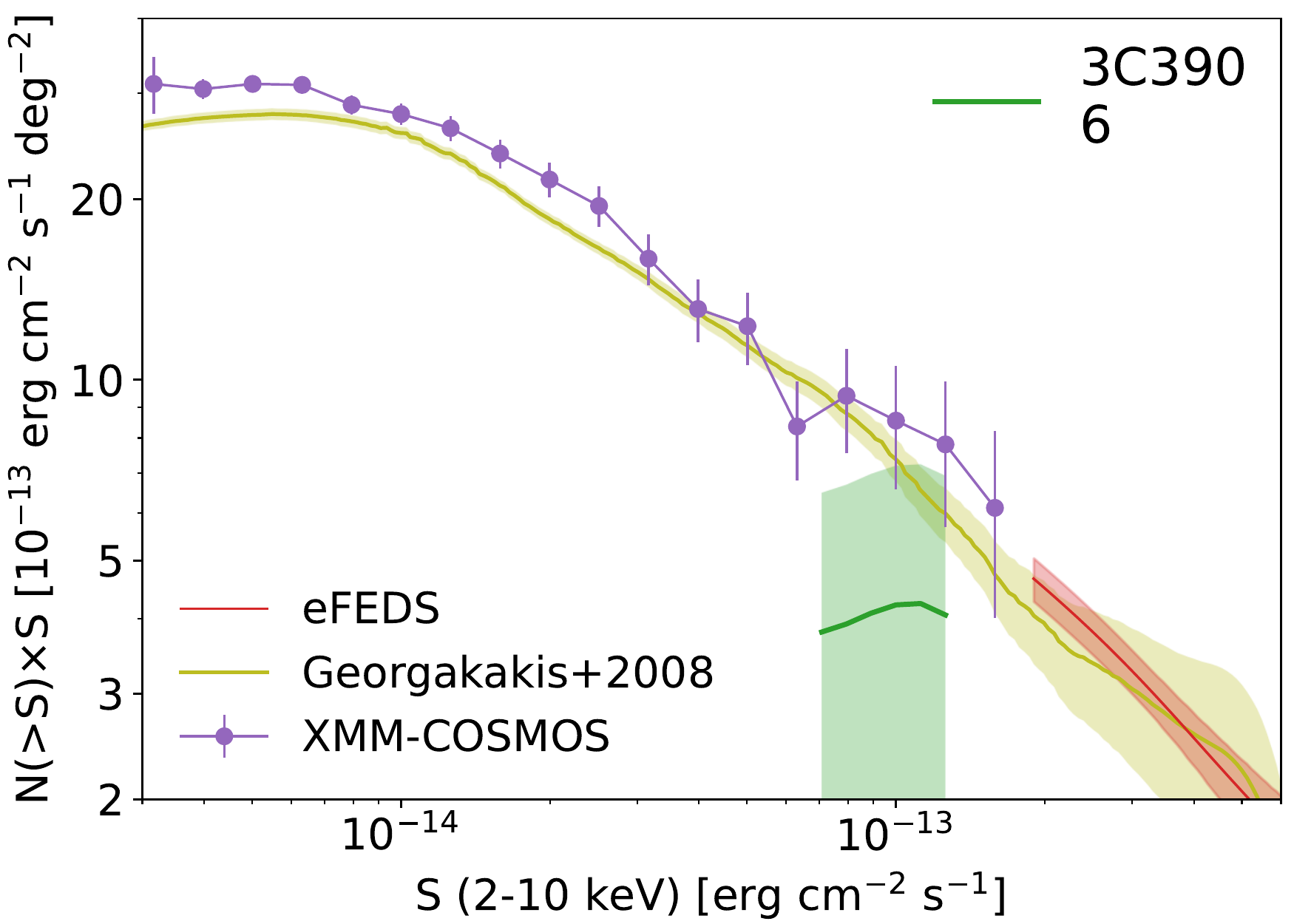}
\includegraphics[width=0.24\textwidth]{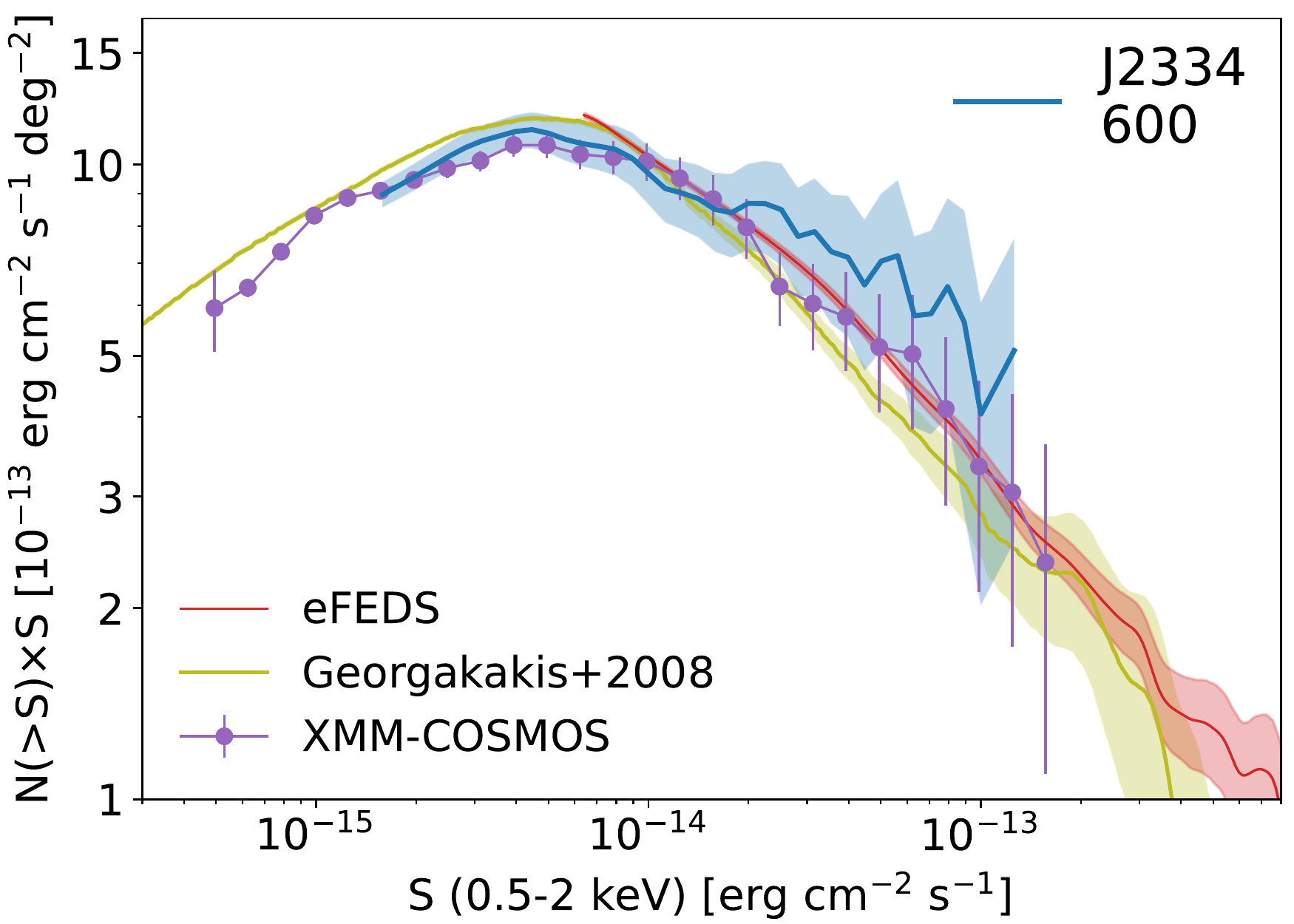}
\includegraphics[width=0.24\textwidth]{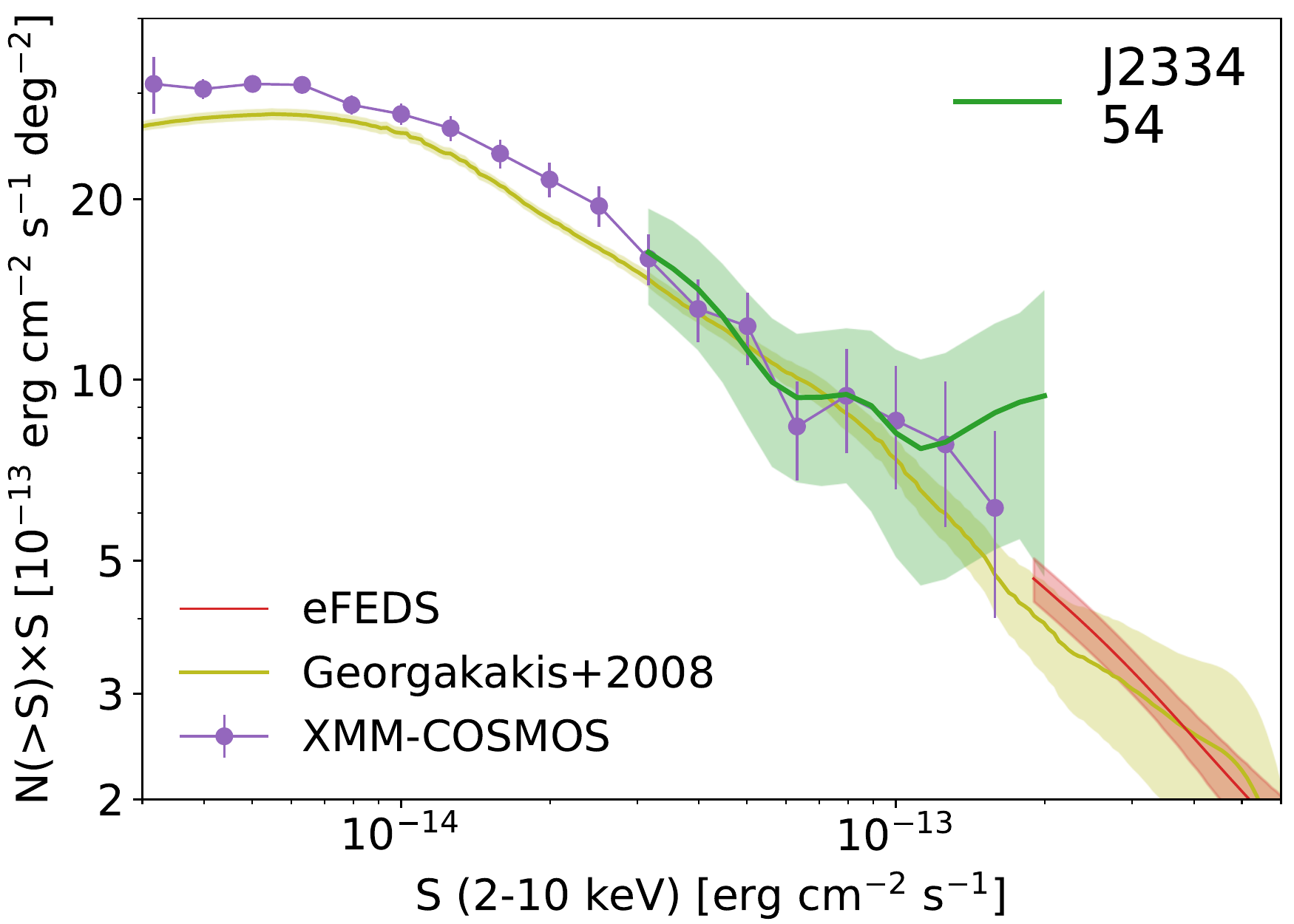}
\includegraphics[width=0.24\textwidth]{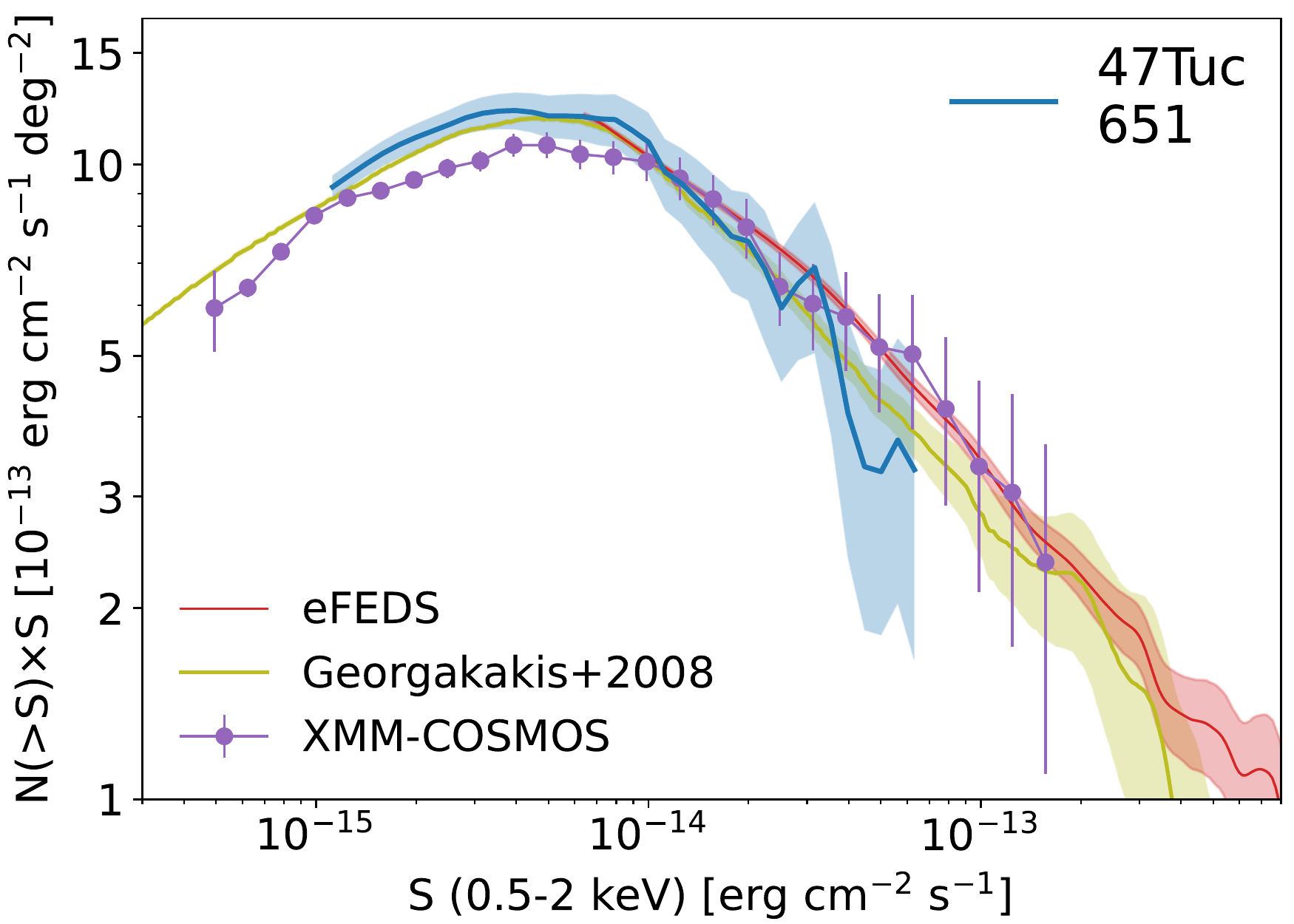}
\includegraphics[width=0.24\textwidth]{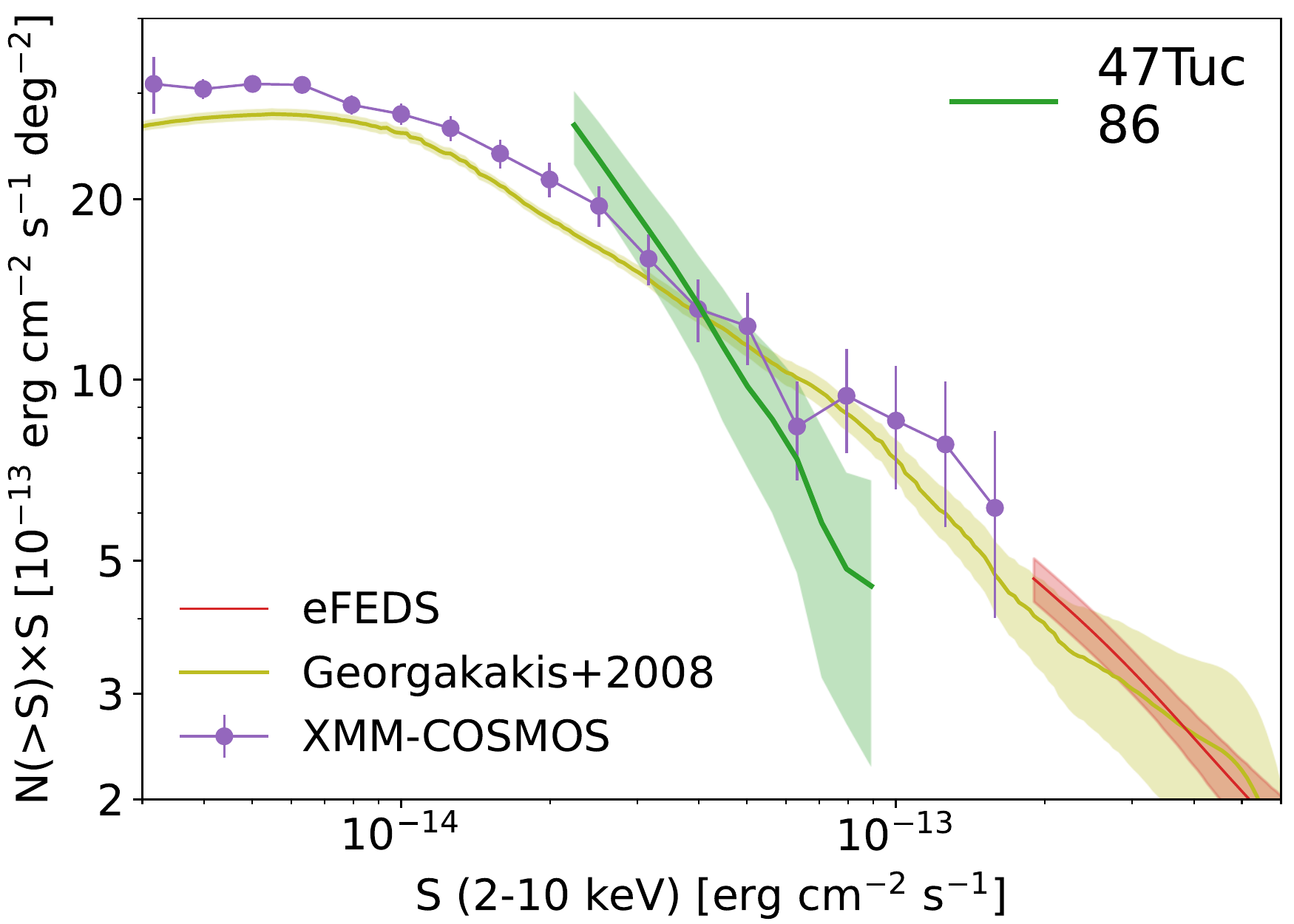}
\includegraphics[width=0.24\textwidth]{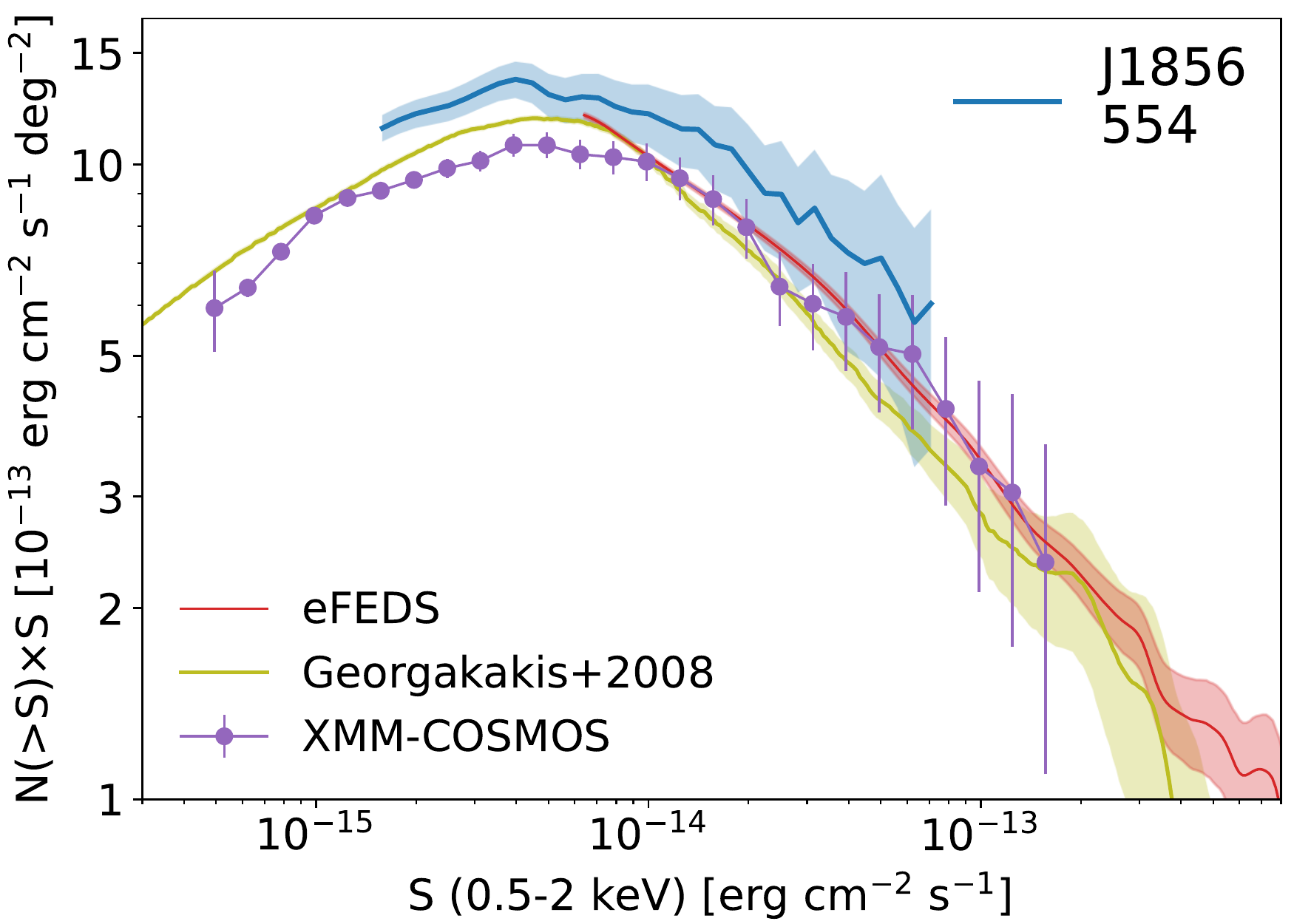}
\includegraphics[width=0.24\textwidth]{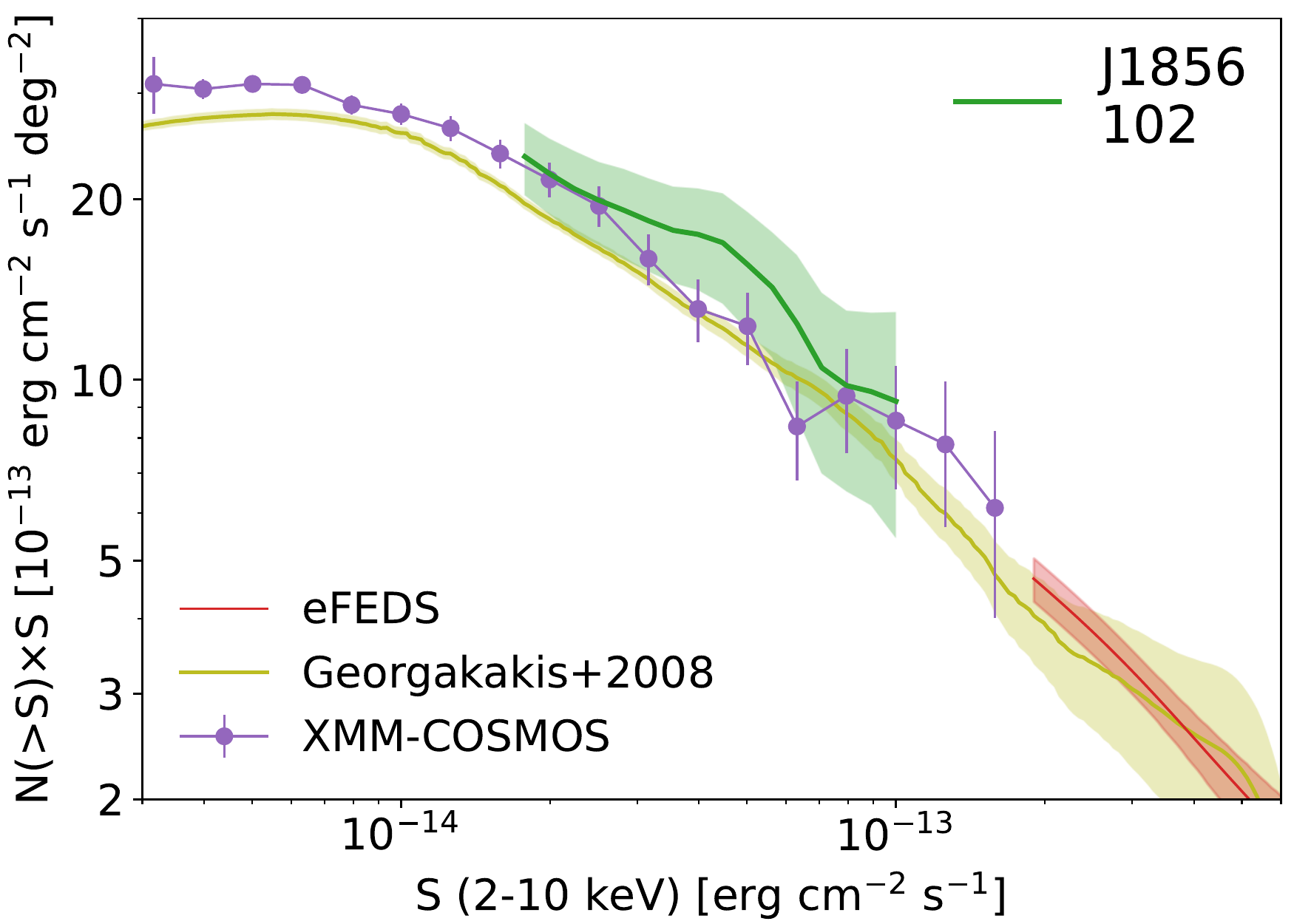}
\includegraphics[width=0.24\textwidth]{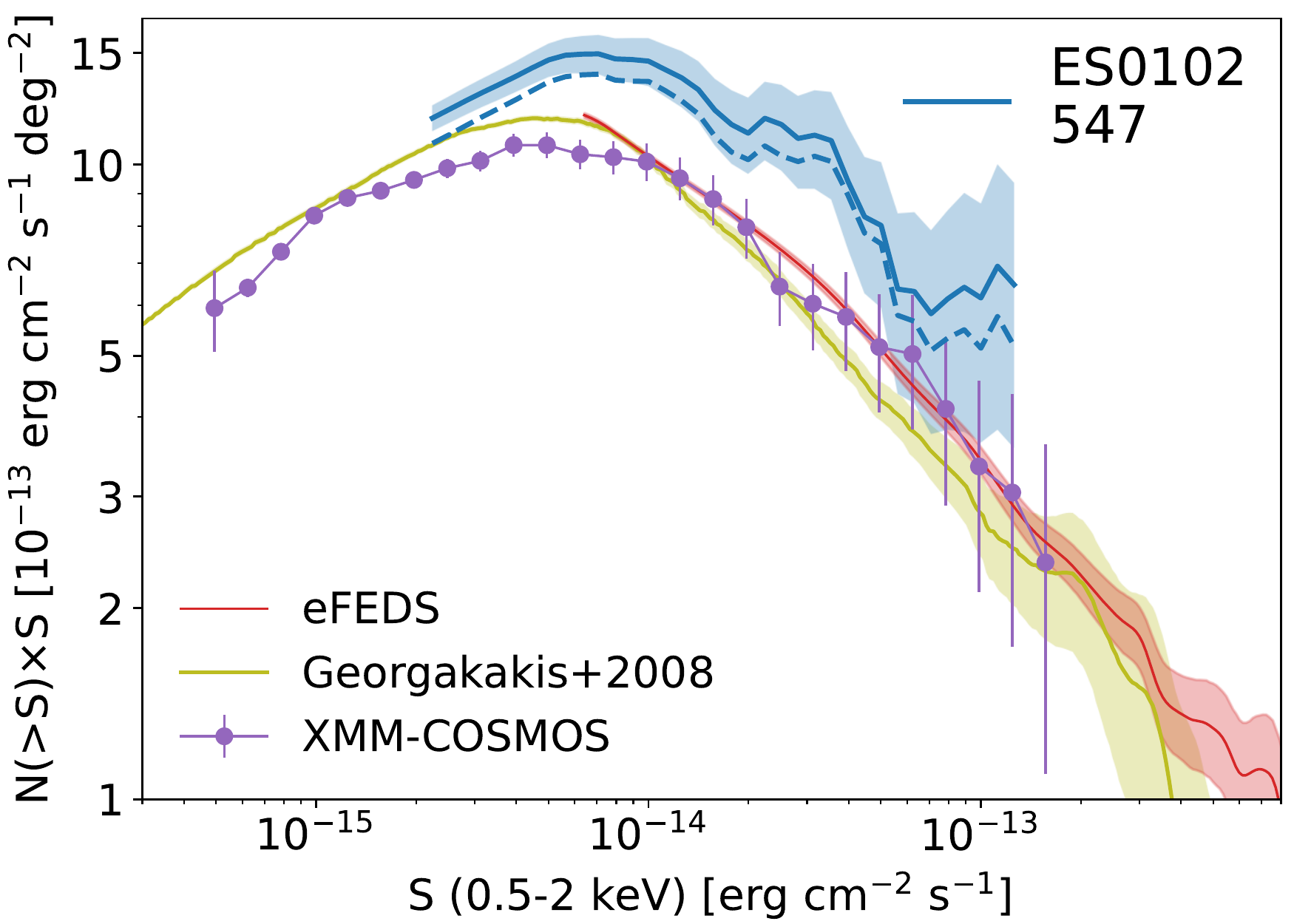}
\includegraphics[width=0.24\textwidth]{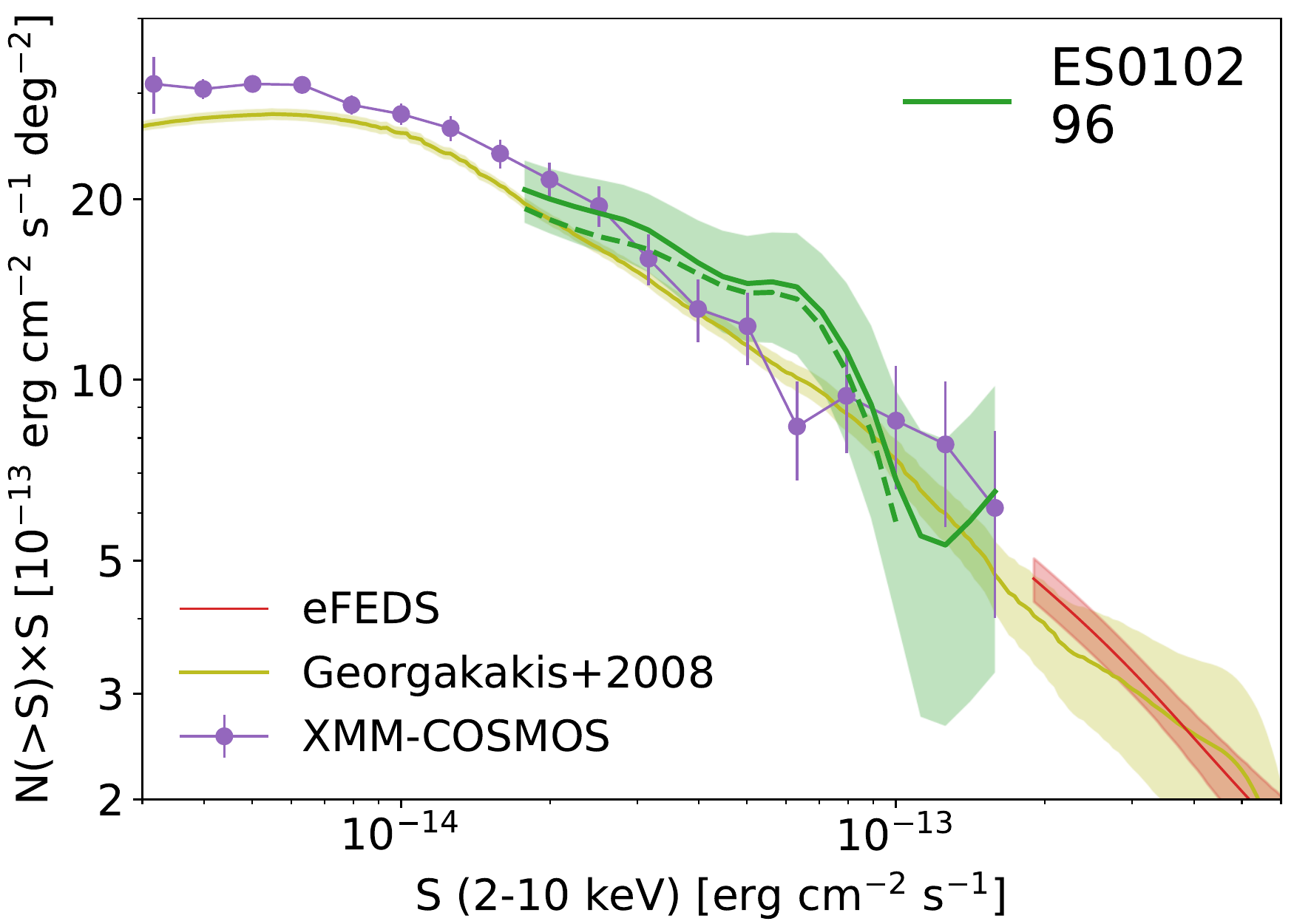}
\includegraphics[width=0.24\textwidth]{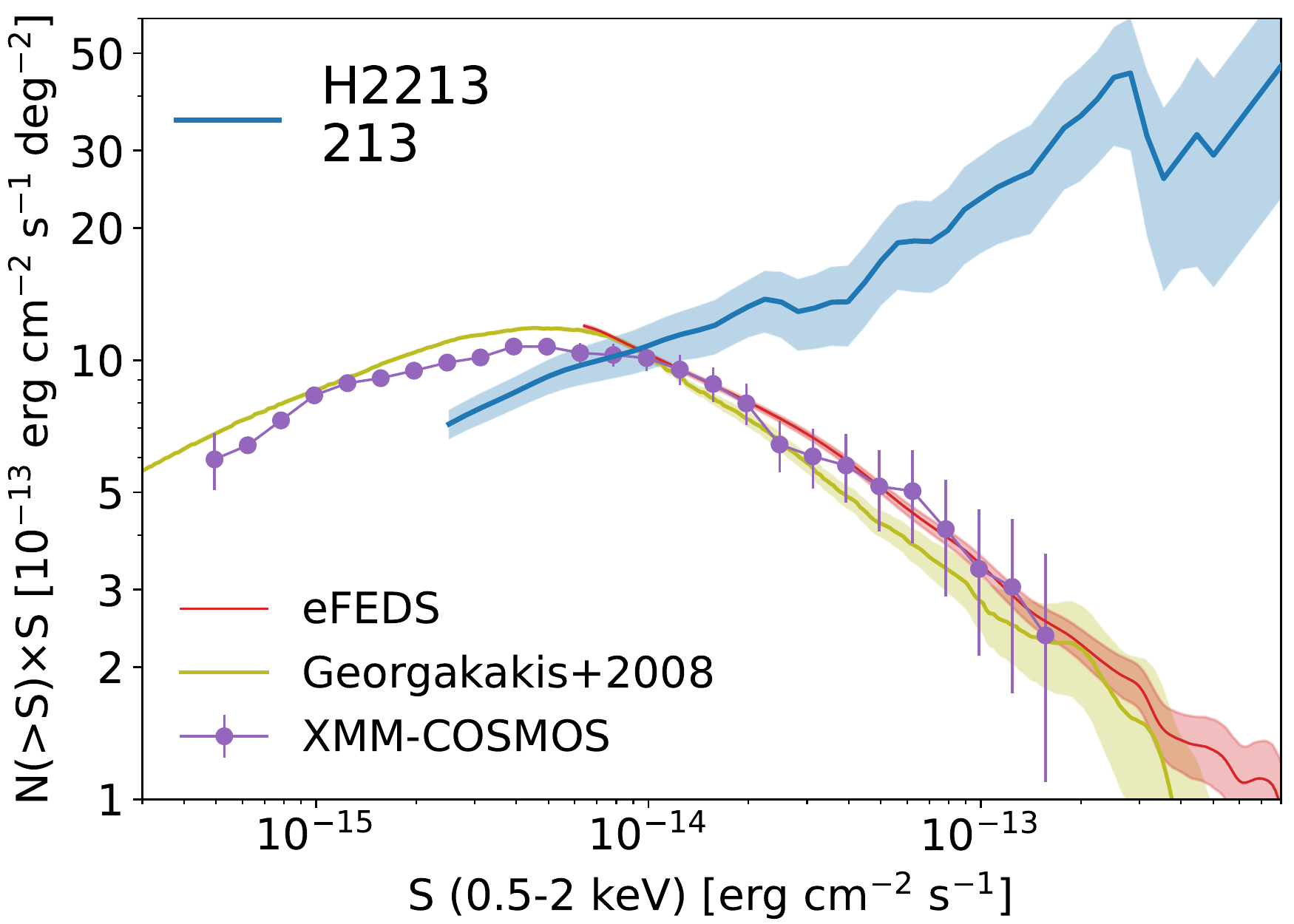}
\includegraphics[width=0.24\textwidth]{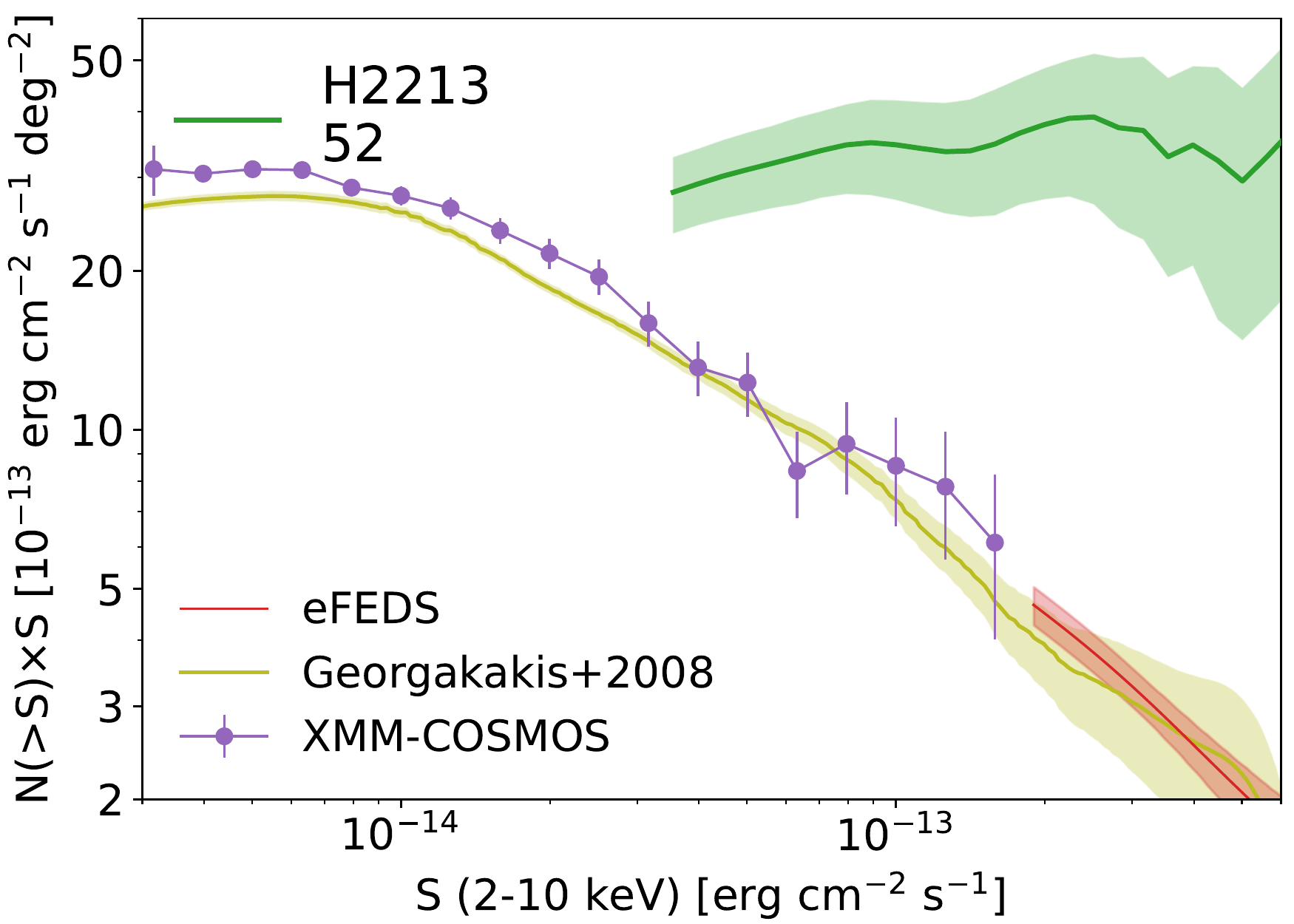}
\includegraphics[width=0.24\textwidth]{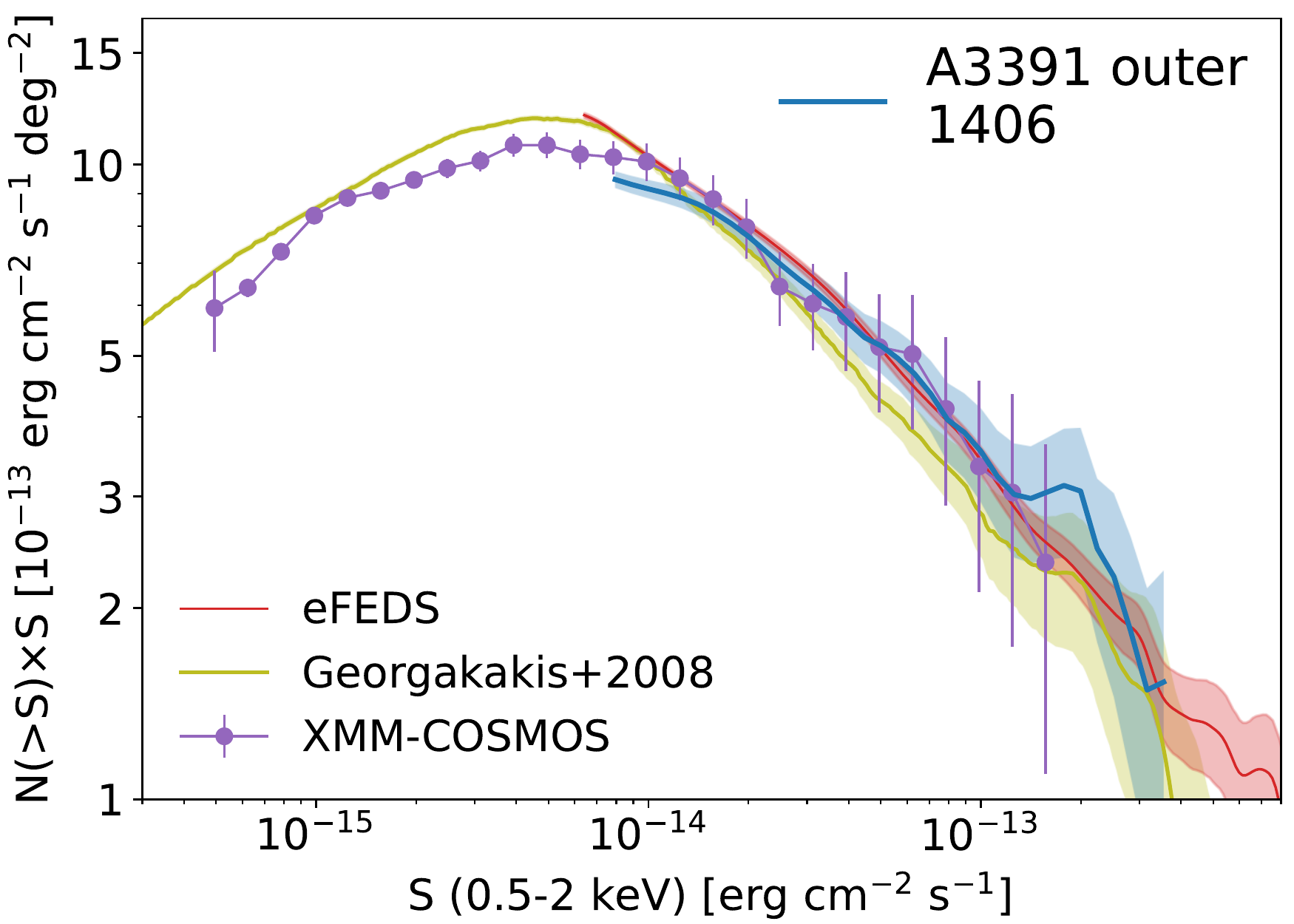}
\includegraphics[width=0.24\textwidth]{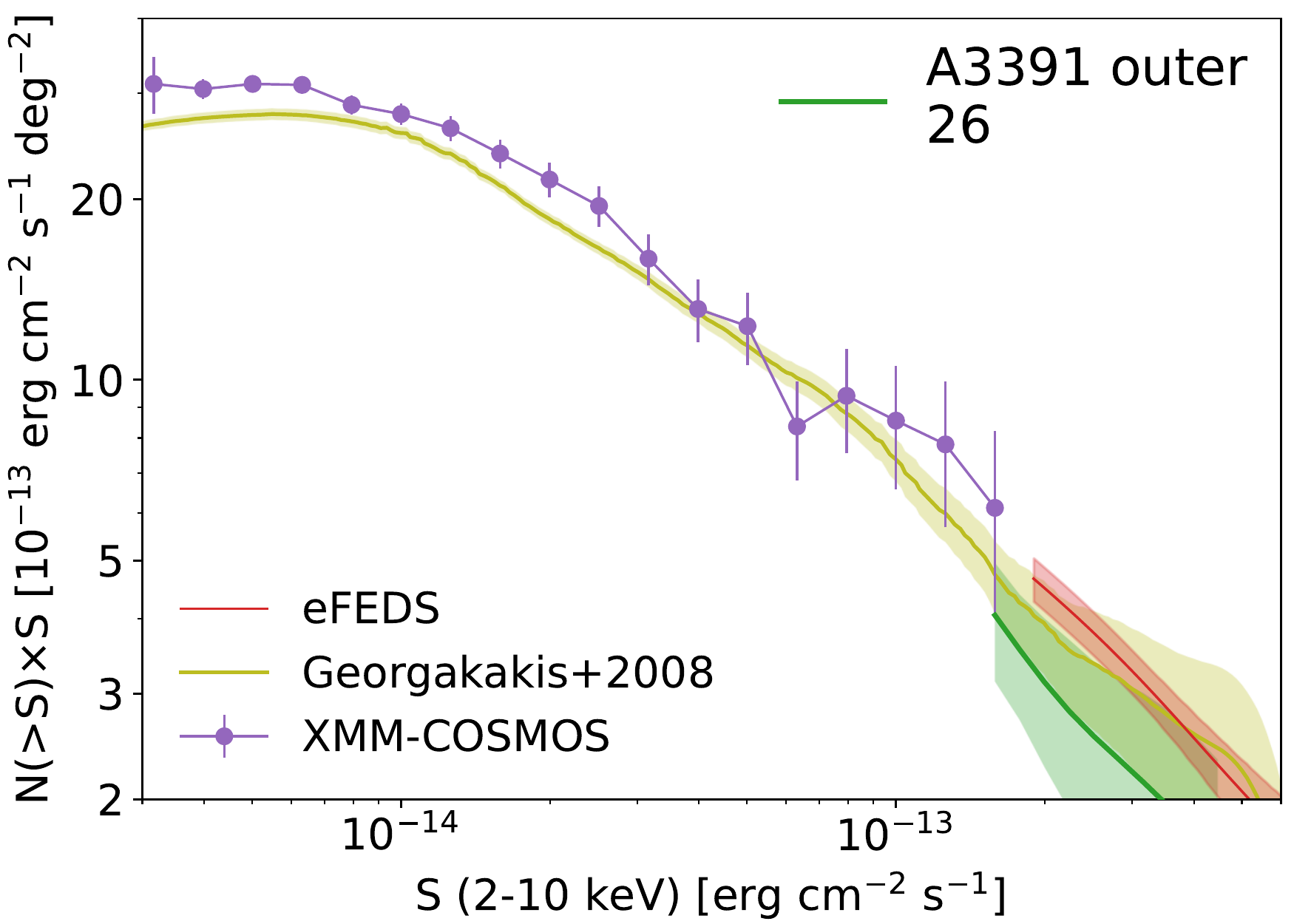}
\includegraphics[width=0.24\textwidth]{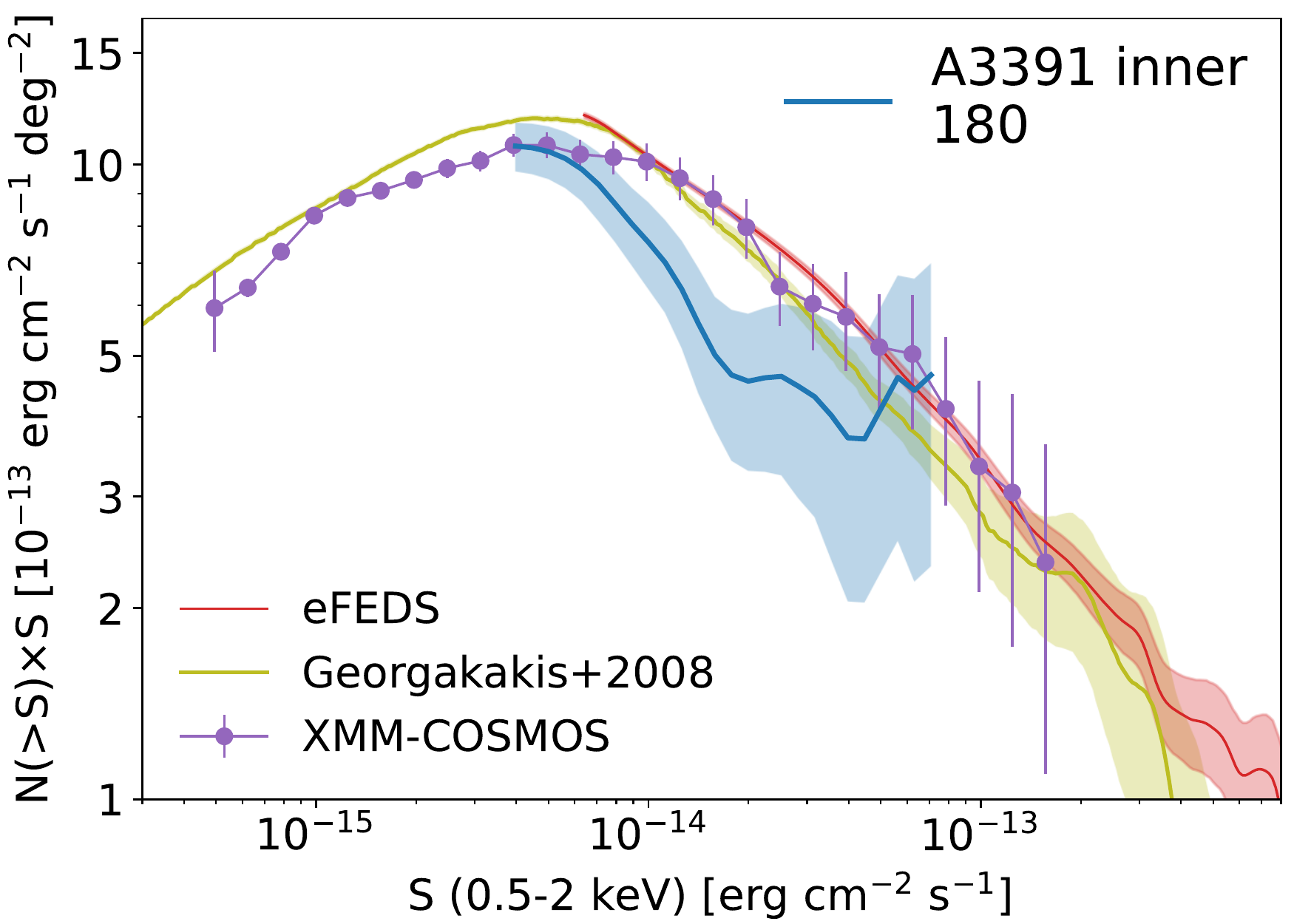}
\includegraphics[width=0.24\textwidth]{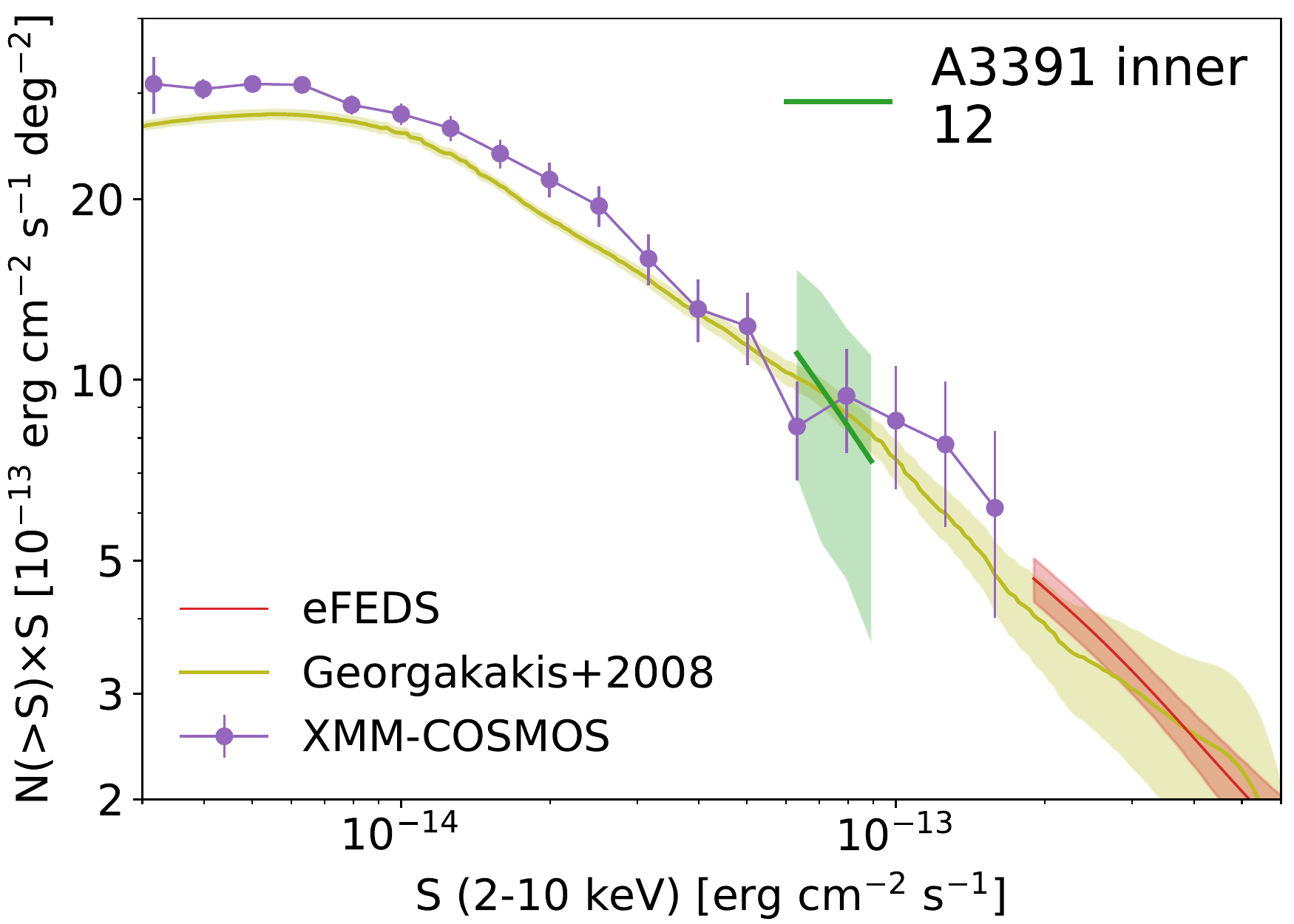}
\caption{\Final{Point} source number counts in the point-source subsurvey region of each field as a function of the 0.5--2~keV (blue) and 2--10~keV (green) Galactic-absorption-corrected fluxes.
  For representation, the cumulative number density (y-axis) is multiplied by the flux limit in units of ($10^{-13}$ erg cm$^{-2}$ s$^{-1}$).
  The lower boundary of the plotted flux range corresponds to 50\% sensitivity (50\% of the total area in the sky coverage curve) in the soft band and 30\% sensitivity in the hard band.
  The shaded region indicates the 1-$\sigma$ uncertainty calculated as the square root of the source number.
  For comparison, we plot the number counts from three previous extragalactic X-ray surveys.
  The A3391 field is divided into two parts, as displayed in the bottom panels.
  For ES0102, the dashed lines indicate the number counts with the SMC stars removed.
\label{fig:logNlogS}}
\end{figure*}

  As displayed in Fig.~\ref{fig:logNlogS}, we calculate the soft- and hard-band number counts of point sources (\texttt{EXT\_LIKE}=0) for each field in the point-source subsurvey region (Sect.~\ref{sec:pntsurvey}) adopting an aperture-photometry likelihood threshold of 10.
Particularly, the field A3391 has three scanning-mode observations covering a large, shallow region and one pointing mode observation making the central region much deeper (Fig.~\ref{fig:images2}). We divide the A3391 selected region into two parts within and outside 28\arcmin from the field center, which is also the center of the pointing-mode observation. We calculate the number counts of the outer region using the three scanning-mode observations and the number counts of the inner region using the pointing-mode observation.
For the outer region, which is shallow and has high Galactic absorption (Fig.~\ref{fig:Exp_NH}), the detection of low-flux sources is more uncertain than in deep, pointing-mode observations because of low counts of photons.
We adopt a higher aperture-photometry likelihood of 15 for this particular region to guarantee high completeness and high purity of the selected subsample.

\section{Data cleaning}
To study the variability of \Final{the} background, we extract a background light curve from each field in a source-free region.
As described in Sect.\ref{sec:bkg3}, using a ``standard detection procedure'', we create a catalog (``PSF-fitting catalog 1'' in Fig.~\ref{fig:flowchart}) through PSF-fitting.
Comparing the best-fit model image (source plus background) of this catalog with the background map adopted in the detection, we select source regions where the former is 10\% higher than the latter.
Then from the point-source subsurvey regions, we mask out the source regions and extract the light curve using the eSASS task \texttt{srctool} adopting a bin size of 10s.
Then we regroup the bins by merging nearby \Final{bins into one, so as to} guarantee at least 100 counts in each bin.
\Edit{The figures of background light curves are presented on the eROSITA EDR website together with the catalog.}

We find that the \eROSITA background is highly stable in each observation. In most cases, the background flux remains constant, having only a few short flares.
Only in one observation (H2213), significant soft proton flares are found.
We adopt a threshold of 54 counts/s/degree$^2$, which is 50\% higher than the flux in the quiescent state, to select the time intervals affected by the flares. These time intervals are removed from the data before source detection.
We also find that at the beginning of a few observations, TM1,2,3,4,6 work in a high-noise state. Generally, this state stops quickly and the cameras enter a quiescent state (e.g., 700174, 700175, 700180, 700181). Only in one observation (700173 of 47Tuc) did the high-noise state last for a few ks.
We removed these high-noise time intervals from all these observations before source detection.

The data used in this work were processed using the EDR version of eSASS. In a few cases, the data show artifacts \Final{such as} hot pixels or bright columns caused by imperfect event filtering, or bright \Final{corners} at soft energies (mostly below 0.3~keV) caused by light leak. We manually removed such artifacts.
\Final{
Some additional manual cleaning are done as follows.
For A3391, a short abnormal time interval was found and removed \citep{Reiprich2021}.
For 1H0707, in the observation 300003, in addition to TM5,6,7, which were active through the whole observation, TM1 and TM2 were active for a short time ($20\%\sim 30\%$ of the observation). We adopted only TM5,6,7.
For 3C390, in one of the four observations (900070), in addition to TM6, which was active through the whole observations, TM5 and TM7 were active for a short time, during which they suffered from light leak. We removed the TM5 and TM7 data of 900070 so that only TM6 was used in all four observations.
For N7793, in observation 300011, we removed TM7 because strong light leak caused abnormal data.
For ES0102, in observation 710000, we removed TM5 because strong light leak caused abnormal data.
}
\end{appendix}
\end{document}